\documentclass[english,3p]{elsarticle}
\usepackage[T1]{fontenc}
\usepackage[latin9]{inputenc}
\usepackage{color}
\usepackage{float}
\usepackage{amsmath}
\usepackage{amssymb}
\usepackage{graphicx}
\usepackage{esint}

\makeatletter

\providecommand{\tabularnewline}{\\}

\usepackage{pdflscape}

\newcommand{\be}{\begin{equation}}\newcommand{\ee}{\end{equation}}\newcommand{\ba}{\begin{array}}\newcommand{\ea}{\end{array}}\newcommand{\bea}{\begin{eqnarray}}\newcommand{\eea}{\end{eqnarray}}

\newcommand{\ket}[1]{|#1\rangle}

\numberwithin{lemma}{section}\numberwithin{corol}{section}\numberwithin{prop}{section}\numberwithin{dfn}{section}\numberwithin{equation}{section}

\usepackage{babel}

\usepackage{babel}

\usepackage{babel}

\usepackage{babel}

\usepackage{babel}

\usepackage{babel}

\usepackage{babel}

\usepackage{babel}

\usepackage{babel}

\makeatother

\usepackage{babel}
\begin{document}

\title{Multiport Impedance Quantization}
\begin{abstract}
\textcolor{black}{With the increase of complexity and coherence of
superconducting systems made using the principles of circuit quantum
electrodynamics, more accurate methods are needed for the characterization,
analysis and optimization of these quantum processors. Here we introduce
a new method of modelling that can be applied to superconducting structures
involving multiple Josephson junctions, high-Q superconducting cavities,
external ports, and voltage sources. Our technique, an extension of
our previous work on single-port structures \citep{brune-quantization-paper},
permits the derivation of system Hamiltonians that are capable of
representing every feature of the physical system over a wide frequency
band and the computation of $T_{1}$ times for qubits. We begin with
a ``black box'' model of the linear and passive part of the system.
Its response is given by its multiport impedance function $\mathbf{Z}_{sim}\left(\omega\right)$,
which can be obtained using a finite-element electormagnetics simulator.
The ports of this black box are defined by the terminal pairs of Josephson
junctions, voltage sources, and $50\Omega$ connectors to high-frequency
lines. We fit $\mathbf{Z}_{sim}\left(\omega\right)$ to a positive-real
(PR) multiport impedance matrix $\mathbf{Z}\left(s\right)$, a function
of the complex Laplace variable $s$.  We then use state-space techniques
to synthesize a finite electric circuit admitting exactly the same
impedance $\mathbf{Z}\left(s\right)$ across its ports; the PR property
ensures the existence of this finite physical circuit. We compare
the performance of state-space algorithms to classical frequency domain
methods, justifying their superiority in numerical stability. The
Hamiltonian of the multiport model circuit is obtained by using existing
lumped element circuit quantization formalisms \citep{BKD,Burkard}.
Due to the presence of ideal transformers in the model circuit, these
quantization methods must be extended, requiring the introduction
of an extension of the Kirchhoff voltage and current laws. }
\end{abstract}

\author{Firat Solgun and David P. DiVincenzo}

\maketitle
\pagebreak{}\tableofcontents{}\pagebreak{}

\section{Introduction}

Superconducting electronics is one of the most promising candidates
for the realization of the hardware of a quantum computer. Small scale
superconducting quantum processors have been demonstrated using coplanar
waveguide (CPW) circuits and 3D microwave cavities \citep{blais,3D,RigettiCu}.
They typically consist of Josephson junctions coupled to cavity resonators,
which in turn are coupled to microwave feedlines that provide the
functions of readout and quantum gate operations. Josephson junctions
are lossless nonlinear circuit elements providing the anharmonicity
needed to have a quantum energy spectrum in which two unique low-lying
energy levels can be picked out to define the qubit.

From a classical point of view, superconducting quantum processors
are microwave systems having multiple resonant modes. The significant
increase in quality factors (Q factors) of both qubit and cavity modes
in the last decade requires highly accurate models for the design
and optimization of those systems. \textcolor{black}{Previous approaches
to model multi-mode superconducting qubit systems encountered problems
in estimating loss rates \citep{Blaisunpub}.} \textcolor{black}{Our
work has been stimulated by a recent study \citep{Nigg}} which introduced
the ``black box\textquotedbl{} concept for separating the modeling
of these systems into the linear, passive, distributed part, connected
via ports to the nonlinear or active parts of the system. We have
improved upon the circuit modeling used in \textcolor{black}{\citep{Nigg,PMC}},
which is based on the so-called Foster approach \textcolor{black}{\citep{Foster}}
which is not fully justified for lossy systems.

In our first work in this direction \citep{brune-quantization-paper}
we have introduced a method for the accurate characterization of superconducting
microwave circuits involving a single Josephson junction connected
to a one-port black box. We have shown that the ad-hoc extension of
Foster synthesis can be replaced by the exact impedance synthesis
technique of Brune \citep{Brune}, which permits the full response
of the black box, both reactive and lossy, to be matched to very high
accuracy over a large frequency range. Applying previous formalisms
\citep{BKD,Burkard} developed for the quantization of lumped element
circuits, we have shown that one can derive highly accurate Hamiltonians
for single qubit systems. We have also estimated relaxation rates
of the qubits, showing that the results are systematically improved
compared with \citep{Nigg}.

\begin{figure}[H]
\begin{centering}
\includegraphics[scale=0.4]{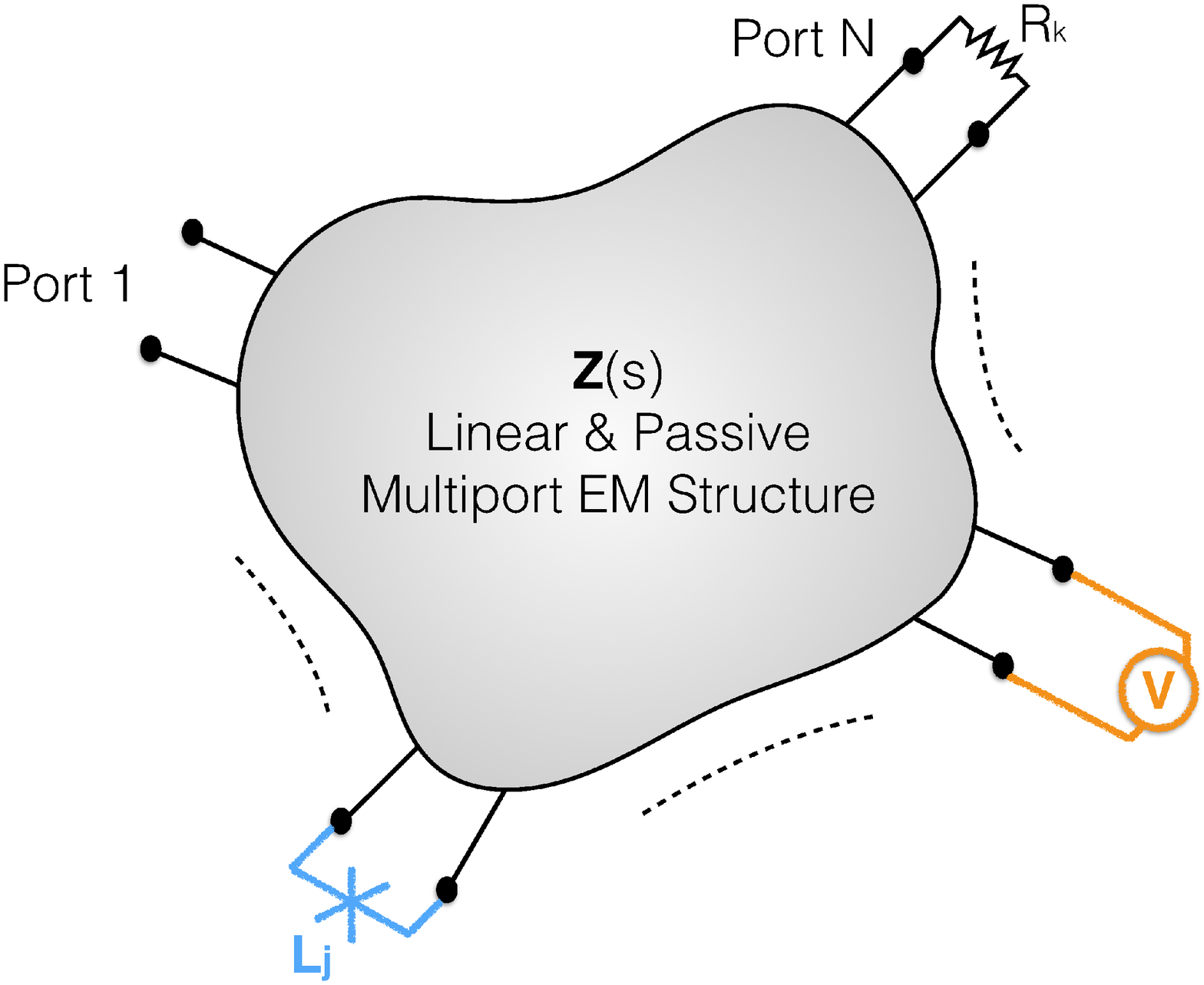} 
\par\end{centering}

\centering{}\caption{\label{fig:Multiport-Blackbox-Impedance.}Multiport Blackbox Impedance.
The port electromagnetic response of a linear and passive structure
is represented by a multiport blackbox having impedance $\mathbf{Z}\left(s\right)$.
The $N$ ports of this black box can be shunted by multiple Josephson
junctions $L_{j}$'s, voltage sources $V$ or resistors $R_{k}$.
In this paper we present systematic techniques for deriving a highly
accurate, lumped-circuit description of this black box response, and
we obtain the system Hamiltonian and relaxation rates for the full
system.}
\end{figure}

In the present paper, we extend our method to handle the task of modeling
and quantizing black-box systems involving multiple ports, with multiple
Josephson junctions along with voltage sources and external impedances.
We again follow the black box approach by representing the linear
and passive part of the microwave circuit by a multiport impedance
$\mathbf{Z}\left(s\right)$ ($s$ is the complex Laplace variable)
as shown in Fig. \eqref{fig:Multiport-Blackbox-Impedance.} with the
Josephson junctions, voltage sources and resistors shunting the ports
of this multiport structure. The first step in the modeling is to
perform a synthesis of the multiport impedance matrix. {\em Synthesis}
is a technical term from theoretical electrical engineering, meaning
the systematic determination of a finite lumped element electrical
circuit which admits the impedance matrix $\mathbf{Z}\left(s\right)$
across its ports. Often, synthesis is to be done approximately, or
within certain design specification. Here the goal is simple: the
synthesis should be {\em exact}. With certain caveats, we do indeed
achieve such an exact synthesis. With the synthesized circuit, we
can then apply lumped-element circuit quantization formalisms \citep{BKD,Burkard}
(with some new adaptations to be described in this paper) to derive
a Hamiltonian for the circuit and to compute relaxation rates.

To obtain the impedance $\mathbf{Z}\left(s\right)$ of a physical
structure, one can imagine a variety of approaches. Ideally this function
would be extracted from experiment, but it is hard to probe (e.g.,
by a spectrum analyser) the response at the Josephson-junction ports,
which are inaccessible to external noninvasive contacting. But it
is felt that electromagnetic response calculators are quite reliable
for reliably simulating this response. The simulation involves the
drawing of a 3D model of the structure to be represented in the blackbox
in Fig. \eqref{fig:Multiport-Blackbox-Impedance.} using a finite-element
simulator such as HFSS \citep{HFSS}. In such programs, ports may
be conveniently defined across the metal nanostructures defining the
Josephson junctions, or where the coaxial inputs enter the cavities
as shown in Fig. \eqref{fig:Cavity-Fig1}. The simulated impedance
response $\mathbf{Z}_{sim}\left(\omega\right)$ of the system is then
computed by solving Maxwell's equations at a discrete set of frequencies
over a finite frequency band. If $N$ ports are defined $\mathbf{Z}_{sim}\left(\omega\right)$
will be a $N\times N$ matrix, with entries being complex numbers
at every frequency.

Next we fit $\mathbf{Z}_{sim}\left(\omega\right)$ to an impedance
matrix $\mathbf{Z}\left(s\right)$ with a finite number of poles $N_{P}$.
For that purpose, an extensively developed formalism and software
package known as Vector Fitting (VF) is available, and we make use
of it here \citep{Vector Fitting} . VF takes the number of poles
$N_{P}$ as input and outputs an impedance matrix $\mathbf{Z}\left(s\right)$
with $N_{P}$ poles as in Eq. \eqref{eq:Vector-Fit} which is a least-squares
fit to $\mathbf{Z}_{sim}\left(\omega\right)$ over the given frequency
band. The $\mathbf{R}_{k}$'s are the residues of the poles at finite
(complex) frequency $s_{k}$, and $\mathbf{E}$ is the residue at
infinity. Passivity of $\mathbf{Z}\left(s\right)$ is enforced by
a subroutine of VF \citep{VF Passivity Enforcement}.

\begin{equation}
\mathbf{Z}\left(s\right)=\underset{k=1}{\overset{N_{P}}{\sum}}\frac{\mathbf{R}_{k}}{s-s_{k}}+\mathbf{D}+\mathbf{E}s\label{eq:Vector-Fit}
\end{equation}

$\mathbf{Z}\left(s\right)$ must be Positive-Real (PR) \citep{Brune,Newcomb}
for a finite physical lumped element circuit to exist having the exact
multiport impedance $\mathbf{Z}\left(s\right)$ across its ports.
We have recapitulated the PR conditions in the $s$-domain for a one-port
impedance function $z\left(s\right)$ in \citep{brune-quantization-paper}.
In Section \eqref{sub:Positive-Real-Property-in-SS} we state the
PR conditions in {\em state space} for a multiport impedance $\mathbf{Z}\left(s\right)$.
$\mathbf{Z}\left(s\right)$ generated by VF is PR since VF produces
stable poles and enforces the passivity.

Below we start by revisiting the one-port problem from the state-space
perspective. The state-space approach is central to modern circuit
modeling and to control theory in a wide range of engineering disciplines\textcolor{cyan}{{}}
\citep{Anderson-Vongpanitlerd,Antoulas}. In short, this approach
represents the dynamics of the system in the time domain, identifying
a sufficient set of variables so that the full dynamics is described
by differential equations that are first-order in time. In Section
\eqref{sec:State-Space-Formalism} we give an introduction to the
state-space theory, keeping it at a level that will be enough for
our purposes.\textcolor{cyan}{{} } Working in state space allows for
numerically more stable circuit synthesis algorithms and leads to
a straightforward algorithm for the generation of multiport circuits
treatable in existing circuit quantization formalisms \citep{BKD,Burkard},
when extended in the way described shortly. In Section \eqref{sub:Positive-Real-Property-in-SS}
we re-state the PR conditions in state space and in Section \eqref{sub:Brune's-algorithm-state-space-1-port}
we describe the extension of Brune's method of circuit synthesis in
state space \citep{Anderson-Moylan-1975} for one ports. We then show
how to quantize the one-port state-space Brune circuit in Section
\eqref{sub:Quantization-state-space-Brune-circuit-1-port} with the
help of an ``effective Kirchhoff'' technique which eliminates ideal
transformers from the circuit equations, leading to equations that
can be thought of as generalizations of the Kirchhoff current and
voltage laws. In the main part of this paper in Section \eqref{sec:Multiport-Brune's-method},
we extend our analysis to multiport circuits. We describe the multiport
Brune algorithm in state space in Section \eqref{sec:Multiport-Brune-Algorithm}
and apply the effective Kirchhoff method to quantize the multiport
Brune circuit and compute loss rates of qubits in Section \eqref{sec:Quantization-of-the-multiport-Brune-circuit}.
We consider one-by-one the cases of the multiport Brune circuit being
shunted by Josephson junctions in Section \eqref{sec:Quantization-of-the-multiport-Brune-circuit},
external resistors in Section \eqref{sub:Resistors-shunting-the-ports}
and voltage sources in Section \eqref{sub:Voltage-sources-shunting-the-ports}.
We show the explicit application of our method on a 2-port 1-stage
example circuit in Section \eqref{sub:2-port-1-stage-generic-cct-example}
and study a numerical example in Section \eqref{sub:3D-Transmon-example}
corresponding to a realistic 3D transmon setup, which is a 3-port
circuit for which the Brune synthesis gives a 12-stage circuit.

\section{One-Port Brune Circuit Quantization in State Space\label{sec:Quantization-of-the-Brune-circuit}}

\subsection{Introduction}

In Section II of \citep{brune-quantization-paper} we presented Brune's
algorithm in $s$-domain (or frequency domain), $s$ being the complex
Laplace variable. In $s$-domain the impedance is given as a rational
function of $s$ (or as a rational matrix for multiports). \textcolor{black}{The
Brune algorithm in the $s$-domain requires a partial-fraction expansion
for the determination of circuit parameters.} Partial fraction expansion
is an ill-conditioned operation since it requires finding roots of
polynomials. Root finding is a numerically unstable problem: the roots
become very sensitive to small perturbations in the coefficients as
the degree of the polynomial increases. This is illustrated by Wilkinson's
polynomial \citep{Wilkinson}. The problem becomes even more severe
if one wants to apply multiport generalizations of synthesis algorithms
\citep{Newcomb} since the degrees of polynomials increase with the
number of ports. Synthesis methods given in $s$-domain are usually
referred to as classical network synthesis. See \citep{Newcomb} for
a comprehensive summary of classical synthesis algorithms. Classical
synthesis methods appeared first in the historical development of
the subject and played a key role in building the theory and expressing
synthesis procedures. However they are not suitable for computer implementation
due to the stability issues mentioned above.

The situation however is not so hopeless. Network synthesis algorithms
can also be expressed in the state space. The state-space approach
can be seen as a reformulation of the synthesis problem in the time-domain.
See \citep{Anderson-Vongpanitlerd} for a comprehensive coverage of
network theory in state space. Most synthesis methods reexpressed
in state space requires the solution of a type of Riccati equation
\citep{Anderson-Vongpanitlerd}. Solving the Riccati equation when
the system's poles approach the imaginary axis is a numerically ``hard''
problem. Since superconducting circuits have very little loss we usually
encounter hard instances of Riccati equations in our models. Brune's
algorithm expressed in state space provides a method for solving such
hard Riccati equations. Reducing the complexity of the problem by
a small amount at each step it avoids numerical instabilities appearing
in more direct methods which try to complete the synthesis in fewer
steps. See \citep{Anderson-Riccati-Brune-1999} for a discussion of
how Brune's method in state space might help in solving hard Riccati
equations. In the following section we briefly review state-space
formalism and present Brune's impedance synthesis algorithm expressed
in state-space terms.

In Section \eqref{sub:Quantization-state-space-Brune-circuit-1-port}
we introduce a new technique for the quantization of the one-port
Brune circuit with ideal transformers. We call this new technique
the ``effective Kirchhoff'' method. We will first see how much the
effective Kirchhoff method simplifies the analysis of the Brune circuit
presented in \citep{brune-quantization-paper}. However the full power
of the effective Kirchhoff method will be apparent in the last section
of this paper when we will apply it to quantize the multiport Brune
circuit. The earliest appearance (and the only one that we could find)
of this technique in the literature is in \citep{ideal-transformers-thesis}
where ideal transformer variables are eliminated from mesh equations
to compute some effective mesh impedance matrices.

\subsection{\label{sec:State-Space-Formalism}State-Space Formalism}

In the state-space formalism (see Chapter 3 of \citep{Anderson-Vongpanitlerd}
for more details on the state-space formalism in the context of network
synthesis or \citep{Antoulas} in the context of dimensionality reduction
theory) the state of a linear time-invariant system with $m$ inputs
and $n$ outputs is given by a real vector $\mathbf{x}$ of length
$N$. The time evolution of the state is described by a first-order
differential equation

\begin{equation}
\mathbf{\dot{x}}=\mathbf{A}\mathbf{x}+\mathbf{B}\mathbf{u}\label{eq:state-space-time-evolution}
\end{equation}
where $\mathbf{u}$ is the input vector of length $m$, $\mathbf{A}$
a $\left(N\times N\right)$ matrix and $\mathbf{B}$ a $\left(N\times m\right)$
matrix. The output vector $\mathbf{y}$ is related to the input vector
$\mathbf{u}$ by the following algebraic relation which involves also
the state vector $\mathbf{x}$

\begin{equation}
\mathbf{y}=\mathbf{C}\mathbf{x}+\mathbf{D}\mathbf{u}\label{eq:state-space-input-output-relation}
\end{equation}
The output vector $\mathbf{y}$ is of length $n$, $\mathbf{C}$ is
a $\left(n\times N\right)$ matrix and $\mathbf{D}$ a $\left(n\times m\right)$
matrix. If $\mathbf{u}$ holds the currents and $\mathbf{y}$ holds
voltages at the ports of a network then $m=n$ and the multiport impedance
is given by

\begin{equation}
\mathbf{Z}\left(s\right)=\mathbf{D}+\mathbf{C}\left(s\mathbf{I}-\mathbf{A}\right)^{-1}\mathbf{B}\label{eq:state-space-to-impedance-in-s-domain}
\end{equation}
We will only consider real realizations here such that the matrices
$\left\{ \mathbf{A},\mathbf{B},\mathbf{C},\mathbf{D}\right\} $ are
all real.

Now let's assume that we transform the state $\mathbf{x}$ by a non-singular
transformation $\mathbf{T}$ such that the new state $\mathbf{x}_{1}$
is given by

\begin{equation}
\mathbf{x}_{1}=\mathbf{T}\mathbf{x}
\end{equation}

Then using Eqs. \eqref{eq:state-space-time-evolution} and \eqref{eq:state-space-input-output-relation}
the state-space description for the state $\mathbf{x}_{1}$ is given
by

\begin{eqnarray}
\mathbf{\dot{x}}_{1} & = & \mathbf{A}_{1}\mathbf{x}_{1}+\mathbf{B}_{1}\mathbf{u}\\
\mathbf{y} & = & \mathbf{C}_{1}\mathbf{x}_{1}+\mathbf{D}_{1}\mathbf{u}
\end{eqnarray}
where

\begin{eqnarray}
\mathbf{A}_{1} & = & \mathbf{T}\mathbf{A}\mathbf{T}^{-1}\label{eq:realization-new-coordinates-1}\\
\mathbf{B}_{1} & = & \mathbf{T}\mathbf{B}\\
\mathbf{C}_{1} & = & \mathbf{C}\mathbf{T}^{-1}\\
\mathbf{D}_{1} & = & \mathbf{D}\label{eq:realization-new-coordinates-4}
\end{eqnarray}
The important point to note here is that the input-output relationship
is unchanged that is $\left\{ \mathbf{A}_{1},\mathbf{B}_{1},\mathbf{C}_{1},\mathbf{D}_{1}\right\} $
is another state-space realization for the impedance $\mathbf{Z}\left(s\right)$,
if $\mathbf{u}$ and $\mathbf{y}$ are the currents and voltages at
the ports of the network corresponding to $\mathbf{Z}\left(s\right)$,
respectively. To show this let $\mathbf{Z}_{1}\left(s\right)$ be
the impedance corresponding to the realization $\left\{ \mathbf{A}_{1},\mathbf{B}_{1},\mathbf{C}_{1},\mathbf{D}_{1}\right\} $
then by Eq. \eqref{eq:state-space-to-impedance-in-s-domain}

\begin{eqnarray}
\mathbf{Z}_{1}\left(s\right) & = & \mathbf{D}_{1}+\mathbf{C}_{1}\left(s\mathbf{I}-\mathbf{A}_{1}\right)^{-1}\mathbf{B}_{1}\\
 & = & \mathbf{D}+\mathbf{C}\mathbf{T}^{-1}\left(s\mathbf{I}-\mathbf{T}\mathbf{A}\mathbf{T}^{-1}\right)^{-1}\mathbf{T}\mathbf{B}\\
 & = & \mathbf{D}+\mathbf{C}\left(s\mathbf{I}-\mathbf{A}\right)^{-1}\mathbf{B}\\
 & = & \mathbf{Z}\left(s\right)
\end{eqnarray}
where in the second line above we used Eqs. \eqref{eq:realization-new-coordinates-1}-\eqref{eq:realization-new-coordinates-4}.
For more details see Theorem (3.3.9) of \citep{Anderson-Vongpanitlerd}.

\subsection{Minimal Realizations}

Given the impedance $\mathbf{Z}\left(s\right)$, a fundamental question
in state-space theory is how to find a set of real matrices $\left\{ \mathbf{A,B,C,D}\right\} $
such that

\begin{equation}
\mathbf{Z}\left(s\right)=\mathbf{D}+\mathbf{C}\left(s\mathbf{I}-\mathbf{A}\right)^{-1}\mathbf{B}\label{eq:impedance-computation-state-space}
\end{equation}
is satisfied with the dimension $N$ of the state space being minimum.
In state-space theory minimal realizations are defined in a more abstract
way. The set $\left\{ \mathbf{A,B,C,D}\right\} $ is called a \emph{minimal
realization} for the impedance $\mathbf{Z}\left(s\right)$ if $\left[\mathbf{A},\mathbf{B}\right]$
is \emph{completely controllable} and $\left[\mathbf{A},\mathbf{C}\right]$
is \emph{completely observable}.

Given the time evolution equation

\begin{equation}
\mathbf{\dot{x}}=\mathbf{A}\mathbf{x}+\mathbf{B}\mathbf{u}\label{eq:time-evolution-complete-controllability}
\end{equation}
$\left[\mathbf{A},\mathbf{B}\right]$ is said to be \emph{completely
controllable}, if given the system is at state $\mathbf{x}\left(t_{0}\right)$
at time $t_{0}$, there exists a control $\mathbf{u}\left(t\right)$
defined over $\left[t_{0},t_{1}\right]$ such that the system can
be brought to the zero state $\mathbf{x}\left(t_{1}\right)=0$ at
time $t_{1}$ under the driven time evolution in Eq. \eqref{eq:time-evolution-complete-controllability}.

\textcolor{black}{Given} the state-space equations

\begin{equation}
\mathbf{\dot{x}}=\mathbf{A}\mathbf{x}+\mathbf{B}\mathbf{u}
\end{equation}

\begin{equation}
\mathbf{y}=\mathbf{C}\mathbf{x}
\end{equation}
The pair $\left[\mathbf{A},\mathbf{C}\right]$ is said to be \emph{completely
observable} if, given the input and ouput functions $\mathbf{u}\left(t\right)$
and $\mathbf{y}\left(t\right)$ over an interval $\left[t_{0},t_{1}\right]$
it is possible to determine $\mathbf{x}\left(t_{0}\right)$ uniquely.

For more details on the properties of state-space realizations and
the minimal realizations we refer the reader to Chapters 3.3 and 3.4
of \citep{Anderson-Vongpanitlerd}.

For a scalar impedance function $z\left(s\right)$ the problem of
finding a minimal state-space realization is relatively easy to answer.
Without loss of generality we can assume that $z\left(\infty\right)=0$.
Let $z\left(s\right)$ be given as

\begin{equation}
z\left(s\right)=\frac{b_{n}s^{n-1}+\ldots+b_{2}s+b_{1}}{s^{n}+a_{n}s^{n-1}+\ldots+a_{2}s+a_{1}}\label{eq:scalar-rational-impedance}
\end{equation}
We assume that the numerator and the denominator polynomials in Eq.
\eqref{eq:scalar-rational-impedance} have no common factors. If we
define $\mathbf{A}$ in \emph{companion matrix} form as

\begin{equation}
\mathbf{A}=\left(\begin{array}{ccccc}
0 & 1 & 0 & \cdots & 0\\
0 & 0 & 1 &  & 0\\
\vdots &  &  & \ddots & \vdots\\
0 & 0 & 0 &  & 1\\
-a_{1} & -a_{2} & -a_{3} & \cdots & -a_{n}
\end{array}\right)\label{eq:A-companion-matrix}
\end{equation}
together with the following definitions for $\mathbf{B}$ and $\mathbf{C}$

\begin{equation}
\mathbf{B}=\left(\begin{array}{c}
0\\
0\\
\vdots\\
0\\
1
\end{array}\right)\label{eq:B-definition-scalar}
\end{equation}

\begin{equation}
\mathbf{C}^{T}=\left(\begin{array}{c}
b_{1}\\
b_{2}\\
\vdots\\
b_{n-1}\\
b_{n}
\end{array}\right)\label{eq:C-definition-scalar}
\end{equation}
then $\left\{ \mathbf{A,B,C}\right\} $ is a minimal realization for
$z\left(s\right)$. This is equivalent to $[\mathbf{A,B}]$ being
completely controllable and $[\mathbf{A,C}]$ completely observable.

Finding a minimal realization corresponding to a multiport impedance
matrix $\mathbf{Z}\left(s\right)$ is more involved and there are
many procedures to find one. We will follow a physically motivated
approach to find a minimal realization. The fitted impedance $\mathbf{Z}\left(s\right)$
obtained by VF in Eq. \eqref{eq:Vector-Fit} contains most of the
time numerical noise which makes its residue matrices $\mathbf{R}_{k}$
full rank. This is generically unphysical since a full rank residue
matrix would correspond to a degenerate mode at a finite frequency.
Finding a minimal representation for such an impedance would introduce
unphysical degrees of freedom. To cure this problem we will apply
the ``compacting'' technique described in \citep{Compacting} to
reduce the rank of residue matrices and to obtain a minimal realization
for our models.

Model-order reduction techniques are also used to reduce the dimension
of non-minimal realizations. In applying order reduction procedures
one should make sure that the passivity and reciprocity of the system
is preserved \citep{Antoulas,IC-Interconnect-analysis}.

\subsection{Positive-Real Property in the State Space \label{sub:Positive-Real-Property-in-SS}}

Given an impedance function $z\left(s\right)$ (or an impedance matrix
$\mathbf{Z}\left(s\right)$) an important question is whether it corresponds
to a finite physical circuit. For the one-port case Brune answered
this question by introducing in \citep{Brune} the so called ``Positive
Real (PR)'' conditions. There exists a finite physical circuit having
the impedance $z\left(s\right)$ across its terminals if $z\left(s\right)$
satisfies the PR conditions. Here we state PR conditions given in
\citep{brune-quantization-paper} in frequency domain for a one-port
impedance function in the state-space language for the most general
case of a multiport impedance matrix $\mathbf{Z}\left(s\right)$.

\emph{Positive Real Lemma}

Given an $m\times m$ impedance matrix $\mathbf{Z}\left(s\right)$
corresponding to an $m$-port network with $\mathbf{Z}\left(\infty\right)<\infty$
and with a minimal realization $\left\{ \mathbf{A,B,C,D}\right\} $.
$\mathbf{Z}\left(s\right)$ is positive real if and only if there
exist real matrices $\mathbf{P}$, $\mathbf{L}$ and $\mathbf{W}_{0}$
with $\mathbf{P}$ being positive definite symmetric satisfying

\begin{eqnarray}
\mathbf{P}\mathbf{A}+\mathbf{A^{T}}\mathbf{P} & = & -\mathbf{LL^{T}}\label{eq:Positive-Real-Lemma}\\
\mathbf{PB} & = & \mathbf{C^{T}-LW_{0}}\\
\mathbf{W_{0}^{T}W_{0}} & = & \mathbf{D+D^{T}}\label{eq:Positive-Real-Lemma-3}
\end{eqnarray}

The Positive Real Lemma stated above goes also under the name ``Kalman
\textendash{} Yakubovich \textendash{} Popov Lemma'' in control theory
literature which refers to names involved in its development \citep{Yakubovic,Kalman-Positive-Real-Lemma-1,Kalman-Positive-Real-Lemma-2,Popov,Anderson-Positive-Real-Lemma}.
For more details on Positive Real lemma see Chapter 5 of \citep{Anderson-Vongpanitlerd}.

Most of the synthesis algorithms stated in state space \citep{Anderson-Vongpanitlerd}
are based on the determination of the matrix $\mathbf{P}$ which is
usually done by solving a Riccati equation. The algorithm we present
in the following will identify $\mathbf{P}$ in a recursive way which
avoids numerical difficulties appearing in more direct methods presented
in \citep{Anderson-Vongpanitlerd}.

We will now present the Brune algorithm in state-space terms as described
in \citep{Anderson-Moylan-1975}.

\subsection{Brune's Algorithm in the State Space (One-Port Case)\label{sub:Brune's-algorithm-state-space-1-port}}

\begin{figure}
\begin{centering}
\includegraphics{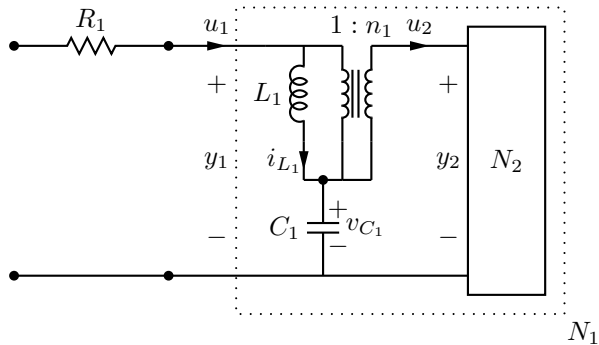} 
\par\end{centering}

\caption{Brune circuit extraction in state space.\label{fig:Brune-circuit-extraction-state-space}}
\end{figure}

Here we assume that we have a one-port positive real impedance function
$z\left(s\right)$ with the minimal realization $\left\{ \mathbf{A},\mathbf{B},\mathbf{C},D\right\} $
(We note that $\mathbf{D}=D$ is a scalar in this case). As shown
in Fig. \eqref{fig:Brune-circuit-extraction-state-space} Brune's
algorithm in state space starts with the extraction of the series
resistance $R_{1}$. Using Eq. \eqref{eq:state-space-to-impedance-in-s-domain}
we can evaluate the real part of the impedance over the imaginary
axis as follows

\begin{eqnarray}
Re[z\left(j\omega\right)] & = & \frac{1}{2}\left(z\left(j\omega\right)+z\left(-j\omega\right)\right)\label{eq:hermitian-part-state-space}\\
 & = & D+\frac{1}{2}\mathbf{C}\left(j\omega\mathbf{I}-\mathbf{A}\right)^{-1}\mathbf{B}+\frac{1}{2}\mathbf{C}\left(-j\omega\mathbf{I}-\mathbf{A}\right)^{-1}\mathbf{B}\\
 & = & D-\mathbf{C}\mathbf{A}\left(\omega^{2}\mathbf{I}+\mathbf{A}^{2}\right)^{-1}\mathbf{B}
\end{eqnarray}
Then the extracted resistance $R_{1}$ is given by

\begin{equation}
R_{1}=\underset{\omega}{\min\,}Re[z\left(j\omega\right)]\label{eq:Extracted-resistance}
\end{equation}
for some frequency $\omega_{0}$ with

\begin{equation}
Re[z\left(j\omega_{0}\right)]=R_{1}\label{eq:R-extraction-min-freq}
\end{equation}

Let the network $N_{2}$ in Fig. \eqref{fig:Brune-circuit-extraction-state-space}
be described by the state-space equations

\begin{eqnarray}
\mathbf{\dot{x}}_{2} & = & \mathbf{A}_{2}\mathbf{x}_{2}+\mathbf{B}_{2}\mathbf{u}_{2}\label{eq:N2-state-space}\\
\mathbf{y}_{2} & = & \mathbf{C}_{2}\mathbf{x}_{2}+D_{2}\mathbf{u}_{2}
\end{eqnarray}
so that the realization $\left\{ \mathbf{A}_{2},\mathbf{B}_{2},\mathbf{C}_{2},D_{2}\right\} $
corresponds to the impedance $z_{2}\left(s\right)=D_{2}+\mathbf{C}_{2}\left(s\mathbf{I}-\mathbf{A}_{2}\right)^{-1}\mathbf{B}_{2}$
seen at the terminals of the network $N_{2}$ ($D_{2}$ is a scalar).

Then the state-space equations for the network $N_{1}$ are given
by

\begin{eqnarray}
\left(\begin{array}{c}
\mathbf{\dot{x}}_{2}\\
\dot{x}_{C_{1}}\\
\dot{x}_{L_{1}}
\end{array}\right) & = & \left(\begin{array}{ccc}
\mathbf{A}_{2} & 0 & -\frac{\mathbf{B}_{2}}{n_{1}\sqrt{L_{1}}}\\
0 & 0 & \frac{1}{n_{1}\sqrt{L_{1}C_{1}}}\\
\frac{\mathbf{C}_{2}}{n_{1}\sqrt{L_{1}}} & -\frac{1}{n_{1}\sqrt{L_{1}C_{1}}} & -\frac{D_{2}}{n_{1}^{2}L_{1}}
\end{array}\right)\left(\begin{array}{c}
\mathbf{x}_{2}\\
x_{C_{1}}\\
x_{L_{1}}
\end{array}\right)+\left(\begin{array}{c}
\frac{\mathbf{B}_{2}}{n_{1}}\\
\frac{\left(1-1/n_{1}\right)}{\sqrt{C_{1}}}\\
\frac{D_{2}}{n_{1}^{2}\sqrt{L_{1}}}
\end{array}\right)\mathbf{u}_{1}\nonumber \\
\mathbf{y}_{1} & = & \left(\begin{array}{ccc}
\frac{\mathbf{C}_{2}}{n_{1}} & \frac{\left(1-1/n_{1}\right)}{\sqrt{C_{1}}} & -\frac{D_{2}}{n_{1}^{2}\sqrt{L_{1}}}\end{array}\right)\left(\begin{array}{c}
\mathbf{x}_{2}\\
x_{C_{1}}\\
x_{L_{1}}
\end{array}\right)+\frac{D_{2}}{n_{1}^{2}}\mathbf{u}_{1}\label{eq:Brune-stage-state-space-equations}
\end{eqnarray}
where $x_{C_{1}}=\sqrt{C_{1}}v_{C_{1}}$ and $x_{L_{1}}=\sqrt{L_{1}}i_{L_{1}}$.
Hence the state-space equations for the network $N_{1}$ are of the
form

\begin{eqnarray}
\mathbf{\dot{x}}_{1} & = & \mathbf{A}_{1}\mathbf{x}_{1}+\mathbf{B}_{1}\mathbf{u}_{1}\label{eq:Brune-stage-state-space-eqs-N1}\\
\mathbf{y}_{1} & = & \mathbf{C}_{1}\mathbf{x}_{1}+D_{1}\mathbf{u}_{1}
\end{eqnarray}
with

\begin{equation}
\mathbf{x}_{1}=\left(\begin{array}{ccc}
\mathbf{x}_{2}^{T} & x_{C_{1}} & x_{L_{1}}\end{array}\right)^{T}\label{eq:State-space-eqs-N1-x1}
\end{equation}

\begin{equation}
\mathbf{A}_{1}=\left(\begin{array}{ccc}
\mathbf{A}_{2} & 0 & -\frac{\mathbf{B}_{2}}{n_{1}\sqrt{L_{1}}}\\
0 & 0 & \frac{1}{n_{1}\sqrt{L_{1}C_{1}}}\\
\frac{\mathbf{C}_{2}}{n_{1}\sqrt{L_{1}}} & -\frac{1}{n_{1}\sqrt{L_{1}C_{1}}} & -\frac{D_{2}}{n_{1}^{2}L_{1}}
\end{array}\right)\label{eq:State-space-eqs-N1-A1}
\end{equation}

\begin{equation}
\mathbf{B}_{1}=\left(\begin{array}{c}
\frac{\mathbf{B}_{2}}{n_{1}}\\
\frac{\left(1-1/n_{1}\right)}{\sqrt{C_{1}}}\\
\frac{D_{2}}{n_{1}^{2}\sqrt{L_{1}}}
\end{array}\right)\label{eq:State-space-eqs-N1-B1}
\end{equation}

\begin{equation}
\mathbf{C}_{1}=\left(\begin{array}{ccc}
\frac{\mathbf{C}_{2}}{n_{1}} & \frac{\left(1-1/n_{1}\right)}{\sqrt{C_{1}}} & -\frac{D_{2}}{n_{1}^{2}\sqrt{L_{1}}}\end{array}\right)\label{eq:State-space-eqs-N1-C1}
\end{equation}

\begin{equation}
D_{1}=\frac{D_{2}}{n_{1}^{2}}\label{eq:State-space-eqs-N1-D1}
\end{equation}
The realization $\left\{ \mathbf{A}_{1},\mathbf{B}_{1},\mathbf{C}_{1},D_{1}\right\} $
then corresponds to the impedance function $z_{1}\left(s\right)=D_{1}+\mathbf{C}_{1}\left(s\mathbf{I}-\mathbf{A}_{1}\right)^{-1}\mathbf{B}_{1}$
seen at the terminals of the network $N_{1}$ ($D_{1}$ is a scalar)
which is related to $z\left(s\right)$ by

\begin{equation}
z_{1}\left(s\right)=z\left(s\right)-R_{1}
\end{equation}
We note that $\mathrm{Re}\left[z_{1}\left(j\omega_{0}\right)\right]=0$.

The following lemma stated in \citep{Anderson-Moylan-1975} shows
that if $z_{1}\left(j\omega_{0}\right)+z_{1}\left(-j\omega_{0}\right)=0$
is satisfied for some $\omega_{0}>0$ for a positive-real impedance
function $z_{1}\left(s\right)$ with a minimal realization $\left\{ \mathbf{A}_{a},\mathbf{B}_{a},\mathbf{C}_{a},D_{a}\right\} $
then there exists a coordinate transformation $\mathbf{T}$ which
would give an equivalent state-state description $\left\{ \mathbf{A}_{1},\mathbf{B}_{1},\mathbf{C}_{1},D_{1}\right\} $
for the impedance $z_{1}\left(s\right)$ in the form given in Eqs.
(\ref{eq:Brune-stage-state-space-eqs-N1}-\ref{eq:State-space-eqs-N1-D1})
with

\begin{eqnarray}
\mathbf{A}_{1} & = & \mathbf{T}\mathbf{A}_{a}\mathbf{T}^{-1}\\
\mathbf{B}_{1} & = & \mathbf{T}\mathbf{B}_{a}\\
\mathbf{C}_{1} & = & \mathbf{C}_{a}\mathbf{T}^{-1}\\
D_{1} & = & D_{a}
\end{eqnarray}
See Section \eqref{sec:State-Space-Formalism} for why $\left\{ \mathbf{A}_{1},\mathbf{B}_{1},\mathbf{C}_{1},D_{1}\right\} $
is an equivalent realization for the same impedance $z_{1}\left(s\right)$.

An explicit procedure is presented in \citep{Anderson-Moylan-1975}
to compute $\mathbf{T}$. We now state the lemma and describe the
algorithm to compute $\mathbf{T}$.

\emph{The Fundamental Lemma (one-port case)}

Let $z_{1}\left(s\right)$ be a positive-real impedance function with
the minimal realization $\left\{ \mathbf{A}_{a},\mathbf{B}_{a},\mathbf{C}_{a},D_{a}\right\} $
satisfying $z_{1}\left(j\omega_{0}\right)+z_{1}\left(-j\omega_{0}\right)=0$
for some finite frequency $\omega_{0}$ (with $j\omega_{0}$ not being
an eigenvalue of $\mathbf{A}_{a}$). Then there exists a coordinate
transformation matrix $\mathbf{T}$ such that $\mathbf{A}_{1}=\mathbf{T}\mathbf{A}_{a}\mathbf{T}^{-1}$,
$\mathbf{B}_{1}=\mathbf{T}\mathbf{B}_{a}$, $\mathbf{C}_{1}=\mathbf{C}_{a}\mathbf{T}^{-1}$
and $D_{1}=D_{a}$ are of the form given in Eqs. (\ref{eq:Brune-stage-state-space-eqs-N1}-\ref{eq:State-space-eqs-N1-D1}).

Now we show how to construct $\mathbf{T}$.

1) Construct a nonsingular matrix $\mathbf{T}_{a}$ with the last
two columns of $\mathbf{T}_{a}^{-1}$ being $\left(\omega_{0}^{2}\mathbf{I}+\mathbf{A}_{a}^{2}\right)^{-1}\mathbf{B}_{a}$
and $-\mathbf{A}_{a}\left(\omega_{0}^{2}\mathbf{I}+\mathbf{A}_{a}^{2}\right)^{-1}\mathbf{B}_{a}$.

2) Set $\mathbf{A}_{b}=\mathbf{T}_{a}\mathbf{A}_{a}\mathbf{T}_{a}^{-1}$,
$\mathbf{B}_{b}=\mathbf{T}_{a}\mathbf{B}_{a}$ and $\mathbf{C}_{b}=\mathbf{C}_{a}\mathbf{T}_{a}^{-1}$
and compute

\begin{equation}
\left(\begin{array}{c}
\mathbf{C}_{b}\left(\omega_{0}^{2}\mathbf{I}+\mathbf{A}_{b}^{2}\right)^{-1}\\
\mathbf{C}_{b}\left(\omega_{0}^{2}\mathbf{I}+\mathbf{A}_{b}^{2}\right)^{-1}\mathbf{A}_{b}
\end{array}\right)=\left(\begin{array}{cc}
\mathbf{K}_{12} & \mathbf{K}_{22}\end{array}\right)
\end{equation}
where $\mathbf{K}_{22}$ is a $2\times2$ matrix. Define

\begin{equation}
\mathbf{T}_{b}=\left(\begin{array}{cc}
\mathbf{I} & \boldsymbol{0}\\
\mathbf{K}_{22}^{-1}\mathbf{K}_{12} & \mathbf{I}
\end{array}\right)
\end{equation}

3) Set $\mathbf{A}_{c}=\mathbf{T}_{b}\mathbf{A}_{b}\mathbf{T}_{b}^{-1}$,
$\mathbf{B}_{c}=\mathbf{T}_{b}\mathbf{B}_{b}$ and $\mathbf{C}_{c}=\mathbf{C}_{b}\mathbf{T}_{b}^{-1}$
then

\begin{equation}
\left(\begin{array}{c}
\mathbf{C}_{c}\left(\omega_{0}^{2}\mathbf{I}+\mathbf{A}_{c}^{2}\right)^{-1}\\
\mathbf{C}_{c}\left(\omega_{0}^{2}\mathbf{I}+\mathbf{A}_{c}^{2}\right)^{-1}\mathbf{A}_{c}
\end{array}\right)=\left(\begin{array}{ccccc}
0 & \cdots & 0 & \alpha^{2} & 0\\
0 & \cdots & 0 & 0 & \beta^{2}
\end{array}\right)
\end{equation}
for non-zero $\alpha,\beta$. Define

\begin{equation}
\mathbf{T}_{c}=\left(\begin{array}{ccc}
\mathbf{I} & 0 & 0\\
0 & \alpha & 0\\
0 & 0 & \beta
\end{array}\right)
\end{equation}

Then $\mathbf{T}=\mathbf{T}_{c}\mathbf{T}_{b}\mathbf{T}_{a}$.

\subsubsection{\label{sub:The-Capacitive-Degenerate-Stage-1-port-ss}The Capacitive
Degenerate Stage}

\begin{figure}
\begin{centering}
\includegraphics{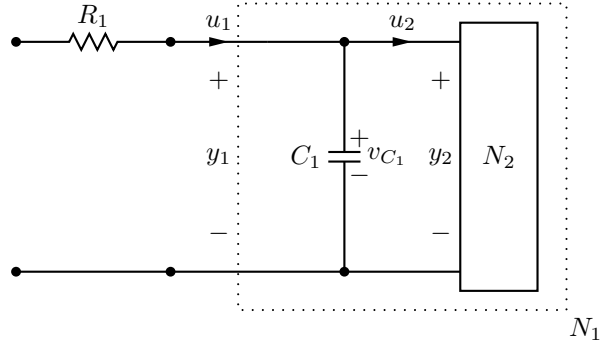} 
\par\end{centering}

\caption{\label{fig:Brune-algorithm-1-port-degenerate}The capacitive degenerate
Brune stage corresponding to the extraction of the resistor $R_{1}$
in Eq. \eqref{eq:Extracted-resistance} at inifinite frequency $\omega_{0}=\infty$.}
\end{figure}

It is possible that the frequency $\omega_{0}$ in Eq. \eqref{eq:R-extraction-min-freq}
where the minimum in Eq. \eqref{eq:Extracted-resistance} is reached
occurs at infinity $\omega_{0}=\infty$. In such a case we need to
extract a capacitive degenerate stage which doesn't involve the inductive
circuit as shown in Fig. \eqref{fig:Brune-algorithm-1-port-degenerate}.
\textcolor{black}{(The case $\omega_{0}=0$ requires the extraction
of an inductive degenerate stage as shown in Fig. \eqref{fig:The-inductive-degenerate-Brune-stage},
see Section \eqref{sub:The-Inductive-Degenerate-Stage} below).}

Let the network $N_{2}$ in Fig. \eqref{fig:Brune-circuit-extraction-state-space}
be described again by the state-space equations

\begin{eqnarray}
\mathbf{\dot{x}}_{2} & = & \mathbf{A}_{2}\mathbf{x}_{2}+\mathbf{B}_{2}\mathbf{u}_{2}\label{eq:N2-state-space-1}\\
\mathbf{y}_{2} & = & \mathbf{C}_{2}\mathbf{x}_{2}+D_{2}\mathbf{u}_{2}
\end{eqnarray}
for some real matrices $\left\{ \mathbf{A}_{2},\mathbf{B}_{2},\mathbf{C}_{2},D_{2}\right\} $
so that the realization $\left\{ \mathbf{A}_{2},\mathbf{B}_{2},\mathbf{C}_{2},D_{2}\right\} $
corresponds to the impedance $z_{2}\left(s\right)=D_{2}+\mathbf{C}_{2}\left(s\mathbf{I}-\mathbf{A}_{2}\right)^{-1}\mathbf{B}_{2}$
seen at the terminals of the network $N_{2}$.

Then the state-space equations for the network $N_{1}$ are given
by

\begin{eqnarray}
\left(\begin{array}{c}
\mathbf{\dot{x}}_{2}\\
\dot{x}_{C_{1}}
\end{array}\right) & = & \left(\begin{array}{cc}
\mathbf{A}_{2}-\frac{\mathbf{B}_{2}\mathbf{C}_{2}}{D_{2}} & \frac{\mathbf{B}_{2}}{D_{2}\sqrt{C_{1}}}\\
\frac{\mathbf{C}_{2}}{D_{2}\sqrt{C_{1}}} & -\frac{1}{D_{2}C_{1}}
\end{array}\right)\left(\begin{array}{c}
\mathbf{x}_{2}\\
x_{C_{1}}
\end{array}\right)+\left(\begin{array}{c}
0\\
\frac{1}{\sqrt{C_{1}}}
\end{array}\right)\mathbf{u}_{1}\label{eq:Brune-stage-state-space-equations-degenerate}\\
\mathbf{y}_{1} & = & \left(\begin{array}{cc}
0 & \frac{1}{\sqrt{C_{1}}}\end{array}\right)\left(\begin{array}{c}
\mathbf{x}_{2}\\
x_{C_{1}}
\end{array}\right)
\end{eqnarray}
where $x_{C_{1}}=\sqrt{C_{1}}v_{C_{1}}$ and $D_{2}$ is a scalar
in the one-port case. Hence the state-space equations for the network
$N_{1}$ are of the form

\begin{eqnarray}
\mathbf{\dot{x}}_{1} & = & \mathbf{A}_{1}\mathbf{x}_{1}+\mathbf{B}_{1}\mathbf{u}_{1}\label{eq:Brune-stage-state-space-eqs-N1-degenerate}\\
\mathbf{y}_{1} & = & \mathbf{C}_{1}\mathbf{x}_{1}+D_{1}\mathbf{u}_{1}
\end{eqnarray}
with

\begin{equation}
\mathbf{x}_{1}=\left(\begin{array}{cc}
\mathbf{x}_{2}^{T} & x_{C_{1}}\end{array}\right)^{T}\label{eq:State-space-eqs-N1-x1-degenerate}
\end{equation}

\begin{equation}
\mathbf{A}_{1}=\left(\begin{array}{cc}
\mathbf{A}_{2}-\frac{\mathbf{B}_{2}\mathbf{C}_{2}}{D_{2}} & \frac{\mathbf{B}_{2}}{D_{2}\sqrt{C_{1}}}\\
\frac{\mathbf{C}_{2}}{D_{2}\sqrt{C_{1}}} & -\frac{1}{D_{2}C_{1}}
\end{array}\right)\label{eq:State-space-eqs-N1-A1-degenerate}
\end{equation}

\begin{equation}
\mathbf{B}_{1}=\left(\begin{array}{c}
0\\
\frac{1}{\sqrt{C_{1}}}
\end{array}\right)\label{eq:State-space-eqs-N1-B1-degenerate}
\end{equation}

\begin{equation}
\mathbf{C}_{1}=\left(\begin{array}{cc}
0 & \frac{1}{\sqrt{C_{1}}}\end{array}\right)\label{eq:State-space-eqs-N1-C1-degenerate}
\end{equation}

\begin{equation}
D_{1}=0\label{eq:State-space-eqs-N1-D1-degenerate}
\end{equation}
The realization $\left\{ \mathbf{A}_{1},\mathbf{B}_{1},\mathbf{C}_{1},D_{1}\right\} $
then corresponds to the impedance function $z_{1}\left(s\right)$
seen at the terminals of the network $N_{1}$ which is related to
$z\left(s\right)$ by

\begin{equation}
z_{1}\left(s\right)=z\left(s\right)-R_{1}
\end{equation}

In such a degenerate case we should also modify \emph{the Fundamental
Lemma} as follows:

\emph{The Fundamental Lemma (one-port capacitive degenerate case)}

Let $z_{1}\left(s\right)$ be a positive-real impedance function with
the minimal realization $\left\{ \mathbf{A}_{a},\mathbf{B}_{a},\mathbf{C}_{a},D_{a}\right\} $
satisfying $z_{1}\left(j\omega_{0}\right)+z_{1}\left(-j\omega_{0}\right)=0$
for $\omega_{0}=\infty$. Then there exists a coordinate transformation
matrix $\mathbf{T}$ such that $\mathbf{A}_{1}=\mathbf{T}\mathbf{A}_{a}\mathbf{T}^{-1}$,
$\mathbf{B}_{1}=\mathbf{T}\mathbf{B}_{a}$, $\mathbf{C}_{1}=\mathbf{C}_{a}\mathbf{T}^{-1}$
and $D_{1}=D_{a}=0$ are of the form given in Eqs. (\ref{eq:Brune-stage-state-space-eqs-N1-degenerate}-\ref{eq:State-space-eqs-N1-D1-degenerate}).

Now we show how to construct $\mathbf{T}$.

1) Construct a nonsingular matrix $\mathbf{T}_{a}$ with the last
column of $\mathbf{T}_{a}^{-1}$ being $\mathbf{B}_{a}$.

2) Set $\mathbf{A}_{b}=\mathbf{T}_{a}\mathbf{A}_{a}\mathbf{T}_{a}^{-1}$,
$\mathbf{B}_{b}=\mathbf{T}_{a}\mathbf{B}_{a}$ and $\mathbf{C}_{b}=\mathbf{C}_{a}\mathbf{T}_{a}^{-1}$
and make the partitioning

\begin{equation}
\begin{array}{c}
\mathbf{C}_{b}\end{array}=\left(\begin{array}{cc}
\mathbf{K}_{12} & \mathbf{K}_{22}\end{array}\right)
\end{equation}
where $\mathbf{K}_{22}$ is a scalar. Define

\begin{equation}
\mathbf{T}_{b}=\left(\begin{array}{cc}
\mathbf{I} & 0\\
\mathbf{K}_{22}^{-1}\mathbf{K}_{12} & \mathbf{I}
\end{array}\right)
\end{equation}

3) Set $\mathbf{A}_{c}=\mathbf{T}_{b}\mathbf{A}_{b}\mathbf{T}_{b}^{-1}$,
$\mathbf{B}_{c}=\mathbf{T}_{b}\mathbf{B}_{b}$ and $\mathbf{C}_{c}=\mathbf{C}_{b}\mathbf{T}_{b}^{-1}$
then

\begin{equation}
\begin{array}{c}
\mathbf{C}_{c}\end{array}=\left(\begin{array}{cccc}
0 & \cdots & 0 & \alpha^{2}\end{array}\right)
\end{equation}
for a non-zero $\alpha$. Define

\begin{equation}
\mathbf{T}_{c}=\left(\begin{array}{cc}
\mathbf{I} & 0\\
0 & \alpha
\end{array}\right)
\end{equation}

Then $\mathbf{T}=\mathbf{T}_{c}\mathbf{T}_{b}\mathbf{T}_{a}$.

\subsubsection{\label{sub:The-Inductive-Degenerate-Stage}The Inductive Degenerate
Stage}

\begin{figure}
\begin{centering}
\includegraphics{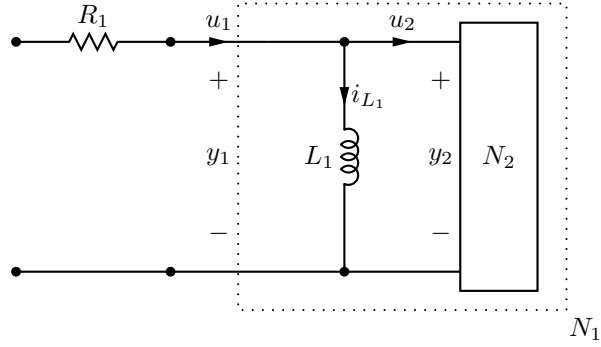}
\par\end{centering}

\caption{\label{fig:The-inductive-degenerate-Brune-stage}The inductive degenerate
Brune stage corresponding to the extraction of the resistor $R_{1}$
in Eq. \eqref{eq:Extracted-resistance} at zero frequency $\omega_{0}=0$.}
\end{figure}

It is possible that the frequency $\omega_{0}$ in Eq. \eqref{eq:R-extraction-min-freq}
where the minimum in Eq. \eqref{eq:Extracted-resistance} is reached
occurs at zero, $\omega_{0}=0$. In that case one needs to extract
an inductive degenerate Brune stage as shown in Fig. \eqref{fig:The-inductive-degenerate-Brune-stage}.
It is straightforward to modify the state-space Brune algorithm to
synthesize an inductive degenerate stage in a way similar to the treatment
we did for the capacitive degenerate case above; we do not present
this algorithm here. See Section \eqref{sub:Inductive-degenerate-case}
for how one can treat such a degenerate stage in the circuit quantization
and dissipation analysis with the assumption that the loss introduced
by the resistor $R_{1}$ in Fig. \eqref{fig:The-inductive-degenerate-Brune-stage}
is small.

\subsection{\label{sub:Quantization-state-space-Brune-circuit-1-port}Quantization
of the One-Port State-Space Brune Circuit}

\begin{figure}
\begin{centering}
\includegraphics{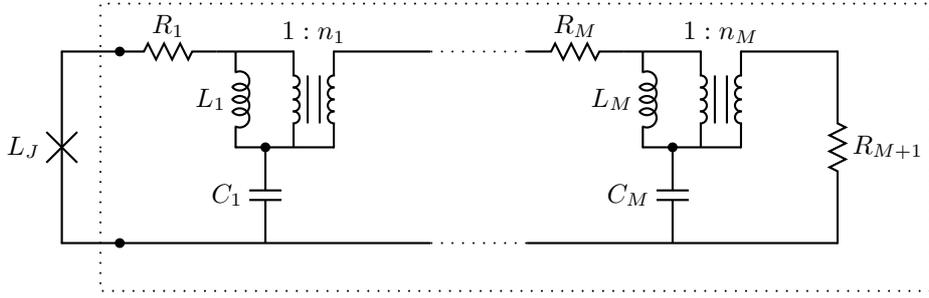} 
\par\end{centering}

\caption{\label{fig:Brune-circuit-state-space}Brune circuit obtained from
the \textcolor{black}{application of }the state-space Brune algorithm
is shown in the dotted box. We note that the coupled inductors at
each stage of the classical Brune circuit in Fig. (1) of \citep{brune-quantization-paper}
are replaced by ordinary inductors shunting ideal transformers as
shown in Fig. \eqref{fig:it-ci-equiv}.}
\end{figure}

\begin{figure}[H]
\begin{centering}
\includegraphics{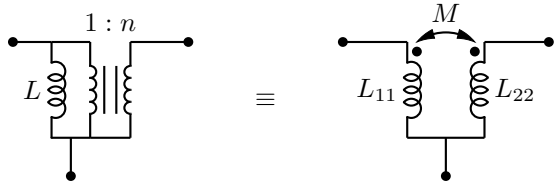} 
\par\end{centering}

\caption{\label{fig:it-ci-equiv}Equivalence of the inductive circuit generated
by the state-space Brune algorithm to the tightly-coupled inductor
pairs appearing at each stage of the original Brune circuit in \citep{brune-quantization-paper}.
Circuit parameters are related by $L=L_{11}$ and $n=\sqrt{\frac{L_{22}}{L_{11}}}$.}
\end{figure}

We call ``the state-space Brune circuit'' the circuit obtained by
the application of the state-space Brune algorithm described in the
previous section. An $M$ stage state-space Brune circuit is shown
in the dotted box in Fig. \eqref{fig:Brune-circuit-state-space}.\textcolor{black}{{}
We note that the coupled inductors at each stage of the classical
Brune circuit in Fig. (1) of \citep{brune-quantization-paper} are
replaced by ordinary inductors shunting ideal transformers in Fig.
\eqref{fig:Brune-circuit-state-space}, which is justified by the
equivalence shown in Fig. \eqref{fig:it-ci-equiv}.}

\textcolor{black}{Ideal transformers were not in the toolbox of any
previous circuit-quantization analysis \citep{BKD,Burkard,Devoret-Les-Houches};}
to treat them we will introduce here a new technique which will eliminate
them by generating effective loop matrices involving turns ratios
in their entries. We will see how this technique will simplify significantly
the analysis of the one-port Brune circuit presented in \citep{brune-quantization-paper}.
It will allow us to skip the transformation defined in Eq. (A9) of
\citep{brune-quantization-paper}. We will however see the full power
of this technique in Section \eqref{sec:Quantization-of-the-multiport-Brune-circuit}
when we will use it to quantize the multiport Brune circuit. To analyze
this circuit we need to modify it to make it treatable in the formalism
of \citep{Burkard}.

\begin{figure}[H]
\begin{centering}
\includegraphics{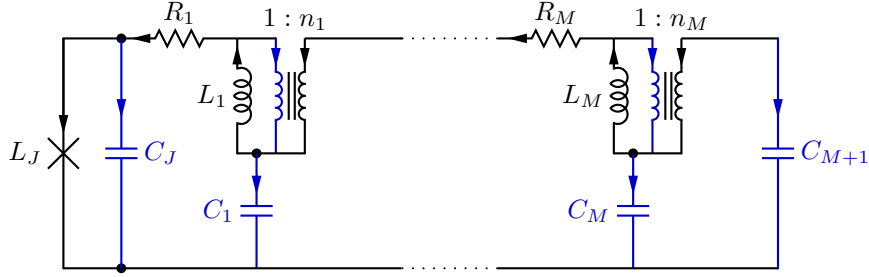} 
\par\end{centering}

\caption{\label{fig:Modified-state-space-Brune-circuit}Modified state-space
Brune circuit. Tree branches are shown in black and chord branches
are shown in blue. Formal capacitance $C_{M+1}$ is introduced for
a technical reason: with the substitution $C_{M+1}=\frac{1}{i\omega R_{M+1}}$
we are able to compute dissipation rate due to \textcolor{black}{shunt
resistor} $R_{M+1}$ in the formalism of \citep{BKD}. After the coordinate
transformation (see below) we take $C_{J}\rightarrow0$ limit.}
\end{figure}

An augmented form of the state-space Brune circuit is shown in Fig.
\eqref{fig:Modified-state-space-Brune-circuit}. The last resistor
$R_{M+1}$ is replaced with a capacitor $C_{M+1}$ which is included
in our analysis later through the substitution $C_{M+1}\leftarrow1/(i\omega R_{M+1})$.
Its contribution to the dissipation rate will be computed referring
to the equation of motion Eq. (61) in \citep{BKD}.

\begin{figure}[H]
\centering{}\includegraphics{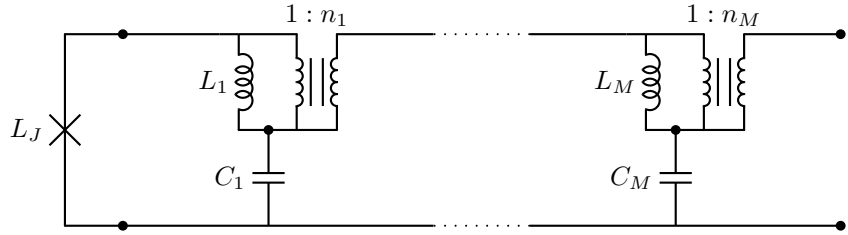}\caption{\label{fig:Lossless-Brune-circuit} Lossless part of the state-space
Brune circuit. It is this circuit that corresponds to the system Hamiltonian
derived below. As discussed above, we take the limit $C_{J}\rightarrow0$
so that this element is removed. The lossless circuit is obtained
from Fig. \eqref{fig:Modified-state-space-Brune-circuit} by taking
$R_{1}$, $R_{2},...R_{M}\rightarrow0$ and $R_{M+1}\rightarrow\infty$.
It is these different limiting treatments that require the descriptions
of the in-series resistors $R_{1}-R_{M}$ follow the low-impedance
treatment as in \citep{Burkard}, while the description of the shunt
resistor $R_{M+1}$ needs the high-impedance treatment as in \citep{BKD}.}
\end{figure}

\begin{figure}[H]
\begin{centering}
\includegraphics{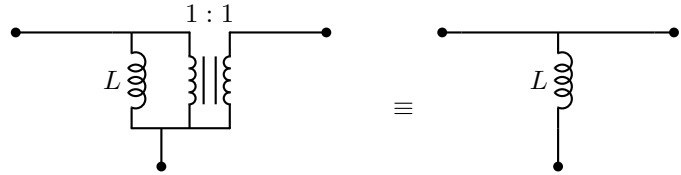} 
\par\end{centering}

\caption{\label{oneturn} Circuit identity showing that inductor-ideal transformer
pairs appearing in state-space Brune stages simplify in the case of
turns ratio equal to one; in this case Fig. \eqref{fig:Lossless-Brune-circuit}
becomes identical to one of the classic lossless Foster canonical
forms in \citep{Foster}.}
\end{figure}

The lossless part of the state-space Brune circuit which corresponds
to the system Hamiltonian derived below is shown in Fig. \eqref{fig:Lossless-Brune-circuit}.
As shown in Fig. \eqref{oneturn} in the special case of unity turns
ratio, this circuit is one of the lossless Foster forms in \citep{Foster}.
We again add a formal capacitance $C_{J}$ shunting the Josephson
junction. This is required for a non-singular capacitance matrix if
there are no degenerate stages.

We now show how to treat ideal transformers by extending the loop
analysis in \citep{Burkard}. Kirchhoff's laws are given by Eqs. (4-5)
in \citep{Burkard}

\begin{equation}
\mathbf{F}\mathbf{I}_{\mathrm{ch}}=-\mathbf{I}_{\mathrm{tr}}\label{eq:Kirchhoff-current-law}
\end{equation}

\begin{equation}
\mathbf{F}^{\mathrm{T}}\mathbf{V}_{\mathrm{tr}}=\mathbf{V}_{\mathrm{ch}}\label{eq:Kirchhoff-voltage-law}
\end{equation}
where we have assumed that there \textcolor{black}{are no external
fluxes in circuit loops. $\mathbf{F}$ is the loop matrix with entries
being $0$, $1$ or $-1$ derived by a graph theoretical analysis
of the circuit \citep{Burkard}. After the effective Kirchhoff analysis
done below $\mathbf{F}$ will be replaced by the effective loop matrix
$\mathbf{F}^{eff}$ with real-valued entries.} $\mathbf{I}_{\mathrm{tr}}$
and $\mathbf{I}_{\mathrm{ch}}$ are the tree and chord branch current
vectors respectively partitioned as follows

\begin{equation}
\mathbf{I}_{\mathrm{tr}}=\left(I_{J},\mathbf{I}_{L},\mathbf{I}_{Z},\mathbf{I}_{T}^{\left(tr\right)}\right)\label{eq:tree-current-vector}
\end{equation}

\begin{equation}
\mathbf{I}_{\mathrm{ch}}=\left(\mathbf{I}_{C},\mathbf{I}_{T}^{\left(ch\right)}\right)\label{eq:chord-current-vector}
\end{equation}
Here labels $J$, $L$, $Z$, $C$, $T$ correspond to Josephson junction,
inductor, resistor, capacitor and ideal transformer branches, respectively.
$\mathbf{I}_{T}^{\left(tr\right)}$ and $\mathbf{I}_{T}^{\left(ch\right)}$
are the current vectors for the ideal transformer branches in the
tree and chords respectively. We also partition loop matrix $\mathbf{F}$
according to the partitioning of current vectors

\begin{equation}
\mathbf{F}=\left(\begin{array}{cc}
\mathbf{F}_{JC} & \mathbf{F}_{JT}\\
\mathbf{F}_{LC} & \mathbf{F}_{LT}\\
\mathbf{F}_{ZC} & \mathbf{F}_{ZT}\\
\mathbf{F}_{TC} & \mathbf{F}_{TT}
\end{array}\right)\label{eq:F-matrix-one-port-ss}
\end{equation}

We will eliminate ideal transformer branches from Kirchhoff laws in
Eqs. \eqref{eq:Kirchhoff-current-law}-\eqref{eq:Kirchhoff-voltage-law}
to get an effective loop matrices $\mathbf{F}^{eff}$ and $\left(\mathbf{F}^{T}\right)^{eff}$
such that we have a new set of effective Kirchhoff relations

\begin{eqnarray}
\mathbf{F}^{eff}\mathbf{I}_{\mathrm{ch}}^{eff} & = & -\mathbf{I}_{\mathrm{tr}}^{eff}\label{eq:effective-Kirchhoff-current-law-1-port}\\
\left(\mathbf{F}^{T}\right)^{eff}\mathbf{V}_{\mathrm{tr}}^{eff} & = & \mathbf{V}_{\mathrm{ch}}^{eff}\label{eq:effective-Kirchhoff-voltage-law-1-port}
\end{eqnarray}
where

\begin{eqnarray*}
\mathbf{I}_{\mathrm{tr}}^{eff} & = & \left(I_{J},\mathbf{I}_{L},\mathbf{I}_{Z}\right)\\
\mathbf{I}_{\mathrm{ch}}^{eff} & = & \mathbf{I}_{C}
\end{eqnarray*}
and

\begin{eqnarray}
\mathbf{F}^{eff} & = & \left(\begin{array}{c}
\mathbf{F}_{JC}^{eff}\\
\mathbf{F}_{LC}^{eff}\\
\mathbf{F}_{ZC}^{eff}
\end{array}\right)\label{eq:effective-F-1-port}\\
\left(\mathbf{F}^{T}\right)^{eff} & = & \left(\begin{array}{ccc}
\left(\mathbf{F}_{JC}^{T}\right)^{eff} & \left(\mathbf{F}_{LC}^{T}\right)^{eff} & \left(\mathbf{F}_{ZC}^{T}\right)^{eff}\end{array}\right)
\end{eqnarray}
\textcolor{black}{We note here that the entries of the effective loop
matrix $\mathbf{F}^{eff}$ are real numbers, being functions of ideal
transformer turn ratios as we will see below.}

In this section for simplicity reasons we will derive only the effective
Kirchhoff's current law in Eq. \eqref{eq:effective-Kirchhoff-current-law-1-port}
by computing $\mathbf{F}^{eff}$. We postpone the derivation of the
effective Kirchhoff's voltage law and the computation of the matrix
$\left(\mathbf{F}^{T}\right)^{eff}$ to the Appendix \eqref{sub:Effective-Kirchhoff's-voltage-law-one-port}.
However we note here that

\begin{equation}
\left(\mathbf{F}^{T}\right)^{eff}=\left(\mathbf{F}^{eff}\right)^{T}
\end{equation}
should be verified to hold. This ensures the symmetry of various matrices
computed in the formalisms of \citep{BKD,Burkard} like the capacitance
matrix $\mathcal{C}$ and the stiffness matrix $M_{0}$ for example.

Now we claim that $\mathbf{F}^{eff}$ in Eq. \eqref{eq:effective-F-1-port}
is given by

\begin{equation}
\mathbf{F}_{JC}^{eff}=\left(\begin{array}{ccccc}
1 & 1 & \cdots & 1 & 1\end{array}\right)\label{eq:effective-FJC-matrix}
\end{equation}

\begin{equation}
\mathcal{\mathbf{F}}_{LC}^{eff}=\begin{pmatrix}1 & (1-n_{1}) & \cdots & (1-n_{1}) & (1-n_{1})\\
 & \ddots & \ddots & \vdots & \vdots\\
 &  & 1 & (1-n_{M-1}) & (1-n_{M-1})\\
\boldsymbol{0} &  &  & 1 & (1-n_{M})
\end{pmatrix}\label{eq:effective-FLC-matrix}
\end{equation}

\begin{equation}
\mathbf{F}_{ZC}^{eff}=\left(\begin{array}{ccccc}
1 & 1 & \cdots & 1 & 1\\
 & 1 & \cdots & 1 & 1\\
 &  & \ddots & \vdots & \vdots\\
\boldsymbol{0} &  &  & 1 & 1
\end{array}\right)\label{eq:effective-FZC-matrix}
\end{equation}
where $\mathbf{F}_{JC}^{eff}$ is a row vector of length $\left(M+1\right)$,
$\mathcal{\mathbf{F}}_{LC}^{eff}$ and $\mathbf{F}_{ZC}^{eff}$ are
$M\times(M+1)$ matrices. To see how the matrices in Eqs. \eqref{eq:effective-FJC-matrix}-\eqref{eq:effective-FZC-matrix}
can be computed we first note the following

\begin{equation}
\mathbf{I}_{T}^{\left(tr\right)}=-\mathbf{F}_{TC}\mathbf{I}_{C}\label{eq:it-tree-currents}
\end{equation}
with

\begin{equation}
\mathbf{F}_{TC}=\left(\begin{array}{ccccc}
0 & 1 & 1 & \cdots & 1\\
 & 0 & 1 & \cdots & 1\\
 &  & \ddots & \ddots & \vdots\\
\boldsymbol{0} &  &  & 0 & 1
\end{array}\right)
\end{equation}
where $\mathbf{F}_{TC}$ is a $M\times(M+1)$ matrix. We note that
$\mathbf{F}_{TC}$ doesn't involve any turns ratios. Using the ideal
transformer relations $\mathbf{I}_{T}^{\left(ch\right)}=-\mathbf{N}\mathbf{I}_{T}^{\left(tr\right)}$
with $\mathbf{N}$ being the diagonal matrix of turns ratios

\begin{equation}
\mathbf{N}=\left(\begin{array}{ccc}
n_{1} &  & \boldsymbol{0}\\
 & \ddots\\
\boldsymbol{0} &  & n_{M}
\end{array}\right)\label{eq:turns-ratio-matrix-one-port-ss}
\end{equation}
and Eq. \eqref{eq:it-tree-currents} we get

\begin{equation}
\mathbf{I}_{T}^{\left(ch\right)}=\mathbf{N}\mathbf{F}_{TC}\mathbf{I}_{C}\label{eq:it-branch-elimination}
\end{equation}
Inductor currents are given by

\begin{equation}
\mathbf{I}_{L}=-\mathbf{F}_{LC}\mathbf{I}_{C}-\mathbf{F}_{LT}\mathbf{I}_{T}^{\left(ch\right)}\label{eq:inductor-currents-relation-one-port-ss}
\end{equation}
where

\begin{equation}
\mathbf{F}_{LC}=\left(\begin{array}{ccccc}
1 & 1 & \cdots & 1 & 1\\
 & 1 & \cdots & 1 & 1\\
 &  & \ddots & \vdots & \vdots\\
\boldsymbol{0} &  &  & 1 & 1
\end{array}\right)
\end{equation}
and $\mathbf{F}_{LT}=-\mathbb{I}$. Using Eqs. \eqref{eq:it-branch-elimination}
and \eqref{eq:inductor-currents-relation-one-port-ss} we get

\begin{equation}
\mathbf{I}_{L}=-\left(\mathbf{F}_{LC}-\mathbf{N}\mathbf{F}_{TC}\right)\mathbf{I}_{C}
\end{equation}
which gives the effective loop matrix $\mathbf{F}_{LC}^{eff}$

\begin{align}
\mathbf{F}_{LC}^{eff} & =\mathbf{F}_{LC}-\mathbf{N}\mathbf{F}_{TC}\label{eq:effective-FLC-one-port-ss}\\
= & \begin{pmatrix}1 & (1-n_{1}) & \cdots & (1-n_{1}) & (1-n_{1})\\
 & \ddots & \ddots & \vdots & \vdots\\
 &  & 1 & (1-n_{M-1}) & (1-n_{M-1})\\
\boldsymbol{0} &  &  & 1 & (1-n_{M})
\end{pmatrix}
\end{align}
We note that $\mathbf{F}_{LC}^{eff}$ is no longer a binary matrix
as we have turns ratios appearing in its entries.

$\mathbf{F}_{JC}^{eff}$ is simply given by

\begin{eqnarray}
\mathbf{F}_{JC}^{eff} & = & \mathbf{F}_{JC}\label{eq:effective-FJC-one-port-ss}\\
 & = & \left(\begin{array}{ccccc}
1 & 1 & \cdots & 1 & 1\end{array}\right)
\end{eqnarray}
Since the current through the Josephson junction depends only on chord
capacitor currents

\begin{equation}
I_{J}=-\mathbf{F}_{JC}\mathbf{I}_{C}
\end{equation}
Note that $\mathbf{F}_{JC}^{eff}$ does not depend on turns ratios.
Similarly the currents through the resistors $R_{j}$ for $1\leq j\leq M$
depend only on chord capacitor currents

\[
\mathbf{I}_{Z}=-\mathbf{F}_{ZC}\mathbf{I}_{C}
\]
Hence

\begin{eqnarray}
\mathbf{F}_{ZC}^{eff} & = & \mathbf{F}_{ZC}\label{eq:effective-FZC-one-port-ss}\\
 & = & \left(\begin{array}{ccccc}
1 & 1 & \cdots & 1 & 1\\
 & 1 & \cdots & 1 & 1\\
 &  & \ddots & \vdots & \vdots\\
\boldsymbol{0} &  &  & 1 & 1
\end{array}\right)\nonumber 
\end{eqnarray}

\textcolor{black}{Now one can write an equation of motion for the
one-port state-space Brune circuit in Fig. \eqref{fig:Brune-circuit-state-space}
in the form of Eq. (29) in \citep{Burkard}}

\textcolor{black}{{} 
\begin{equation}
\left(\mathcal{C}+\mathcal{C}_{Z}\right)*\ddot{\mathbf{\Phi}}=-\frac{\partial U}{\partial\mathbf{\Phi}}\label{eq:brune-ss-eq-of-motion}
\end{equation}
However effective loop matrices derived above have to replace the
ordinary loop matrices while computing quantities appearing in the
equation of motion Eq. \eqref{eq:brune-ss-eq-of-motion} like the
capacitance matrix $\mathcal{C}$ and the dissipation matrix $\mathcal{C}_{Z}$,{}
as we now show.}

We compute the capacitance matrix $\mathcal{C}_{0}$ for the Brune
circuit in Fig. \eqref{fig:Brune-circuit-state-space} using the Eq.
(22) of \citep{Burkard} with the effective loop matrix $\mathcal{F}_{C}^{eff}$

\begin{equation}
\mathcal{C}_{0}=\left(\begin{array}{cc}
C_{J} & \boldsymbol{0}\\
\boldsymbol{0} & \boldsymbol{0}
\end{array}\right)+\mathcal{F}_{C}^{eff}\mathbf{C}\left(\mathcal{F}_{C}^{eff}\right)^{T}\label{eq:C0-ss}
\end{equation}
where $\mathbf{C}$ is the diagonal matrix of capacitances

\begin{equation}
\mathbf{C}=\left(\begin{array}{ccc}
C_{1} &  & \boldsymbol{0}\\
 & \ddots\\
\boldsymbol{0} &  & C_{M+1}
\end{array}\right)
\end{equation}
and

\begin{equation}
\mathcal{F}_{C}^{eff}=\left(\begin{array}{c}
\mathbf{F}_{JC}^{eff}\\
\mathbf{F}_{LC}^{eff}
\end{array}\right)\label{eq:effective-FC-one-port-ss}
\end{equation}

$\mathbf{L}_{t}$ in Eq. (15) of \citep{Burkard} is a diagonal matrix
of inductances

\begin{equation}
\mathbf{L}_{t}=\left(\begin{array}{ccc}
L_{1} &  & \boldsymbol{0}\\
 & \ddots\\
\boldsymbol{0} &  & L_{M}
\end{array}\right)
\end{equation}

With

\begin{equation}
\mathcal{G}=\begin{pmatrix}\boldsymbol{0}\\
1_{M\times M}
\end{pmatrix}
\end{equation}
since there are no chord inductors. We get using Eq. (31) of \citep{Burkard}

\begin{align}
M_{0} & =\mathcal{G}\mathbf{L}_{t}^{-1}\mathcal{G}^{t}\\
 & =\begin{pmatrix}0 & \boldsymbol{0}\\
\boldsymbol{0} & \mathbf{L}_{t}^{-1}
\end{pmatrix}\label{eq:M0-ss}
\end{align}

We skip the first transformation defined in Eq. (A9) of \citep{brune-quantization-paper}
and the truncation afterwards since we are already in the low dimensional
subspace with $\left(M+1\right)$ degrees of freedom. We again define
a local transformation matrix $T$ which makes the Langrangian description
(i.e., both $\mathcal{C}_{0}$ and $M{}_{0}$) of the system band-diagonal:

\begin{equation}
T=\begin{pmatrix}1\\
-1/(1-n_{1}) & -1/(1-n_{1}) &  & \text{{\huge0}}\\
 & 1/(1-n_{2}) & 1/(1-n_{2})\\
 &  & \ddots & \ddots\\
 & \text{{\huge0}} &  & \left(-1\right)^{M}/(1-n_{M}) & \left(-1\right)^{M}/(1-n_{M})
\end{pmatrix}\label{eq:T-tr-ss}
\end{equation}
Applying $T$ to $\mathcal{C}_{0}$ and $M_{0}$ we get

\begin{align}
\mathcal{C} & =T^{t}\mathcal{C}_{0}T\label{eq:capacitance}\\
 & \mathcal{=}\begin{pmatrix}C_{J}+n_{1}^{2}C_{1}^{'} & n_{1}C_{1}^{'}\\
n_{1}C_{1}^{'} & C_{1}^{'}+n_{2}^{2}C_{2}^{'} & \ddots & \text{{\huge0}}\\
 & \ddots & \ddots\\
 & \text{{\huge0}} &  & C_{M-1}^{'}+n_{M}^{2}C_{M}^{'} & n_{M}C_{M}^{'}\\
 &  &  & n_{M}C_{M}^{'} & C_{M}^{'}+C_{M+1}
\end{pmatrix}
\end{align}

\begin{align}
\mathbf{M}_{0} & =T^{t}M_{0}T\\
 & =\begin{pmatrix}\frac{1}{L'_{1}} & \frac{1}{L'_{1}}\\
\frac{1}{L'_{1}} & \frac{1}{L'_{1}}+\frac{1}{L'_{2}} & \frac{1}{L'_{2}} &  & \text{{\huge0}}\\
 & \frac{1}{L'_{2}} & \frac{1}{L'_{2}}+\frac{1}{L'_{3}} & \ddots\\
 &  & \ddots & \ddots\\
 & \text{{\huge0}} &  &  & \frac{1}{L'_{M-1}}+\frac{1}{L'_{M}} & \frac{1}{L'_{M}}\\
 &  &  &  & \frac{1}{L'_{M}} & \frac{1}{L'_{M}}
\end{pmatrix}\label{eq:M0}
\end{align}
where $C_{j}^{'}=C_{j}/\left(1-n_{j}\right)^{2}$, $L'_{j}=L_{j}\left(1-n_{j}\right)^{2}$.

A Lagrangian $\mathcal{L}_{0}$ (and equivalently a Hamiltonian $\mathcal{H_{S}}$)
can be written as

\begin{equation}
\mathcal{L}_{0}=\frac{1}{2}\mathbf{\boldsymbol{\dot{\mathbf{\Phi}}}}^{T}\mathcal{C}\boldsymbol{\dot{\Phi}}-U\left(\boldsymbol{\Phi}\right),\;\mathcal{H_{S}}=\frac{1}{2}\boldsymbol{Q}^{T}\mathcal{C}^{-1}\boldsymbol{Q}+U\left(\boldsymbol{\Phi}\right)\label{eq:Lagrangian}
\end{equation}
where

\begin{equation}
U\left(\boldsymbol{\Phi}\right)=-\left(\frac{\Phi_{0}}{2\pi}\right)^{2}L_{J}^{-1}\cos\left(\varphi_{J}\right)+\frac{1}{2}\boldsymbol{\Phi}^{T}\mathbf{M}_{0}\boldsymbol{\Phi}\label{eq:Potential-energy-function-1}
\end{equation}
$\mathbf{\Phi}$ is the vector of transformed coordinates of length
$\left(M+1\right)$ and $\Phi_{1}=\left(\frac{\Phi_{0}}{2\pi}\right)\varphi_{J}$.
We note that the transformation $T$ in Eq. \eqref{eq:T-tr-ss} introduces
a local relationship between the final coordinates $\mathbf{\Phi}$
and the branch fluxes $\Phi_{L_{j}}$'s across the ordinary inductors
$L_{j}$'s for $1\leq j\leq M$ in the Brune circuit in Fig. \eqref{fig:Brune-circuit-state-space}
in the sense that the flux $\Phi_{L_{j}}$ is only a superposition
of two consecutive coordinates $\Phi_{j}$ and $\Phi_{j+1}$ by the
relation $\mathbf{\Phi}_{L}=T\mathbf{\Phi}$ which gives

\begin{equation}
\Phi_{L_{j}}=\frac{\left(-1\right)^{j}}{(1-n_{j})}\left(\Phi_{j}+\Phi_{j+1}\right)
\end{equation}
for $1\leq j\leq M$. $\mathbf{\Phi}_{L}$ is the vector holding the
fluxes across the inductors in the Brune circuit in Fig. \eqref{fig:Brune-circuit-state-space}
such that

\begin{equation}
\mathbf{\Phi}_{L}=\left(\begin{array}{cccc}
\Phi_{J} & \Phi_{L_{1}} & \ldots & \Phi_{L_{M}}\end{array}\right)^{T}
\end{equation}
with $\Phi_{J}=\left(\frac{\Phi_{0}}{2\pi}\right)\varphi_{J}$ being
the flux across the Josephson junction.

\subsection{\label{sub:Dissipation-Analysis-one-port-ss}Dissipation Analysis}

\textcolor{black}{In this section our aim is to compute relaxation
rates. We follow here the treatment given in \citep{brune-quantization-paper}
with the exception that we are going to use effective loop matrices
derived in the previous section instead of the ordinary ones. }

\textcolor{black}{We treat resistors in the{} Caldeira-Leggett formalism
with each resistor representing a bath of harmonic oscillators with
a smooth frequency spectrum. Couplings of the baths to the circuit
degrees of freedom are given by $\mathbf{\bar{m}}$ matrices as defined
in Eqs. (65) and (27) of \citep{BKD} and \citep{Burkard}, respectively.}

\textcolor{black}{We will interpret the equation of motion in Eq.
(29) of \citep{Burkard} as an equation of motion in Eq. (61) of \citep{BKD}.{}
We start by rearranging the equation of motion Eq. (29) of \citep{Burkard}}

\textcolor{black}{{} 
\begin{equation}
\mathcal{C}*\mathbf{\ddot{\Phi}}=-\frac{\partial U}{\partial\mathbf{\Phi}}-\mathcal{C}_{Z}*\mathbf{\ddot{\Phi}}\label{eq:Burkard-equation of motion}
\end{equation}
}

\textcolor{black}{$\mathcal{C}_{Z}$ is given in frequency domain
in Eq. (26) of \citep{Burkard} as}

\textcolor{black}{{} 
\begin{equation}
\mathcal{C}_{Z}(\omega)=\mathbf{\bar{m}}\mathbf{\bar{C}}_{Z}\left(\omega\right)\mathbf{\bar{m}}^{T}\label{eq:Dissipation-matrix-Burkard}
\end{equation}
with}

\textcolor{black}{{} 
\begin{equation}
\mathbf{\bar{m}}=\mathcal{F}_{C}^{eff}\mathbf{C}\left(\mathbf{F}_{ZC}^{eff}\right)^{T}\label{eq:m-vector-formula-Burkard-1}
\end{equation}
}

\textcolor{black}{{} 
\begin{equation}
\bar{\mathbf{C}}_{Z}\left(\omega\right)=-i\omega\mathbf{Z}\left(\omega\right)\left(\mathbf{I}+\mathbf{F}_{ZC}^{eff}\mathbf{C}\left(\mathbf{F}_{ZC}^{eff}\right)^{T}i\omega\mathbf{Z}\left(\omega\right)\right)^{-1}\label{eq:CbarZ-eff-Burkard}
\end{equation}
where we used Eqs. (27) and (28) of \citep{Burkard} (correcting a
typo in the sign of $\bar{\mathbf{C}}_{Z}\left(\omega\right)$) with
effective loop matrices $\mathcal{F}_{C}^{eff}$ and $\mathbf{F}_{ZC}^{eff}$
computed in the previous section. We note that $\mathbf{F}_{ZC}^{eff}$
is independent of ideal transformer turns ratios. }

\textcolor{black}{Comparing Eq. \eqref{eq:Burkard-equation of motion}
to the equation of motion Eq. (61) in \citep{BKD} we identifiy in
frequency domain }

\textcolor{black}{{} 
\begin{equation}
\mathbf{M}_{d}\left(\omega\right)=-\omega^{2}\mathcal{C}_{Z}\left(\omega\right)\label{eq:BKD-Md-matrix-related-to-CZ-in-Burkard}
\end{equation}
}

\textcolor{black}{Using Eq. \eqref{eq:Dissipation-matrix-Burkard}
and \eqref{eq:BKD-Md-matrix-related-to-CZ-in-Burkard}}

\textcolor{black}{{} 
\begin{equation}
\mathbf{M}_{d}\left(\omega\right)=-\omega^{2}\mathbf{\bar{m}}\mathbf{\bar{C}}_{Z}\left(\omega\right)\mathbf{\bar{m}}^{T}\label{eq:Md-matrix-related-to-CZ-in-Burkard-2}
\end{equation}
}

\textcolor{black}{Comparing Eq. \eqref{eq:Md-matrix-related-to-CZ-in-Burkard-2}
to the Eq. (64) in \citep{BKD} we make the identification}

\textcolor{black}{{} 
\begin{equation}
\mathbf{\bar{L}}_{Z}^{-1}\left(\omega\right)=-\omega^{2}\mathbf{\bar{C}}_{Z}\left(\omega\right)\label{eq:LZ-matrix-BKD-dissipation-matrix}
\end{equation}
with the $\mathbf{\bar{m}}$ matrix being given by Eq. \eqref{eq:m-vector-formula-Burkard-1}}

\textcolor{black}{
\begin{equation}
\mathbf{\bar{m}}=T^{t}\mathcal{F}_{C}^{eff}\mathbf{C}\left(\mathbf{F}_{ZC}^{eff}\right)^{T}\label{eq:m-tr-T-ss}
\end{equation}
where we have also taken into account the coordinate transformation
$T$ defined in Eq. \eqref{eq:T-tr-ss}.}

\textcolor{black}{We will treat resistors one at a time{} \citep{Burkard-Brito}.
Hence for the series resistors $R_{j}$ with $1\leq j\leq M$, $\mathbf{\bar{C}}_{Z}\left(\omega\right)$
defined in Eq. \eqref{eq:CbarZ-eff-Burkard} is a scalar function
$\mathbf{\bar{C}}_{Z,j}\left(\omega\right)$ which allows us to write
the kernel defined in Eq. (73) of \citep{BKD} using Eq. \eqref{eq:LZ-matrix-BKD-dissipation-matrix}
above}

\textcolor{black}{{} 
\begin{equation}
K_{j}\left(\omega\right)=-\omega^{2}\mathbf{\bar{C}}_{Z,j}\left(\omega\right)\label{eq:BKD-kernel}
\end{equation}
for $1\leq j\leq M$. Also we will use the columns $\mathbf{\bar{m}}_{j}$
for $1\leq j\leq M$ of the matrix $\mathbf{\bar{m}}$ defined in
Eq. \eqref{eq:m-tr-T-ss}; $\mathbf{\bar{m}}_{j}$ giving the coupling
of the system degrees of freedom to the bath of the resistor $R_{j}$.
For the last resistor $R_{M+1}$ we will read off the kernel $K_{M+1}\left(\omega\right)$
and the coupling vector $\mathbf{\bar{m}}_{M+1}$ directly from the
equation of motion after the replacement $C_{M+1}\leftarrow1/(i\omega R_{M+1})$
as shown below. }

\textcolor{black}{We compute the spectral density of the bath corresponding
to the resistor $R_{j}$ for all resistors $1\leq j\leq M+1$ using
the Eq. (93) of \cite{BKD}}

\textcolor{black}{
\begin{equation}
J_{j}\left(\omega\right)=Im\left[K_{j}\left(\omega\right)\right]\label{eq:BKD-spectral-density}
\end{equation}
where we corrected a sign typo and dropped the factor $\mu\left(\frac{\Phi_{0}}{2\pi}\right)^{2}$
which will be justified down below. }

\textcolor{black}{Applying Eq. (124) of \citep{BKD} we get the contribution
to the relaxation rate from the resistor $R_{j}$ $\left(1\leq j\leq M+1\right)$:}

\textcolor{black}{{} 
\begin{equation}
\frac{1}{T_{1,j}}=\frac{4}{\hbar}\left|\left\langle 0\left|\mathbf{\bar{m}}_{j}\cdot\mathbf{\Phi}\right|1\right\rangle \right|^{2}J_{j}\left(\omega_{01}\right)\coth\left(\frac{\hbar\omega_{01}}{2k_{B}T}\right)\label{eq:T1-formula}
\end{equation}
$\left|0,1\right\rangle $ are the qubit eigenstates of the system
Hamiltonian in Eq. \eqref{eq:Lagrangian} and $\omega_{01}$ is the
transition frequency between them. Calculating these quantities requires
solving the Schrödinger equation for the system Hamiltonian in Eq.
\eqref{eq:Lagrangian} above; this can be a difficult task, but many
effective accurate methods have been developed for doing this, in
many works right up to the present \citep{BKD,Burkard,blais,Brito,moreHcalculations}.
The vector $\mathbf{\bar{m}}_{j}$ represents the coupling of the
system to the bath of the resistor $R_{j}$.}

\textcolor{black}{To treat the $j^{th}$ in-series resistor $R_{j}$
for $1\leq j\leq M$, we imagine that all other in-series resistors
are short circuited such that $R_{k}=0$ for $1\leq k\leq M$ and
$k\neq j$ and the last shunt resistor is open circuited such that
$R_{M+1}=\infty$. Using Eqs. \eqref{eq:m-tr-T-ss} and \eqref{eq:CbarZ-eff-Burkard}
we write}

\textcolor{black}{
\begin{equation}
\mathbf{\bar{m}}=T^{t}\mathcal{F}_{C}^{eff}\mathbf{C}\left(\mathbf{F}_{R_{j},C}^{eff}\right)^{T}\label{eq:m-vec-ss}
\end{equation}
}

\textcolor{black}{
\[
\bar{\mathbf{C}}_{Z,j}\left(\omega\right)=-i\omega R_{j}\left(\mathbf{I}+\mathbf{F}_{R_{j},C}^{eff}\mathbf{C}\left(\mathbf{F}_{R_{j},C}^{eff}\right)^{T}i\omega R_{j}\right)^{-1}
\]
where we made the replacements $\mathbf{Z}\left(\omega\right)\leftarrow R_{j}$
and $\mathbf{F}_{ZC}^{eff}\leftarrow\mathbf{F}_{R_{j},C}^{eff}$ with
$\mathbf{F}_{R_{j},C}^{eff}$ being the $j^{th}$ row of $\mathbf{F}_{ZC}^{eff}$.}

\textcolor{black}{
\begin{equation}
\mathbf{\bar{m}}_{j}=\begin{pmatrix}0\\
\vdots\\
0\\
j^{th}\, entry\rightarrow\frac{\left(-1\right)^{j}n_{j}C{}_{j}}{\left(1-n_{j}\right)}\\
\frac{\left(-1\right)^{j+1}n_{j+1}C{}_{j+1}}{\left(1-n_{j+1}\right)}+\frac{\left(-1\right)^{j}C{}_{j}}{\left(1-n_{j}\right)}\\
\vdots\\
\frac{\left(-1\right)^{M}n_{M}C{}_{M}}{\left(1-n_{M}\right)}+\frac{\left(-1\right)^{M-1}C{}_{M-1}}{\left(1-n_{M-1}\right)}\\
\frac{\left(-1\right)^{M}C{}_{M}}{\left(1-n_{M}\right)}
\end{pmatrix}\label{lastm}
\end{equation}
where $\mathbf{\bar{m}}_{j}$ are vectors of length $(M+1)$ and}

\textcolor{black}{
\begin{equation}
\bar{\mathbf{C}}_{Z,j}\left(\omega\right)=\frac{-i\omega R_{j}}{1+i\omega R_{j}\left(\underset{k=j}{\overset{M}{\sum}}C_{k}\right)}
\end{equation}
We then have using Eq. \eqref{eq:BKD-kernel}}

\textcolor{black}{
\begin{align}
K_{j}\left(\omega\right) & =-\omega^{2}\bar{\mathbf{C}}_{Z,j}\left(\omega\right)\\
 & =\frac{i\omega^{3}R_{j}}{1+i\omega R_{j}\left(\underset{k=j}{\overset{M}{\sum}}C_{k}\right)}
\end{align}
Hence we obtain using Eq. \eqref{eq:BKD-spectral-density}}

\textcolor{black}{
\begin{align}
J_{j}\left(\omega\right) & =Im\left[K_{j}\left(\omega\right)\right]\\
 & =\frac{\omega^{3}R_{j}}{1+\omega^{2}R_{j}^{2}\left(\underset{k=j}{\overset{M}{\sum}}C_{k}\right)^{2}}\label{eq:spectral-function-state-space}
\end{align}
Note that our use of the non-normalized coupling vector $\mathbf{\bar{m}}_{j}$
and the flux vector $\mathbf{\Phi}$ in Eq. \eqref{eq:T1-formula}
implies removal of the factor $\mu\left(\frac{\Phi_{0}}{2\pi}\right)^{2}$
from the definition of the spectral function of the bath $J\left(\omega\right)$
in Eq. (93) of \citep{BKD} (See also \eqref{eq:spectral-function-last-resistor-1}
below).}

\textcolor{black}{To treat the last resistor $R_{M+1}$ we replace
$C_{M+1}$ in the last row of capacitance matrix by $1/(i\omega R_{M+1})$.
We get the following dissipation matrix for resistor $R_{M+1}$ 
\begin{equation}
\mathbf{M}_{d}=K_{M+1}\left(\omega\right)\mathbf{\bar{m}}_{M+1}\mathbf{\bar{m}}_{M+1}^{T},
\end{equation}
where $K_{M+1}\left(\omega\right)=\frac{i\omega}{R_{M+1}}$ and $\mathbf{\bar{m}}_{M+1}=\begin{pmatrix}0\\
\vdots\\
0\\
1
\end{pmatrix}$ is a vector with $\left(M+1\right)$ rows. We then have}

\textcolor{black}{
\begin{equation}
J_{M+1}\left(\omega\right)=Im\left[K_{M+1}\left(\omega\right)\right]=\frac{\omega}{R_{M+1}},\label{eq:spectral-function-last-resistor-1}
\end{equation}
Compare $\mathbf{\bar{m}}_{M+1}$ which is dimensionless to $\mathbf{\bar{m}}_{j}$
in Eq. \eqref{lastm} which has the dimension of capacitance(Farad)
for $1\leq j\leq M$. Note also the difference in dimensionality between
$J_{M+1}\left(\omega\right)$ in Eq. \eqref{eq:spectral-function-last-resistor-1}
and $J_{j}\left(\omega\right)$ defined in Eq. \eqref{eq:spectral-function-state-space}
for $1\leq j\leq M$.}

\subsection{Capacitive Degenerate Case}

Here we consider only a single capacitive degenerate stage. We consider
a degenerate case appearing at $k^{th}$ stage as described in Section
\eqref{sub:The-Capacitive-Degenerate-Stage-1-port-ss}. In case of
such degeneracy we remove the $(k+1)^{th}$ row in the $\mathcal{F}_{C}^{eff}$
matrix in Eq. \eqref{eq:effective-FC-one-port-ss}:

\begin{equation}
\mathcal{F}_{C}^{eff}=\begin{pmatrix}1 & 1 & \cdots & 1 & 1 & \cdots & 1 & 1\\
1 & (1-n_{1}) & \cdots & (1-n_{1}) & (1-n_{1}) & \cdots & (1-n_{1}) & (1-n_{1})\\
 & \ddots & \ddots & \vdots & \vdots &  & \vdots & \vdots\\
 &  & 1 & (1-n_{k-1}) & (1-n_{k-1}) & \cdots & (1-n_{k-1}) & (1-n_{k-1})\\
 &  &  & 1 & (1-n_{k+1}) & \cdots & (1-n_{k+1}) & (1-n_{k+1})\\
 &  &  &  & \ddots & \ddots & \vdots & \vdots\\
 & \text{\text{{\huge0}}} &  &  &  & 1 & (1-n_{M-1}) & (1-n_{M-1})\\
 &  &  &  &  &  & 1 & (1-n_{M})
\end{pmatrix}\label{eq:FC-degenerate}
\end{equation}

We also modify the $T$ transformation defined in Eq. \eqref{eq:T-tr-ss}
above by dropping its $(k+1)^{th}$ row to obtain

\begin{equation}
T=\begin{pmatrix}1\\
-\frac{1}{(1-n_{1})} & -\frac{1}{(1-n_{1})} &  &  &  &  & \text{{\huge0}}\\
 & \frac{1}{(1-n_{2})} & \frac{1}{(1-n_{2})}\\
 &  & \ddots & \ddots\\
 &  &  & \frac{\left(-1\right)^{k-1}}{(1-n_{k-1})} & \frac{\left(-1\right)^{k-1}}{(1-n_{k-1})}\\
 &  &  &  & \frac{\left(-1\right)^{k}}{(1-n_{k+1})} & \frac{\left(-1\right)^{k}}{(1-n_{k+1})}\\
 & \text{{\huge0}} &  &  &  & \ddots & \ddots\\
 &  &  &  &  &  & \frac{\left(-1\right)^{M-1}}{(1-n_{M})} & \frac{\left(-1\right)^{M-1}}{(1-n_{M})}
\end{pmatrix}\label{eq:T-tr-modified}
\end{equation}

Using again Eqs. \eqref{eq:C0-ss} and \eqref{eq:M0-ss} respectively
we obtain after applying the modified $T$ transformation in Eq. \eqref{eq:T-tr-modified}
above

\begin{align}
\mathcal{C} & =\begin{pmatrix}C_{J}+n_{1}^{2}C_{1}^{'} & n_{1}C_{1}^{'} &  &  & \text{{\huge0}}\\
n_{1}C_{1}^{'} & C_{1}^{'}+n_{2}^{2}C_{2}^{'}\\
 &  & \ddots C'_{k-1}+C_{k}+n_{k+1}^{2}C'_{k+1} & n_{k+1}C'_{k+1}\\
 &  & n_{k+1}C'_{k+1} & C'_{k+1}+n_{k+2}^{2}C'_{k+2}\\
 & \text{{\huge0}} &  &  & \ddots C_{M-1}^{'}+n_{M}^{2}C_{M}^{'} & n_{M}C_{M}^{'}\\
 &  &  &  & n_{M}C_{M}^{'} & C_{M}^{'}+C{}_{M+1}
\end{pmatrix}\nonumber \\
\nonumber \\
\label{eq:degenerate-c-matrix-1}
\end{align}

\begin{equation}
\mathbf{M}_{0}=\begin{pmatrix}\frac{1}{L'_{1}} & \frac{1}{L'_{1}}\\
\frac{1}{L'_{1}} & \frac{1}{L'_{1}}+\frac{1}{L'_{2}} & \frac{1}{L'_{2}}\\
 & \frac{1}{L'_{2}} & \frac{1}{L'_{2}}+\frac{1}{L'_{3}} & \ddots &  &  & \text{{\huge0}}\\
 &  & \ddots & \ddots\\
 &  &  &  & \frac{1}{L'_{k-1}}+\frac{1}{L'_{k+1}} & \frac{1}{L'_{k+1}}\\
 &  &  &  & \frac{1}{L'_{k+1}} & \frac{1}{L'_{k+1}}+\frac{1}{L'_{k+2}} & \ddots\\
 &  & \text{{\huge0}} &  &  & \ddots & \ddots\\
 &  &  &  &  &  &  & \frac{1}{L'_{M-1}}+\frac{1}{L'_{M}} & \frac{1}{L'_{M}}\\
 &  &  &  &  &  &  & \frac{1}{L'_{M}} & \frac{1}{L'_{M}}
\end{pmatrix}
\end{equation}

Note that the matrices above are of size $M\times M$.

In case of degeneracy $\bar{\mathbf{m}}$ vectors are computed again
using the Eq. \eqref{eq:m-vec-ss} for $1\leq j\leq M$ and applying
the $T$ transformation in Eq. \eqref{eq:T-tr-modified}. We define
some auxiliary vectors

\begin{equation}
\mathbf{\bar{m}}_{a}=\begin{pmatrix}-\frac{n_{1}C{}_{1}}{\left(1-n_{1}\right)}\\
\vdots\\
\left(-1\right)^{k-1}\frac{n_{k-1}C{}_{k-1}}{\left(1-n_{k-1}\right)}\\
\left(-1\right)^{k}\frac{n_{k+1}C{}_{k+1}}{\left(1-n_{k+1}\right)}\\
\vdots\\
(-1)^{M-1}\frac{n_{M}C_{M}}{(1-n_{M})}\\
0
\end{pmatrix}
\end{equation}

\begin{equation}
\mathbf{\bar{m}}_{b}=\begin{pmatrix}0\\
-\frac{C{}_{1}}{\left(1-n_{1}\right)}\\
\vdots\\
\left(-1\right)^{k-1}\frac{C{}_{k-1}}{\left(1-n_{k-1}\right)}\\
\left(-1\right)^{k}\frac{C{}_{k+1}}{\left(1-n_{k+1}\right)}\\
\vdots\\
\left(-1\right)^{M-1}\frac{C{}_{M}}{\left(1-n_{M}\right)}
\end{pmatrix}
\end{equation}
and the vectors $\mathbf{\bar{m}}_{a,j}$ and $\mathbf{\bar{m}}_{b,j}$
as

\begin{equation}
\mathbf{\bar{m}}_{a,j}\left(i\right)=\begin{cases}
\begin{array}{c}
0\\
\mathbf{\bar{m}}_{a}\left(i\right)
\end{array} & \begin{array}{c}
1\leq i<j\\
j\leq i\leq M
\end{array}\end{cases}\label{eq:coupling-vector-ma-1}
\end{equation}

\begin{equation}
\mathbf{\bar{m}}_{b,j}\left(i\right)=\begin{cases}
\begin{array}{c}
0\\
\mathbf{\bar{m}}_{b}\left(i\right)
\end{array} & \begin{array}{c}
1\leq i\leq j\\
j<i\leq M
\end{array}\end{cases}\label{eq:coupling-vector-mb-1}
\end{equation}
where $\mathbf{\bar{m}}_{a}\left(i\right)$, $\mathbf{\bar{m}}_{b}\left(i\right)$,
$\mathbf{\bar{m}}_{a,j}\left(i\right)$, $\mathbf{\bar{m}}_{b,j}\left(i\right)$
are the $i^{th}$ entries of the vectors $\mathbf{\bar{m}}_{a}$,
$\mathbf{\bar{m}}_{b}$, $\mathbf{\bar{m}}_{a,j}$, $\mathbf{\bar{m}}_{b,j}$
respectively. Finally we define the vector $\bar{\mathbf{m}}_{C_{k}}$
as

\begin{equation}
\mathbf{\bar{m}}_{C_{k}}\left(i\right)=\begin{cases}
\begin{array}{c}
0\\
C_{k}
\end{array} & \begin{array}{c}
i\neq k\\
i=k
\end{array}\end{cases}\label{eq:coupling-vector-mck-1}
\end{equation}
where $\mathbf{\bar{m}}_{C_{k}}\left(i\right)$ is the $i^{th}$ entry
of the vector $\mathbf{\bar{m}}_{C_{k}}$. We note that the vectors
$\mathbf{\bar{m}}_{a}$, $\mathbf{\bar{m}}_{b}$, $\mathbf{\bar{m}}_{a,j}$,
$\mathbf{\bar{m}}_{b,j}$ and $\mathbf{\bar{m}}_{C_{k}}$ are all
of length $M$.

Now we can write coupling vector $\bar{\mathbf{m}}_{j}$ to the bath
of the resistor $R_{j}$ for $1\leq j\leq M$ as a function of the
vectors defined in Eqs. \eqref{eq:coupling-vector-ma-1}, \eqref{eq:coupling-vector-mb-1},
\eqref{eq:coupling-vector-mck-1} above as

\begin{align}
\mathbf{\bar{m}}_{j} & =\bar{\mathbf{m}}_{a}\left(j\right)+\mathbf{\bar{m}}_{b}\left(j\right)+\bar{\mathbf{m}}_{C_{k}},\qquad if\; j\leq k\\
 & =\bar{\mathbf{m}}_{a}\left(j\right)+\mathbf{\bar{m}}_{b}\left(j\right),\qquad if\; j>k
\end{align}

Spectral densities $J_{j}\left(\omega\right)$ are the same as in
the non-degenerate case (Eqs. \eqref{eq:spectral-function-state-space},\eqref{eq:spectral-function-last-resistor-1})
for all resistors. Note also that dissipation treatment for the last
resistor $R_{M+1}$ is unaffected since $C{}_{M+1}$ is untouched
in Eq. \eqref{eq:degenerate-c-matrix-1}.

\subsection{\label{sub:Inductive-degenerate-case}Inductive Degenerate Case}

\begin{figure}
\begin{centering}
\includegraphics{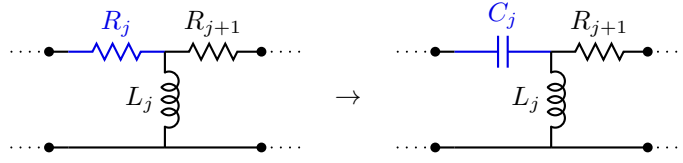}
\par\end{centering}

\caption{\label{fig:The-inductive-degenerate-stage}The inductive degenerate
case. The degenerate inductor $L_{j}$ is in the tree whereas the
resistor $R_{j}$ is replaced by the formal capacitor $C_{j}$ which
is a chord branch. $R_{j}$ can be treated in the same way as the
last shunt inductor $R_{M+1}$ (by making the substitution $C_{j}\leftarrow\frac{1}{i\omega R_{j}}$
to do a dissipation analysis) in the Brune circuit provided that the
loss introduced by $R_{j}$ is small; that is a high impedance treatment
is possible for $R_{j}$.}

\end{figure}

In Section \eqref{sub:The-Inductive-Degenerate-Stage} we noted that
an inductive degenerate stage might appear in the $j^{th}$ stage
of the Brune circuit as shown on the left part of Fig. \eqref{fig:The-inductive-degenerate-stage}.
One can deal with such a degeneracy by first replacing the resistor
$R_{j}$ extracted right before the degenerate inductor $L_{j}$ with
a formal capacitance $C_{j}$ and putting the capacitance $C_{j}$
in a chord branch (as shown on the right in Fig. \eqref{fig:The-inductive-degenerate-stage})
and treat it in the same way as the last shunt resistor $R_{M+1}$;
that is by making the substitution $C_{j}\leftarrow\frac{1}{i\omega R_{j}}$
and doing a dissipation analysis similar to the one done above for
the last shunt resistor $R_{M+1}$ provided that a high impedance
treatment is possible for $R_{j}$. The degenerate inductor $L_{j}$
should be in the tree like the ordinary inductors appearing in the
Brune circuit. We note that the degenerate inductor $L_{j}$ will
introduce a new degree of freedom since it is a tree branch compared
to the degenerate capacitor treated in the previous section which
didn't introduce any new degree of freedom.

\begin{landscape}

\begin{figure}[H]
\begin{centering}
\includegraphics{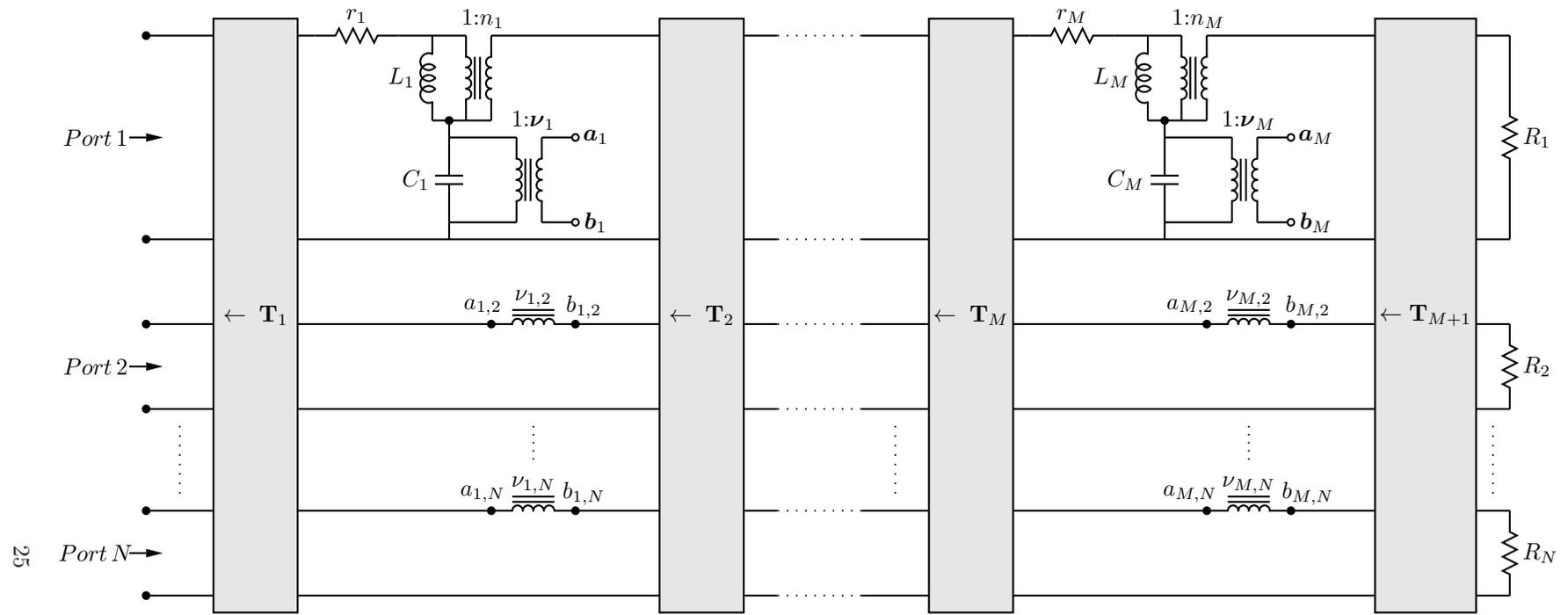} 
\par\end{centering}

\caption{\label{fig:Multiport-Brune-Circuit}The multiport Brune Circuit}
\end{figure}

\end{landscape}

\section{\label{sec:Multiport-Brune's-method}\textcolor{black}{Multiport
Brune Quantization}}

\textcolor{black}{Here we extend our analysis in the previous sections
to multiport circuits.}\textcolor{red}{{} }The multiport Brune circuit
is shown in Fig. \eqref{fig:Multiport-Brune-Circuit}. This circuit
is obtained by applying the multiport Brune's method \citep{Anderson-Moylan-1975}
in state space which we describe below. \textcolor{black}{We again
apply circuit quantization formalisms \citep{BKD,Burkard} with the
help of the effective Kirchhoff technique we introduced for the one-port
Brune circuit to derive a Hamiltonian and compute relaxation rates
for the multiport Brune circuit.}

The circuit in Fig. \eqref{fig:Multiport-Brune-Circuit} consists
of $N$ ports and $M$ stages. On the far left we have $N$ terminal
pairs corresponding to the ports. Each stage starts with the extraction
of a Belevitch transformer $\mathbf{T}_{k}$. Each $\mathbf{T}_{k}$
is a $2N$-port transformer with $N$ ports on the left and $N$ ports
on the right which we describe in detail in the next section. The
circuit representation of the Belevitch transformer is shown in Fig.
\eqref{fig:Belevitch-transformer-reflected}. The multiport Belevitch
tranformer $\mathbf{T}_{k}$ is followed by a resistor $r_{k}$ extracted
only at the first port. We will describe below how to extract Belevitch
transformers and resistors. After the resistor extraction we have
the reactive part of the multiport Brune stage. We observe that the
part of each Brune stage that comes after the Belevitch transformer
at the first port is almost identical to the one-port Brune stage.
We see however an additional transformer $\mathbf{\boldsymbol{\nu}}_{k}$
coupling the reactive circuit in the first port to the remaining ports.
$\mathbf{\boldsymbol{\nu}}_{k}$ is a multiport transformer with the
primary winding connected in parallel across the terminals of the
capacitor $C_{k}$ at stage $k$. Each of the secondary windings are
connected in series between terminals of the Belevitch transformers
at the remaining ports. The last stage consists of the Belevitch transformer
$\mathbf{T}_{M+1}$ shunted by the resistors $R_{1},\ldots,R_{N}$
.

\subsection{\label{sec:Multiport-Brune-Algorithm}Multiport Brune Algorithm}

\begin{figure}
\begin{centering}
\includegraphics{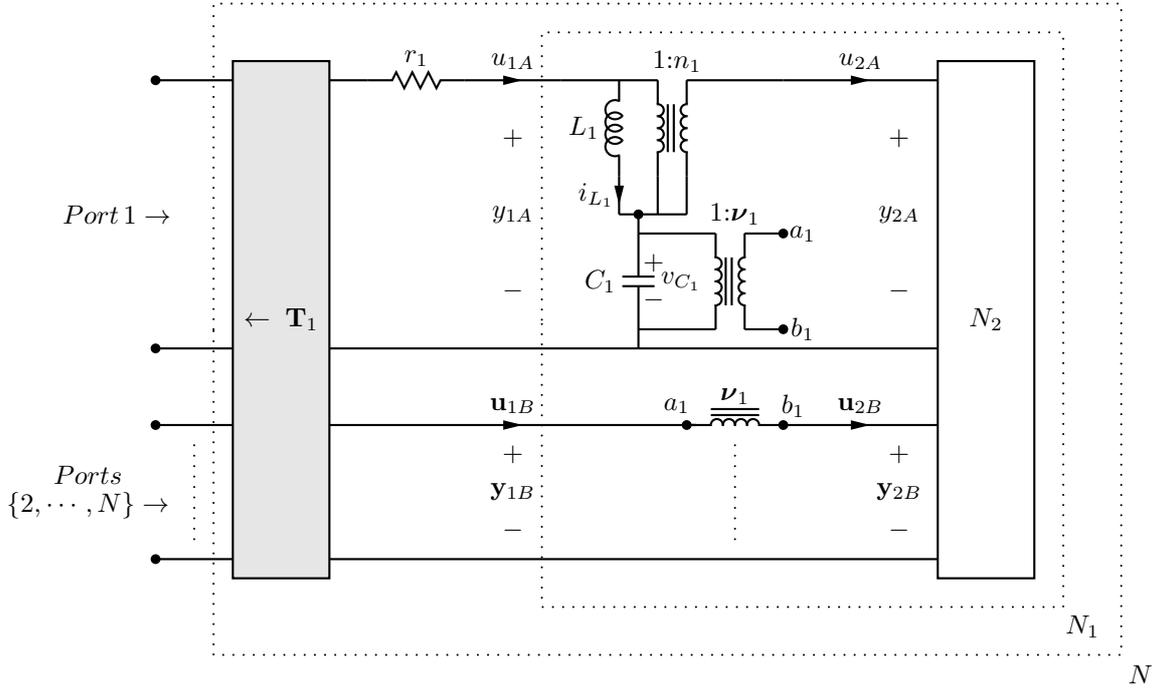} 
\par\end{centering}

\caption{\label{fig:Multiport-Brune-circuit-extraction}Multiport Brune circuit
extraction step. The extraction starts with the Belevitch transformer
$\mathbf{T}_{1}$. The circuit that follows $\mathbf{T}_{1}$ is almost
identical to the one-port Brune stage in Fig. \eqref{fig:Brune-circuit-extraction-state-space}
except the multiport transformer $\boldsymbol{\nu}_{1}$ coupling
the first port to the remaining ones.}
\end{figure}

In this section we will describe multiport generalization of the Brune's
synthesis algorithm described in state space in Section \eqref{sub:Brune's-algorithm-state-space-1-port}
for one-port networks. Fig. \eqref{fig:Multiport-Brune-circuit-extraction}
illustrates extraction of a multiport Brune stage. Since the algorithm
is recursive we will describe it only on the first stage. At each
stage the degree of the network is reduced by two hence the algorithm
terminates once a constant multiport impedance is reached as in the
one-port case.

We will only focus on the reciprocal response case, that is when $\mathbf{Z}=\mathbf{Z}^{T}$.
We will show later how the gyrators appear in the multiport Brune
circuit in case of a non-reciprocal impedance.

The synthesis of a multiport Brune stage starts with the extraction
of the multiport Belevitch transformer $\mathbf{T}_{1}$ together
with the resistor $r_{1}$ at the first port.

\subsubsection{\textcolor{black}{The Belevitch Transformer}}

\begin{figure}
\begin{centering}
\includegraphics{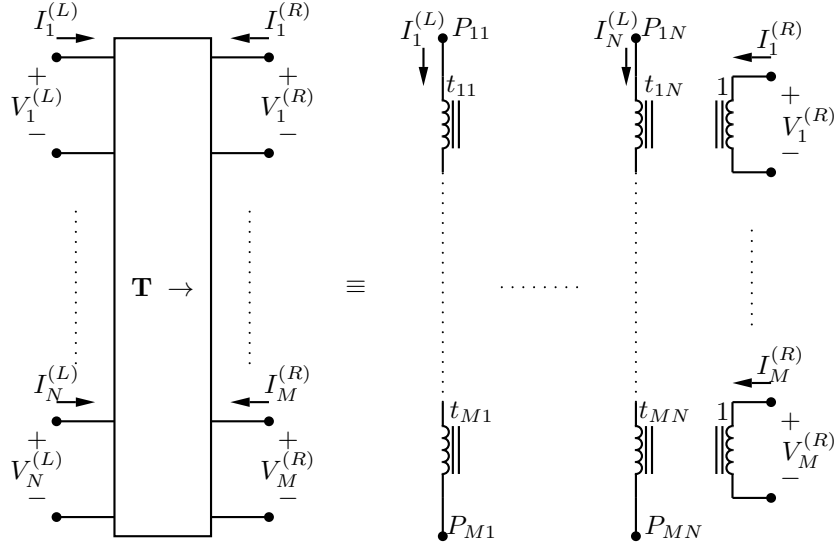} 
\par\end{centering}

\caption{\label{fig:The-Belevitch-Transformer.}The Belevitch Transformer with
$N$ ports on the left and $M$ ports on the right. On the right side
we see the detailed circuit representation of this device. There is
a turns ratio matrix $\mathbf{T=\left(\protect\begin{array}{ccc}
t_{11} & \cdots & t_{1N}\protect\\
\vdots & \ddots & \vdots\protect\\
t_{M1} & \cdots & t_{MN}
\protect\end{array}\right)}$ associated with the Belevitch transformer which relates currents
and voltages on both side of the device as given in Eqs. \eqref{eq:Belevitch-relation-current}
and \eqref{eq:Belevitch-relation-voltage}. The arrow in the box is
used to refer to the asymmetrical character of the Belevitch transformer.}
\end{figure}

A generic multiport Belevitch transformer with $N$ ports on the left
and $M$ ports on the right is shown in Fig. \eqref{fig:The-Belevitch-Transformer.}
on the left. The detailed circuit representation of the Belevitch
transformer $\mathbf{T}$ is shown on the right of Fig. \eqref{fig:The-Belevitch-Transformer.}
which defines a $M\times N$ matrix for $\mathbf{T}$

\begin{equation}
\mathbf{T}=\left(\begin{array}{ccc}
t_{11} & \cdots & t_{1N}\\
\vdots & \ddots & \vdots\\
t_{M1} & \cdots & t_{MN}
\end{array}\right)\label{eq:Belevitch-transformer-matrix}
\end{equation}

Let the current vectors $\mathbf{I}^{(L)}$ and $\mathbf{I}^{(R)}$
be the vectors holding the currents at the ports on the left side
and the right side of the Belevitch transformer $\mathbf{T}$ , respectively,
i.e.

\begin{eqnarray}
\mathbf{I}^{(L)} & = & \left(I_{1}^{(L)},\ldots,I_{N}^{(L)}\right)^{T}\\
\mathbf{I}^{(R)} & = & \left(I_{1}^{(R)},\ldots,I_{M}^{(R)}\right)^{T}
\end{eqnarray}
and let the vectors $\mathbf{V}^{(L)}$ and $\mathbf{V}^{(R)}$ be
the vectors holding the voltages at the ports on the left side and
the right side of the Belevitch transformer $\mathbf{T}$ , respectively,
i.e.

\begin{eqnarray}
\mathbf{V}^{(L)} & = & \left(V_{1}^{(L)},\ldots,V_{N}^{(L)}\right)^{T}\\
\mathbf{V}^{(R)} & = & \left(V_{1}^{(R)},\ldots,V_{M}^{(R)}\right)^{T}
\end{eqnarray}
then we can write the Belevitch transformer relations as

\begin{eqnarray}
\mathbf{I}^{(R)} & = & -\mathbf{T}\mathbf{I}^{(L)}\label{eq:Belevitch-relation-current}\\
\mathbf{V}^{(L)} & = & \mathbf{T}^{T}\mathbf{V}^{(R)}\label{eq:Belevitch-relation-voltage}
\end{eqnarray}

\begin{figure}
\begin{centering}
\includegraphics{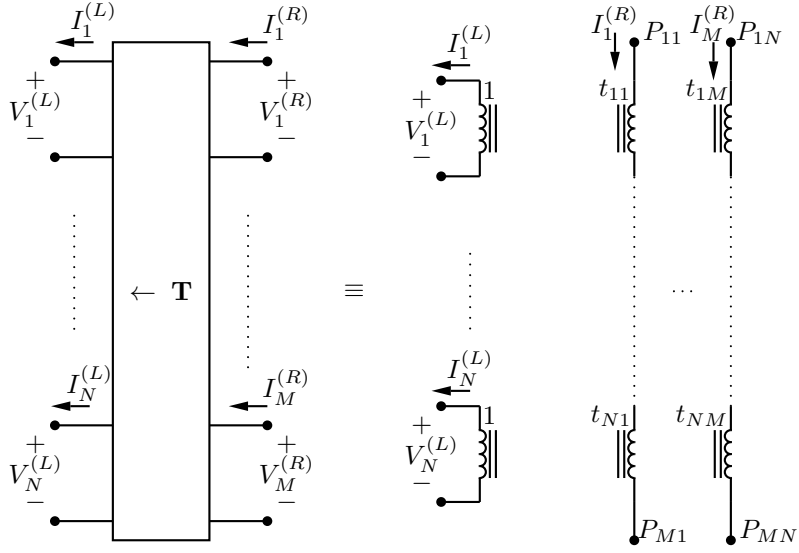} 
\par\end{centering}

\caption{\label{fig:Belevitch-transformer-reflected}The Belevitch transformer
in reflected form as used in\textcolor{red}{{} }\textcolor{black}{the
multiport Brune circuit} in Fig. \eqref{fig:Multiport-Brune-Circuit}.
The current and voltage relations for this transformer are given in
Eqs. \eqref{eq:Belevitch-relation-current-reflected}, \eqref{eq:Belevitch-relation-voltage-reflected}.}
\end{figure}

One should recognize the asymmetrical character of the Belevitch transformer
which we noted by putting an arrow in the box representing the Belevitch
transformer in Fig. \eqref{fig:The-Belevitch-Transformer.}. However
in the multiport Brune circuit in Fig. \eqref{fig:Multiport-Brune-Circuit}
we use the Belevitch transformer in Fig. \eqref{fig:The-Belevitch-Transformer.}
in a reflected form as shown in Fig. \eqref{fig:Belevitch-transformer-reflected}.

The current and voltage relations for the reflected Belevitch transformer
in Fig. \eqref{fig:Belevitch-transformer-reflected} are given by

\begin{eqnarray}
\mathbf{I}^{(L)} & = & \mathbf{T}\mathbf{I}^{(R)}\label{eq:Belevitch-relation-current-reflected}\\
\mathbf{V}^{(R)} & = & \mathbf{T}^{T}\mathbf{V}^{(L)}\label{eq:Belevitch-relation-voltage-reflected}
\end{eqnarray}

It is interesting to note the similarity of the Belevitch transformer
relations in Eqs. \eqref{eq:Belevitch-relation-current} and \eqref{eq:Belevitch-relation-voltage}
to the Kirchhoff's laws given in terms of the loop matrix $\mathbf{F}$
in Eqs. \eqref{eq:Kirchhoff-current-law} and \eqref{eq:Kirchhoff-voltage-law}
given in Section \eqref{sub:Quantization-state-space-Brune-circuit-1-port}.

Let the impedance matrices seen at the ports of the networks labeled
$N$ and $N_{1}$ in Fig. \eqref{fig:Multiport-Brune-circuit-extraction}
be $\mathbf{Z}$ and $\mathbf{Z}_{1}$ respectively. Assuming that
$\mathbf{Z}$ is positive-real the aim of the resistance and Belevitch
transformer extraction is to get $\mathbf{Z}_{1}$ $PR$ together
with $Re\left[\mathbf{Z}_{1,11}\left(j\omega_{1}\right)\right]=0$
for some frequency $\omega_{1}$ where $\mathbf{Z}_{1,11}$ is the
$(1,1)$ entry of the impedance matrix $\mathbf{Z}_{1}$. As we will
see below those are necessary and sufficient conditions for the multiport
Brune algorithm.

\subsubsection{Resistance Extraction}

The most direct way to get $Re\left[\mathbf{Z}_{1,11}\left(j\omega_{1}\right)\right]=0$
is to first make an eigenvalue decomposition for the Hermitian part
$\mathbf{Z}_{H}$ of $\mathbf{Z}$ at each frequency $\omega$

\begin{equation}
\mathbf{Z}_{H}\left(j\omega\right)=U\left(\omega\right)S\left(\omega\right)U^{T}\left(\omega\right)
\end{equation}
where

\begin{equation}
\mathbf{Z}_{H}\left(j\omega\right)=\frac{1}{2}\left(\mathbf{Z}\left(j\omega\right)+\mathbf{Z}^{T}\left(-j\omega\right)\right)\label{eq:Hermitian-part-imaginary-axis}
\end{equation}
$S\left(\omega\right)$ can be assumed to be the diagonal matrix having
the eigenvalues $\alpha_{j}\left(\omega\right)$'s for $1\leq j\leq N$
on its diagonal in increasing order ($\alpha_{j}\left(\omega\right)\geq0$
since $\mathbf{Z}_{H}\left(j\omega\right)$ is positive semi-definite):

\begin{equation}
S\left(\omega\right)=\left(\begin{array}{ccc}
\alpha_{1}\left(\omega\right) &  & \boldsymbol{0}\\
 & \ddots\\
\boldsymbol{0} &  & \alpha_{N}\left(\omega\right)
\end{array}\right)
\end{equation}
and $U\left(\omega\right)$ can be chosen to be orthogonal.

If we then choose the following value for the extracted resistor $r_{1}$

\begin{equation}
r_{1}=\underset{0\leq\omega\leq\infty}{\min}\alpha_{1}\left(\omega\right)\label{eq:multiport-Brune-resistance-extraction-orthogonal-T-increasing-singular-values}
\end{equation}
with $\omega_{1}$ being the frequency at which the minimum in Eq.
\eqref{eq:multiport-Brune-resistance-extraction-orthogonal-T-increasing-singular-values}
occurs

\begin{equation}
\alpha_{1}\left(\omega_{1}\right)=r_{1}\label{eq:multiport-Brune-min-resistor-frequency}
\end{equation}

We have $Re\left[\mathbf{Z}_{1,11}\left(j\omega_{1}\right)\right]=0$
with

\begin{equation}
\mathbf{Z}_{1}=U^{T}\left(\omega_{1}\right)\mathbf{Z}U\left(\omega_{1}\right)-r_{1}\left(\begin{array}{cccc}
1 &  &  & \boldsymbol{0}\\
 & 0\\
 &  & \ddots\\
\boldsymbol{0} &  &  & 0
\end{array}\right)
\end{equation}
One then needs to choose the turns-ratio matrix $\mathbf{T}_{1}$
as

\begin{equation}
\mathbf{T}_{1}=U\left(\omega_{1}\right)
\end{equation}

One can also imagine extracting first the resistor and then the Belevitch
transformer. The value of the resistor $r_{1}$ in that case can be
computed using the following formula \citep{Newcomb-Resistance-Extraction}

\begin{equation}
r_{1}=\underset{0\leq\omega\leq\infty}{\min}\boldsymbol{\Delta}\left(\omega\right)/\boldsymbol{\Delta}{}_{11}\left(\omega\right)\label{eq:Resistance-extraction-multiport-Brune}
\end{equation}
where $\boldsymbol{\Delta}\left(\omega\right)$ is the determinant
and $\boldsymbol{\Delta}_{11}\left(\omega\right)$ is the $\left(1,1\right)$
minor of $\mathbf{Z}_{H}\left(j\omega\right)$.

Let $\omega_{0}$ be the frequency at which the minimum in Eq. \eqref{eq:Resistance-extraction-multiport-Brune}
occurs such that

\begin{equation}
r_{1}=\boldsymbol{\Delta}\left(\omega_{0}\right)/\boldsymbol{\Delta}_{11}\left(\omega_{0}\right)\label{eq:Resistance-extraction-minimum-frequency}
\end{equation}

Then the Belevitch transformer matrix $\mathbf{T}_{1}$ is given by
the matrix that simultaneously diagonalizes \citep{Newcomb-simultaneous-diagonalization}
$\mathbf{Z}_{H}\left(j\omega_{0}\right)$ and $\left(\begin{array}{cccc}
1 &  &  & \boldsymbol{0}\\
 & 0\\
 &  & \ddots\\
\boldsymbol{0} &  &  & 0
\end{array}\right)$ such that

\begin{equation}
\mathbf{Z}_{H}\left(j\omega_{0}\right)=\mathbf{T}_{1}\mathbf{D}\mathbf{T}_{1}^{T}\label{eq:T1-matrix-Hermitian-part-diagonalization}
\end{equation}
where $\mathbf{D}$ is a diagonal matrix with $\mathbf{D}\left(1,1\right)=r_{1}$.
$\mathbf{T}_{1}$ in that case can be found using the Gauss diagonalization
procedure\textcolor{black}{{} \citep{Newcomb-Resistance-Extraction}.}

The formula given in Eq. \eqref{eq:Resistance-extraction-multiport-Brune}
is nice in the sense that it doesn't require an eigenvalue decomposition
for each frequency $\omega$, $0\leq\omega\leq\infty$. However the
Belevitch transformer matrix $\mathbf{T}_{1}$ obtained by the above
method is in general non-orthogonal.

\subsubsection{\label{sub:Extraction-of-the-multiport-Brune-stage}Extraction of
the Reactive Part of the Multiport Brune Stage}

Let the subnetwork $N_{2}$ in Fig. \eqref{fig:Multiport-Brune-circuit-extraction}
be described by the following state-space equations

\begin{eqnarray}
\mathbf{\dot{x}}_{2} & = & \mathbf{A}_{2}\mathbf{x}_{2}+\mathbf{B}_{2}\mathbf{u}_{2}\label{eq:multiport-Brune-N2-ss-description-start}\\
\mathbf{y}_{2} & = & \mathbf{C}_{2}\mathbf{x}_{2}+\mathbf{D}_{2}\mathbf{u}_{2}
\end{eqnarray}
where

\begin{equation}
\mathbf{B}_{2}=\left(\begin{array}{cc}
\mathbf{B}_{2A} & \mathbf{B}_{2B}\end{array}\right)
\end{equation}

\begin{equation}
\mathbf{C}_{2}=\left(\begin{array}{c}
\mathbf{C}_{2A}\\
\mathbf{C}_{2B}
\end{array}\right)
\end{equation}

\begin{equation}
\mathbf{D}_{2}=\left(\begin{array}{cc}
D_{2AA} & \mathbf{D}_{2AB}\\
\mathbf{D}_{2BA} & \mathbf{D}_{2BB}
\end{array}\right)
\end{equation}
and

\begin{equation}
\mathbf{u}_{2}=\left(\begin{array}{c}
u_{2A}\\
\mathbf{u}_{2B}
\end{array}\right)
\end{equation}

\begin{equation}
\mathbf{y}_{2}=\left(\begin{array}{c}
y_{2A}\\
\mathbf{y}_{2B}
\end{array}\right)\label{eq:multiport-Brune-N2-ss-description-end}
\end{equation}
where $u_{2A}$ is the current into the first port of the subnetwork
$N_{2}$ in Fig. \eqref{fig:Multiport-Brune-circuit-extraction} and
$\mathbf{u}_{2B}$ is the vector holding the currents at the remaining
ports (ports $2-N$) of the subnetwork $N_{2}$. Similarly $y_{2A}$
is the voltage across the first port of the subnetwork $N_{2}$ and
$\mathbf{y}_{2B}$ is the vector holding the voltages across the remaining
ports (ports $2-N$) of the subnetwork $N_{2}$. $\left\{ \mathbf{A}_{2},\mathbf{B}_{2},\mathbf{C}_{2},\mathbf{D}_{2}\right\} $
is a state-space realization for the impedance $\mathbf{Z}_{2}\left(s\right)$
seen at the ports of the network $N_{2}$.

Then the network $N_{1}$ is described by the the following equations

\begin{equation}
\left(\begin{array}{c}
\mathbf{\dot{x}}_{2}\\
\dot{x}_{C_{1}}\\
\dot{x}_{L_{1}}
\end{array}\right)=\left(\begin{array}{ccc}
\mathbf{A}_{2} & 0 & -\frac{\mathbf{B}_{2A}}{n_{1}\sqrt{L_{1}}}\\
0 & 0 & \frac{1}{n_{1}\sqrt{L_{1}C_{1}}}\\
\frac{\mathbf{C}_{2A}}{n_{1}\sqrt{L_{1}}} & -\frac{1}{n_{1}\sqrt{L_{1}C_{1}}} & -\frac{D_{2AA}}{n_{1}^{2}L_{1}}
\end{array}\right)\left(\begin{array}{c}
\mathbf{x}_{2}\\
x_{C_{1}}\\
x_{L_{1}}
\end{array}\right)+\left(\begin{array}{cc}
\mathbf{B}_{2A}/n_{1} & \mathbf{B}_{2B}\\
\frac{1-1/n_{1}}{\sqrt{C_{1}}} & \frac{\boldsymbol{\nu}_{1}^{T}}{\sqrt{C_{1}}}\\
\frac{D_{2AA}}{n_{1}^{2}\sqrt{L_{1}}} & \frac{\mathbf{D}_{2AB}}{n_{1}\sqrt{L_{1}}}
\end{array}\right)\left(\begin{array}{c}
u_{1A}\\
\mathbf{u}_{1B}
\end{array}\right)\label{eq:multiport-Brune-state-space-eqs-N1-time-evolution}
\end{equation}

\begin{equation}
\left(\begin{array}{c}
y_{1A}\\
\mathbf{y}_{1B}
\end{array}\right)=\left(\begin{array}{ccc}
\frac{\mathbf{C}_{2A}}{n_{1}} & \frac{1-1/n_{1}}{\sqrt{C_{1}}} & -\frac{D_{2AA}}{n_{1}^{2}\sqrt{L_{1}}}\\
\mathbf{C}_{2B} & \frac{\boldsymbol{\nu}_{1}}{\sqrt{C_{1}}} & -\frac{\mathbf{D}_{2BA}}{n_{1}\sqrt{L_{1}}}
\end{array}\right)\left(\begin{array}{c}
\mathbf{x}_{2}\\
x_{C_{1}}\\
x_{L_{1}}
\end{array}\right)+\left(\begin{array}{cc}
D_{2AA}/n_{1}^{2} & \mathbf{D}_{2AB}/n_{1}\\
\mathbf{D}_{2BA}/n_{1} & \mathbf{D}_{2BB}
\end{array}\right)\left(\begin{array}{c}
u_{1A}\\
\mathbf{u}_{1B}
\end{array}\right)\label{eq:multiport-Brune-state-space-eqs-N1-input-output}
\end{equation}
from which we identify

\begin{equation}
\mathbf{A}_{1}=\left(\begin{array}{ccc}
\mathbf{A}_{2} & 0 & -\frac{\mathbf{B}_{2A}}{n_{1}\sqrt{L_{1}}}\\
0 & 0 & \frac{1}{n_{1}\sqrt{L_{1}C_{1}}}\\
\frac{\mathbf{C}_{2A}}{n_{1}\sqrt{L_{1}}} & -\frac{1}{n_{1}\sqrt{L_{1}C_{1}}} & -\frac{D_{2AA}}{n_{1}^{2}L_{1}}
\end{array}\right)\label{eq:A1-multiport-Brune}
\end{equation}

\begin{equation}
\mathbf{B}_{1}=\left(\begin{array}{cc}
\mathbf{B}_{2A}/n_{1} & \mathbf{B}_{2B}\\
\frac{1-1/n_{1}}{\sqrt{C_{1}}} & \frac{\boldsymbol{\nu}_{1}^{T}}{\sqrt{C_{1}}}\\
\frac{D_{2AA}}{n_{1}^{2}\sqrt{L_{1}}} & \frac{\mathbf{D}_{2AB}}{n_{1}\sqrt{L_{1}}}
\end{array}\right)
\end{equation}

\begin{equation}
\mathbf{C}_{1}=\left(\begin{array}{ccc}
\frac{\mathbf{C}_{2A}}{n_{1}} & \frac{1-1/n_{1}}{\sqrt{C_{1}}} & -\frac{D_{2AA}}{n_{1}^{2}\sqrt{L_{1}}}\\
\mathbf{C}_{2B} & \frac{\boldsymbol{\nu}_{1}}{\sqrt{C_{1}}} & -\frac{\mathbf{D}_{2BA}}{n_{1}\sqrt{L_{1}}}
\end{array}\right)
\end{equation}

\begin{equation}
\mathbf{D}_{1}=\left(\begin{array}{cc}
D_{2AA}/n_{1}^{2} & \mathbf{D}_{2AB}/n_{1}\\
\mathbf{D}_{2BA}/n_{1} & \mathbf{D}_{2BB}
\end{array}\right)\label{eq:D1-multiport-Brune}
\end{equation}
and

\begin{equation}
\mathbf{x}_{1}=\left(\begin{array}{c}
\mathbf{x}_{2}\\
x_{C_{1}}\\
x_{L_{1}}
\end{array}\right)
\end{equation}
where $x_{C_{1}}=\sqrt{C_{1}}v_{C_{1}}$, $x_{L_{1}}=\sqrt{L_{1}}i_{L_{1}}$
and $\boldsymbol{\nu}_{1}=\left(\nu_{12},\ldots,\nu_{1N}\right)^{T}$;
$u_{1A}$ is the current into the first port of the subnetwork $N_{1}$
in Fig. \eqref{fig:Multiport-Brune-circuit-extraction} and $\mathbf{u}_{1B}$
is the vector holding the currents at the remaining ports (ports $2-N$)
of the subnetwork $N_{1}$. Similarly $y_{1A}$ is the voltage across
the first port of the subnetwork $N_{1}$ and $\mathbf{y}_{1B}$ is
the vector holding the voltages across the remaining ports (ports
$2-N$) of the subnetwork $N_{1}$. $\left\{ \mathbf{A}_{1},\mathbf{B}_{1},\mathbf{C}_{1},\mathbf{D}_{1}\right\} $
is then a realization for the impedance $\mathbf{Z}_{1}\left(s\right)$
seen at the ports of the network $N_{1}$.

Now we will state the multiport version of the fundamental lemma stated
in Section \eqref{sub:Brune's-algorithm-state-space-1-port} to show
how to transform state-space equations given for a minimal realization
of the impedance $\mathbf{Z}_{1}$ into the form in Eqs. \eqref{eq:multiport-Brune-state-space-eqs-N1-time-evolution}
and \eqref{eq:multiport-Brune-state-space-eqs-N1-input-output}. For
details refer to \citep{Anderson-Moylan-1975}.

\emph{The Multiport Synthesis Lemma}

Let $\left\{ \mathbf{A}_{a},\mathbf{B}_{a},\mathbf{C}_{a},\mathbf{D}_{a}\right\} $
be a minimal realization corresponding to the positive-real impedance
$\mathbf{Z}_{1}\left(s\right)$ satisfying $\mathbf{Z}_{1,11}\left(j\omega_{0}\right)+\mathbf{Z}_{1,11}\left(-j\omega_{0}\right)=0$
for some frequency $\omega_{0}$, $\mathbf{Z}_{1,11}$ is the $(1,1)$
entry of the impedance matrix $\mathbf{Z}_{1}$ (We also assume that
$j\omega_{0}$ is not an eigenvalue of $\mathbf{A}_{a}$). Then there
exists a coordinate transformation matrix $\mathbf{T}$ such that
$\mathbf{A}_{1}=\mathbf{T}\mathbf{A}_{a}\mathbf{T}^{-1}$, $\mathbf{B}_{1}=\mathbf{T}\mathbf{B}_{a}$,
$\mathbf{C}_{1}=\mathbf{C}_{a}\mathbf{T}^{-1}$ and $\mathbf{D}_{1}=\mathbf{D}_{a}$
are of the form given in Eqs. (\ref{eq:A1-multiport-Brune}-\ref{eq:D1-multiport-Brune}).

To compute $\mathbf{T}$ we will follow the algorithm described in
\emph{Fundamental Lemma} in Section \eqref{sub:Brune's-algorithm-state-space-1-port}.
Before applying the algorithm we set

\begin{eqnarray}
\mathbf{b}_{a} & = & \mathbf{B}_{a}\mathbf{e}_{1}\\
\mathbf{c}_{a}^{T} & = & \mathbf{C}_{a}^{T}\mathbf{e}_{1}
\end{eqnarray}
and apply the one-port algorithm described in the \emph{Fundamental
Lemma} in Section \eqref{sub:Brune's-algorithm-state-space-1-port}
to the set $\left\{ \mathbf{A}_{a},\mathbf{b}_{a},\mathbf{c}_{a},\mathbf{D}_{a}\right\} $
where $\mathbf{e}_{1}=\left(\begin{array}{cccc}
1 & 0 & \ldots & 0\end{array}\right)^{T}$; that is we apply the one-port algorithm by picking up the first
columns of $\mathbf{B}_{a}$ and $\mathbf{C}_{a}^{T}$ matrices.

To see why $\left\{ \mathbf{A}_{1},\mathbf{B}_{1},\mathbf{C}_{1},\mathbf{D}_{1}\right\} $
is an equivalent realization for the impedance $\mathbf{Z}_{1}\left(s\right)$
see Section \eqref{sec:State-Space-Formalism} or Theorem (3.3.9)
in \citep{Anderson-Vongpanitlerd}.

\subsubsection{The Multiport Capacitive Degenerate Stage}

\begin{figure}
\begin{centering}
\includegraphics{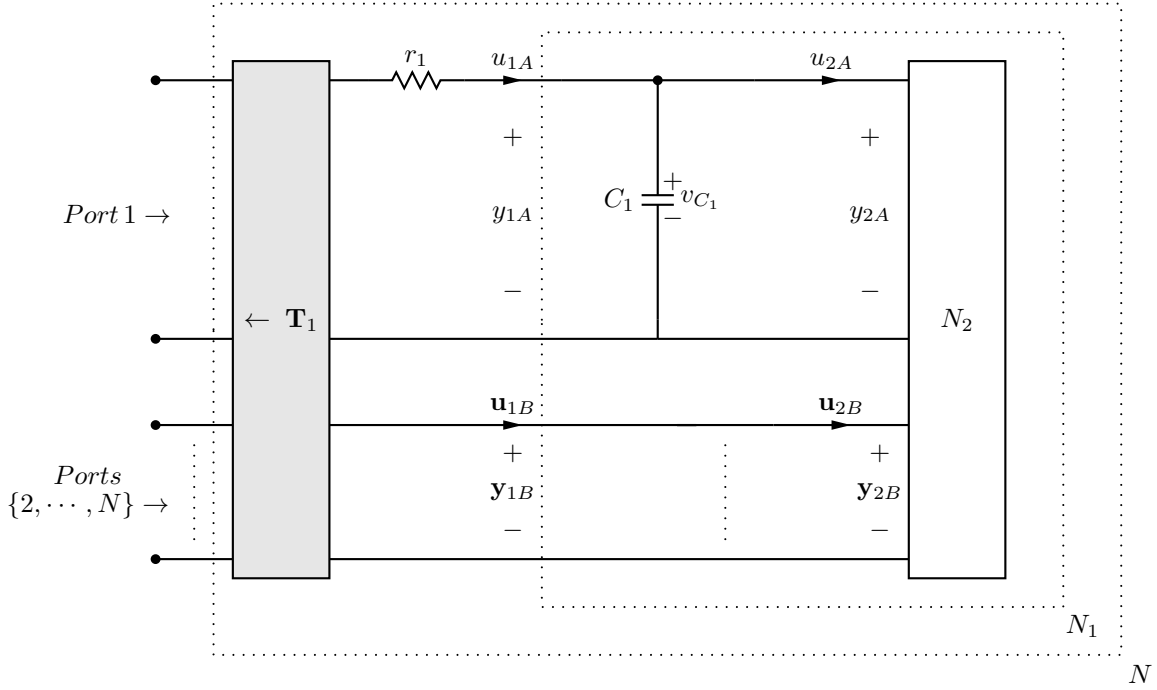} 
\par\end{centering}

\caption{\label{fig:Multiport-capacitive-degenerate-stage}Extraction of the
multiport capacitive denegenerate stage in the multiport Brune circuit
when $\omega_{1}=\infty$ in Eq. \eqref{eq:multiport-Brune-min-resistor-frequency}.
We observe that the $\boldsymbol{\nu}$-type transformer is not necessary
in that case.}
\end{figure}

Similar to our discussion in Section \eqref{sub:The-Capacitive-Degenerate-Stage-1-port-ss}
it is possible that the minimum in Eq. \eqref{eq:multiport-Brune-min-resistor-frequency}
occurs at infinity, $\omega_{1}=\infty$. Such a case needs the extraction
of the degenerate reactive stage shown in Fig. \eqref{fig:Multiport-capacitive-degenerate-stage}.
As in the one-port case the reactive Brune stage doesn't involve any
inductive part. Note also that we don't have the $\boldsymbol{\nu}$-type
transformer coupling the first port to the remaining ports.

To synthesize such a stage we need to modify our treatment in Section
\eqref{sub:Extraction-of-the-multiport-Brune-stage}. Assuming that
the subnetwork $N_{2}$ in Fig. \eqref{fig:Multiport-capacitive-degenerate-stage}
is described by Eqs. \eqref{eq:multiport-Brune-N2-ss-description-start}-\eqref{eq:multiport-Brune-N2-ss-description-end},
the network $N_{1}$ is described by

\begin{equation}
\left(\begin{array}{c}
\mathbf{\dot{x}}_{2}\\
\dot{x}_{C_{1}}
\end{array}\right)=\left(\begin{array}{cc}
\mathbf{A}_{2}-\frac{\mathbf{B}_{2A}\mathbf{C}_{2A}}{D_{2AA}} & \frac{\mathbf{B}_{2A}}{D_{2AA}\sqrt{C_{1}}}\\
\frac{\mathbf{C}_{2A}}{D_{2AA}\sqrt{C_{1}}} & -\frac{1}{D_{2AA}C_{1}}
\end{array}\right)\left(\begin{array}{c}
\mathbf{x}_{2}\\
x_{C_{1}}
\end{array}\right)+\left(\begin{array}{cc}
0 & \mathbf{B}_{2B}-\frac{\mathbf{B}_{2A}\mathbf{D}_{2AB}}{D_{2AA}}\\
\frac{1}{\sqrt{C_{1}}} & \frac{\mathbf{D}_{1AB}}{D_{2AA}\sqrt{C_{1}}}
\end{array}\right)\left(\begin{array}{c}
u_{1A}\\
\mathbf{u}_{1B}
\end{array}\right)\label{eq:multiport-Brune-state-space-eqs-N1-time-evolution-degenerate}
\end{equation}

\begin{equation}
\left(\begin{array}{c}
y_{1A}\\
\mathbf{y}_{1B}
\end{array}\right)=\left(\begin{array}{cc}
0 & \frac{1}{\sqrt{C_{1}}}\\
\mathbf{C}_{2B}-\frac{\mathbf{D}_{2BA}\mathbf{C}_{2A}}{D_{2AA}} & \frac{\mathbf{D}_{2BA}}{D_{2AA}\sqrt{C_{1}}}
\end{array}\right)\left(\begin{array}{c}
\mathbf{x}_{2}\\
x_{C_{1}}
\end{array}\right)+\left(\begin{array}{cc}
0 & 0\\
0 & \mathbf{D}_{2BB}-\frac{\mathbf{D}_{2BA}\mathbf{D}_{2AB}}{D_{2AA}}
\end{array}\right)\left(\begin{array}{c}
u_{1A}\\
\mathbf{u}_{1B}
\end{array}\right)\label{eq:multiport-Brune-state-space-eqs-N1-input-output-degenerate}
\end{equation}
from which we identify

\begin{equation}
\mathbf{A}_{1}=\left(\begin{array}{cc}
\mathbf{A}_{2}-\frac{\mathbf{B}_{2A}\mathbf{C}_{2A}}{D_{2AA}} & \frac{\mathbf{B}_{2A}}{D_{2AA}\sqrt{C_{1}}}\\
\frac{\mathbf{C}_{2A}}{D_{2AA}\sqrt{C_{1}}} & -\frac{1}{D_{2AA}C_{1}}
\end{array}\right)\label{eq:A1-multiport-Brune-degenerate}
\end{equation}

\begin{equation}
\mathbf{B}_{1}=\left(\begin{array}{cc}
0 & \mathbf{B}_{2B}-\frac{\mathbf{B}_{2A}\mathbf{D}_{2AB}}{D_{2AA}}\\
\frac{1}{\sqrt{C_{1}}} & \frac{\mathbf{D}_{1AB}}{D_{2AA}\sqrt{C_{1}}}
\end{array}\right)
\end{equation}

\begin{equation}
\mathbf{C}_{1}=\left(\begin{array}{cc}
0 & \frac{1}{\sqrt{C_{1}}}\\
\mathbf{C}_{2B}-\frac{\mathbf{D}_{2BA}\mathbf{C}_{2A}}{D_{2AA}} & \frac{\mathbf{D}_{2BA}}{D_{2AA}\sqrt{C_{1}}}
\end{array}\right)
\end{equation}

\begin{equation}
\mathbf{D}_{1}=\left(\begin{array}{cc}
0 & 0\\
0 & \mathbf{D}_{2BB}-\frac{\mathbf{D}_{2BA}\mathbf{D}_{2AB}}{D_{2AA}}
\end{array}\right)\label{eq:D1-multiport-Brune-degenerate}
\end{equation}
and

\begin{equation}
\mathbf{x}_{1}=\left(\begin{array}{c}
\mathbf{x}_{2}\\
x_{C_{1}}
\end{array}\right)
\end{equation}
where $x_{C_{1}}=\sqrt{C_{1}}v_{C_{1}}$; $u_{1A}$ is the current
into the first port of the subnetwork $N_{1}$ in Fig. \eqref{fig:Multiport-capacitive-degenerate-stage}
and $\mathbf{u}_{1B}$ is the vector holding the currents at the remaining
ports (ports $2-N$) of the subnetwork $N_{1}$. Similarly $y_{1A}$
is the voltage across the first port of the subnetwork $N_{1}$ and
$\mathbf{y}_{1B}$ is the vector holding the voltages across the remaining
ports (ports $2-N$) of the subnetwork $N_{1}$. $\left\{ \mathbf{A}_{1},\mathbf{B}_{1},\mathbf{C}_{1},\mathbf{D}_{1}\right\} $
is then a realization for the impedance $\mathbf{Z}_{1}\left(s\right)$
seen at the ports of the network $N_{1}$.

One needs to modify also the \emph{Multiport Synthesis Lemma} as follows:

\emph{The Multiport Synthesis Lemma (multiport capacitive degenerate
case)}

Let $\left\{ \mathbf{A}_{a},\mathbf{B}_{a},\mathbf{C}_{a},\mathbf{D}_{a}\right\} $
be a minimal realization corresponding to the positive-real impedance
$\mathbf{Z}_{1}\left(s\right)$ satisfying $\mathbf{Z}_{1,11}\left(j\omega_{0}\right)+\mathbf{Z}_{1,11}\left(-j\omega_{0}\right)=0$
for some $\omega_{0}=\infty$, $\mathbf{Z}_{1,11}$ is the $(1,1)$
entry of the impedance matrix $\mathbf{Z}_{1}$. Then there exists
a coordinate transformation matrix $\mathbf{T}$ such that $\mathbf{A}_{1}=\mathbf{T}\mathbf{A}_{a}\mathbf{T}^{-1}$,
$\mathbf{B}_{1}=\mathbf{T}\mathbf{B}_{a}$, $\mathbf{C}_{1}=\mathbf{C}_{a}\mathbf{T}^{-1}$
and $\mathbf{D}_{1}=\mathbf{D}_{a}$ are of the form given in Eqs.
(\eqref{eq:A1-multiport-Brune-degenerate}-\eqref{eq:D1-multiport-Brune-degenerate}).

To compute $\mathbf{T}$ we will follow the algorithm described in
\emph{The Fundamental Lemma (one-port capacitive degenerate case)}
in Section \eqref{sub:The-Capacitive-Degenerate-Stage-1-port-ss}.
Before applying the algorithm we set

\begin{eqnarray}
\mathbf{b}_{a} & = & \mathbf{B}_{a}\mathbf{e}_{1}\\
\mathbf{c}_{a}^{T} & = & \mathbf{C}_{a}^{T}\mathbf{e}_{1}
\end{eqnarray}
and apply the one-port algorithm described in \emph{The Fundamental
Lemma (one-port capacitive degenerate case)} in \textcolor{black}{Section}
\eqref{sub:The-Capacitive-Degenerate-Stage-1-port-ss} to the set
$\left\{ \mathbf{A}_{a},\mathbf{b}_{a},\mathbf{c}_{a},\mathbf{D}_{a}\right\} $
where $\mathbf{e}_{1}=\left(\begin{array}{cccc}
1 & 0 & \ldots & 0\end{array}\right)^{T}$; that is we apply the one-port algorithm by picking up the first
columns of $\mathbf{B}_{a}$ and $\mathbf{C}_{a}^{T}$ matrices.

\subsubsection{\label{sub:The-Multiport-Inductive-Degenerate-Stage}The Multiport
Inductive Degenerate Stage}

\begin{figure}
\begin{centering}
\includegraphics{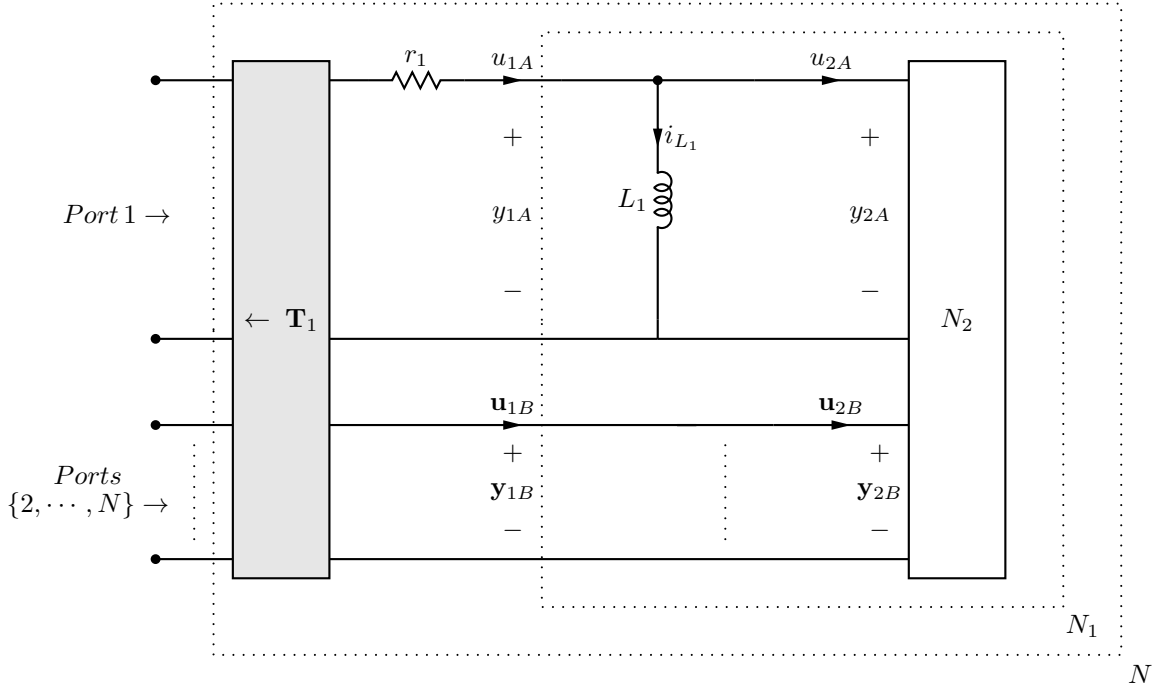}
\par\end{centering}

\caption{\label{fig:Extraction-of-the-multiport-inductive-degenerate-stage}Extraction
of the multiport inductive degenerate stage.}

\end{figure}

Similar to our discussion in Section \eqref{sub:The-Inductive-Degenerate-Stage}
for the one-port Brune circuit extraction there is the possibility
of the minimum in Eq. \eqref{eq:multiport-Brune-min-resistor-frequency}
occuring at $\omega_{1}=0$. Such a case corresponds to the extraction
of an inductive degenerate stage as shown in Fig. \eqref{fig:Extraction-of-the-multiport-inductive-degenerate-stage}.
Again it is straightforward to extend the multiport state-space Brune
algorithm presented in Section \eqref{sub:Extraction-of-the-multiport-Brune-stage}
to extract the multiport inductive degenerate Brune stage in Fig.
\eqref{fig:Extraction-of-the-multiport-inductive-degenerate-stage};
we do not describe this algorithm here. We would like to also note
here that one can extend the treatment in Section \eqref{sub:Inductive-degenerate-case}
of the one-port inductive degenerate case for the quantization and
dissipation analysis to the multiport case in a straightforward way
provided that the resistance $r_{1}$ in Fig. \eqref{fig:Extraction-of-the-multiport-inductive-degenerate-stage}
is high (that is the loss introduced by the resistor $r_{1}$ is small).

\subsection{\label{sec:Quantization-of-the-multiport-Brune-circuit}Quantization
of the Multiport Brune Circuit}

\begin{figure}
\begin{centering}
\includegraphics{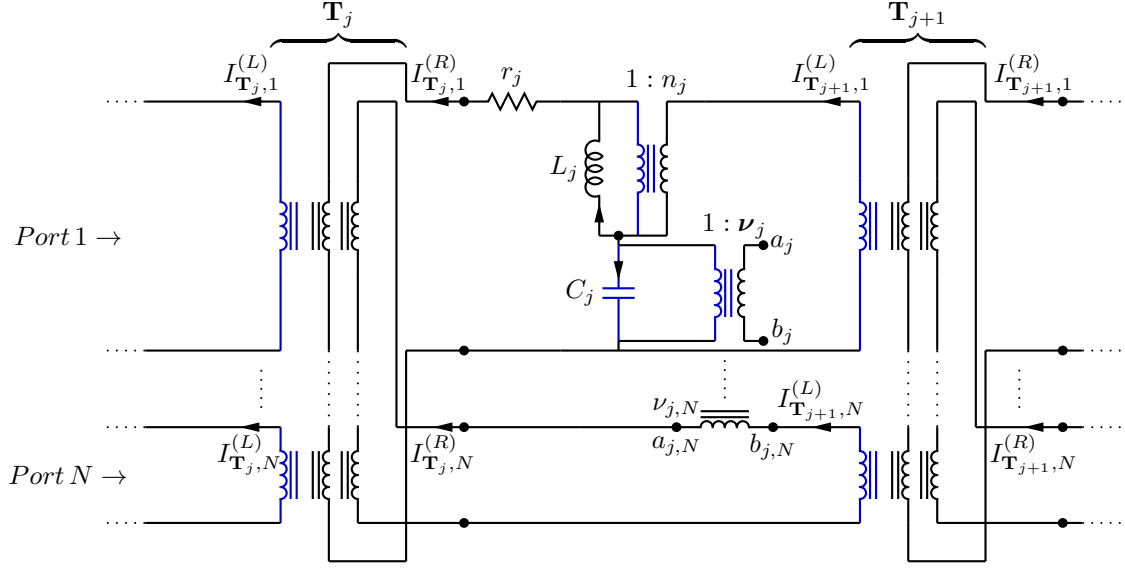} 
\par\end{centering}

\caption{\label{fig:Multiport-Brune-stage}Multiport Brune stage. Tree branches
are shown in black and chord branches are shown in blue. Note that
the Belevitch transformers $\mathbf{T}_{j}$ and $\mathbf{T}_{j+1}$
are reflected compared to the Fig. \eqref{fig:The-Belevitch-Transformer.}.}
\end{figure}

The multiport Brune circuit contains ideal transformers. In this section
we show that one can eliminate transformer branch variables and write
a set of effective Kirchhoff relations for the rest of the branch
currents and voltages. The effective Kirchhoff relations are given
by the loop matrix $\mathbf{F}^{eff}$ involving turn ratios which
we define below. The treatment here is similar to the analysis done
in \textcolor{black}{Section} \eqref{sub:Quantization-state-space-Brune-circuit-1-port}
for the one-port state-space Brune circuit. However as we will see
the addition of Belevitch transformers and $\boldsymbol{\nu}$-type
transformers makes the analysis more involved. We note that we replace
the shunt resistors $R_{j}$'s in the last stage by capacitors $C_{R_{j}}$'s
for $1\leq j\leq N$ as shown in Fig. \eqref{fig:Last-stage-multiport-Brune}
to do the dissipation analysis as discussed in detail in Appendix
\eqref{sub:Treatment-of-resistors-using-BKD-appendix}.

We will follow the same approach of \textcolor{black}{Section} \eqref{sub:Quantization-state-space-Brune-circuit-1-port}.
We write again the Kirchhoff's laws (Eqs. \eqref{eq:KCL-appendix},
\eqref{eq:KVL-appendix}) for the multiport Brune circuit in Fig.
\eqref{fig:Multiport-Brune-Circuit} whose $j^{th}$ stage is shown
in detail in Fig. \eqref{fig:Multiport-Brune-stage}

\begin{equation}
\mathbf{F}\mathbf{I}_{\mathrm{ch}}=-\mathbf{I}_{\mathrm{tr}}\label{eq:Kirchhoff-current-law-multiport}
\end{equation}

\begin{equation}
\mathbf{F}^{\mathrm{T}}\mathbf{V}_{\mathrm{tr}}=\mathbf{V}_{\mathrm{ch}}\label{eq:Kirchhoff-voltage-law-multiport}
\end{equation}
where we have again assumed that there \textcolor{black}{are no external
fluxes in circuit loops. As in the one-port case $\mathbf{F}$ is
again the loop matrix with entries being $0$, $1$ or $-1$ derived
by a graph theoretical analysis of the circuit \citep{Burkard}. After
the effective Kirchhoff analysis done below $\mathbf{F}$ will be
replaced by the effective loop matrix $\mathbf{F}^{eff}$ with real-valued
entries.}

$\mathbf{I}_{\mathrm{tr}}$ and $\mathbf{I}_{\mathrm{ch}}$ are the
tree and chord branch current vectors in Fig. \eqref{fig:Multiport-Brune-Circuit}
respectively partitioned as follows

\begin{equation}
\mathbf{I}_{\mathrm{tr}}=\left(\mathbf{I}_{J},\mathbf{I}_{L},\mathbf{I}_{Z},\mathbf{I}_{T}^{\left(tr\right)}\right)\label{eq:tree-current-vector-multiport}
\end{equation}

\begin{equation}
\mathbf{I}_{\mathrm{ch}}=\left(\mathbf{I}_{C},\mathbf{I}_{T}^{\left(ch\right)}\right)\label{eq:chord-current-vector-multiport}
\end{equation}
and tree and chord branches' voltages are partitioned respectively
as

\begin{equation}
\mathbf{V}_{\mathrm{tr}}=\left(\mathbf{V}_{J},\mathbf{V}_{L},\mathbf{V}_{Z},\mathbf{V}_{T}^{\left(tr\right)}\right)\label{eq:tree-voltage-vector-multiport}
\end{equation}

\begin{equation}
\mathbf{V}_{\mathrm{ch}}=\left(\mathbf{V}_{C},\mathbf{V}_{T}^{\left(ch\right)}\right)\label{eq:chord-voltage-vector-multiport}
\end{equation}
Here labels $J$, $L$, $Z$, $C$, $T$ correspond to Josephson junction,
inductor, resistor, capacitor and ideal transformer branches, respectively.

Our aim here is to write an effective set of Kirchhoff relations as

\begin{equation}
\mathbf{F}^{eff}\mathbf{I}_{\mathrm{ch}}^{eff}=-\mathbf{I}_{\mathrm{tr}}^{eff}\label{eq:effective-Kirchhoff-current-law}
\end{equation}

\begin{equation}
\left(\mathbf{F}^{T}\right)^{eff}\mathbf{V}_{\mathrm{tr}}^{eff}=\mathbf{V}_{\mathrm{ch}}^{eff}\label{eq:effective-Kirchhoff-voltage-law}
\end{equation}
where transformer branches are eliminated such that

\begin{eqnarray}
\mathbf{I}_{\mathrm{tr}}^{eff} & = & \left(\mathbf{I}_{J},\mathbf{I}_{L},\mathbf{I}_{Z}\right)\\
\mathbf{I}_{\mathrm{ch}}^{eff} & = & \mathbf{I}_{C}\\
\mathbf{V}_{\mathrm{tr}}^{eff} & = & \left(\mathbf{V}_{J},\mathbf{V}_{L},\mathbf{V}_{Z}\right)\\
\mathbf{V}_{\mathrm{ch}}^{eff} & = & \mathbf{V}_{C}
\end{eqnarray}

\textcolor{black}{We note here that the entries of the effective loop
matrix $\mathbf{F}^{eff}$ in Eq. \eqref{eq:effective-Kirchhoff-current-law}
are real numbers (as in the one-port case) being functions of ideal
transformer turn ratios as we will see below.}

Here we will do this effective loop matrix analysis for the Kirchhoff's
current law to get the matrix $\mathbf{F}^{eff}$ in Eq. \eqref{eq:effective-Kirchhoff-current-law}.
It is important to note that one should also do a similar analysis
for the Kirchhoff's voltage law to get an effective $\left(\mathbf{F}^{T}\right)^{eff}$
in Eq. \eqref{eq:effective-Kirchhoff-voltage-law} and verify that

\begin{equation}
\left(\mathbf{F}^{T}\right)^{eff}=\left(\mathbf{F}^{eff}\right)^{T}\label{eq:effective-loop-matrices-symmetry}
\end{equation}
holds. This we do in the Appendix \eqref{sub:Effective-Kirchhoff's-voltage-law-multiport}.
Eq. \eqref{eq:effective-loop-matrices-symmetry} is important to keep
the various matrices of interest like the capacitance $\mathcal{C}$
and stiffness $\mathbf{M}_{0}$ matrices symmetric.

To show how one can find such a $\mathbf{F}^{eff}$ matrix we will
further partition transformer current vectors $\mathbf{I}_{T}^{\left(tr\right)}$
and $\mathbf{I}_{T}^{\left(ch\right)}$ in Eqs. \eqref{eq:tree-current-vector-multiport}
and \eqref{eq:chord-current-vector-multiport}. We first note that
left branches of all transformers in the circuit in Fig. \eqref{fig:Multiport-Brune-stage}
are chord branches (colored in blue) and that right branches of all
transformers are in the tree (shown in black in Fig. \eqref{fig:Multiport-Brune-stage}).
Hence we can write

\begin{eqnarray}
\mathbf{I}_{T}^{\left(tr\right)} & = & \left(\mathbf{I}_{n}^{(R)},\mathbf{I}_{\mathrm{\mathbf{T}}}^{(R)},\mathbf{I}_{\boldsymbol{\nu}}^{(R)}\right)\\
\mathbf{I}_{T}^{\left(ch\right)} & = & \left(\mathbf{I}_{n}^{(L)},\mathbf{I}_{\mathrm{\mathbf{T}}}^{(L)},\mathbf{I}_{\boldsymbol{\nu}}^{(L)}\right)
\end{eqnarray}
where

\begin{eqnarray}
\mathbf{I}_{n}^{(R)} & = & \left(I_{n_{1}}^{(R)},\ldots,I_{n_{M}}^{(R)}\right)\\
\mathbf{I}_{\mathrm{\mathbf{T}}}^{(R)} & = & \left(\mathbf{I}_{\mathbf{T}_{1}}^{(R)},\ldots,\mathbf{I}_{\mathbf{T}_{M+1}}^{(R)}\right)\\
\mathbf{I}_{\boldsymbol{\nu}}^{(R)} & = & \left(\mathbf{I}_{\boldsymbol{\nu}_{1}}^{(R)},\ldots,\mathbf{I}_{\boldsymbol{\nu}_{M}}^{(R)}\right)
\end{eqnarray}
and

\begin{eqnarray}
\mathbf{I}_{n}^{(L)} & = & \left(I_{n_{1}}^{(L)},\ldots,I_{n_{M}}^{(L)}\right)\\
\mathbf{I}_{\mathrm{\mathbf{T}}}^{(L)} & = & \left(\mathbf{I}_{\mathbf{T}_{1}}^{(L)},\ldots,\mathbf{I}_{\mathbf{T}_{M+1}}^{(L)}\right)\\
\mathbf{I}_{\boldsymbol{\nu}}^{(L)} & = & \left(I_{\boldsymbol{\nu}_{1}}^{(L)},\ldots,I_{\boldsymbol{\nu}_{M}}^{(L)}\right)
\end{eqnarray}
with

\begin{eqnarray}
\mathbf{I}_{\mathbf{T}_{j}}^{(L)(R)} & = & \left(\begin{array}{c}
I_{\mathbf{T}_{j},1}^{(L)(R)}\\
\vdots\\
I_{\mathbf{T}_{j},N}^{(L)(R)}
\end{array}\right)\\
\mathbf{I}_{\boldsymbol{\nu}_{j}}^{(R)} & = & \left(\begin{array}{c}
I_{\boldsymbol{\nu}_{j},2}^{(R)}\\
\vdots\\
I_{\boldsymbol{\nu}_{j},N}^{(R)}
\end{array}\right)
\end{eqnarray}
where $\mathbf{I}_{\mathbf{T}_{j}}^{(L)(R)}$ are vectors of length
$N$ for $1\leq j\leq M+1$ and $\mathbf{I}_{\boldsymbol{\nu}_{j}}^{(R)}$
are vectors of length $\left(N-1\right)$ for $1\leq j\leq M$.

\begin{figure}
\begin{centering}
\includegraphics[scale=1.2]{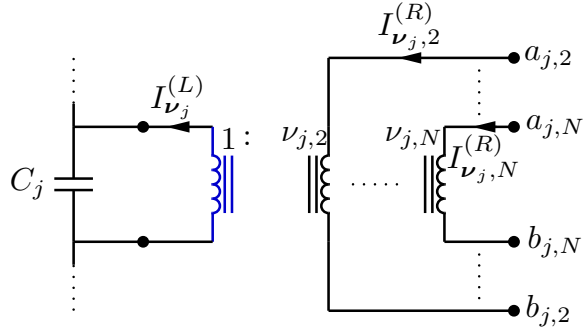} 
\par\end{centering}

\caption{\label{fig:nu-type-transformer-circuit}$\boldsymbol{\nu}$-type transformer
circuit at $j^{th}$ multiport Brune stage in Fig. \eqref{fig:Multiport-Brune-stage}.
We also show the connection of the transformer to the capacitor $C_{j}$
to make the correpondence to the Fig. \eqref{fig:Multiport-Brune-stage}
more clear.}
\end{figure}

Before moving further in the analysis we beriefly review the relations
between the currents through left and right branches of the three
different types of transformers in the multiport Brune circuit. We
also give voltage relations for completeness although we don't need
them for the analysis in this section. However we will refer to the
voltage relations in the Appendix \eqref{sub:Effective-Kirchhoff's-voltage-law-multiport}.

$\mathbf{I}_{\mathbf{T}_{j}}^{\left(L\right)(R)}$ is the vector of
currents through the left(right) branches of the Belevitch multiport
transformer appearing at the $j^{th}$ multiport Brune stage, $1\leq j\leq M+1$
as shown in Fig. \eqref{fig:Multiport-Brune-stage}. Hence by Eqs.
\eqref{eq:Belevitch-relation-current-reflected} and \eqref{eq:Belevitch-relation-voltage-reflected}
we have

\begin{eqnarray}
\mathbf{I}_{\mathbf{T}_{j}}^{\left(L\right)} & = & \mathbf{T}_{j}\mathbf{I}_{\mathbf{T}_{j}}^{\left(R\right)}\label{eq:interstage-T-current-relation}\\
\mathbf{V}_{\mathbf{T}_{j}}^{\left(R\right)} & = & \mathbf{T}_{j}^{T}\mathbf{V}_{\mathbf{T}_{j}}^{\left(L\right)}\label{eq:interstage-T-voltage-relation}
\end{eqnarray}
where $\mathbf{V}_{\mathbf{T}_{j}}^{\left(L\right)(R)}$ is the vector
of voltages across the left(right) branches of the Belevitch multiport
transformer appearing at the $j^{th}$ multiport Brune stage, $1\leq j\leq M+1$
in Fig. \eqref{fig:Multiport-Brune-stage}.

The turns ratio vector of the $\boldsymbol{\nu}$-type transformer
$\boldsymbol{\nu}_{j}$ at the $j^{th}$ multiport Brune stage is
given by

\begin{equation}
\boldsymbol{\nu}_{j}=\left(\begin{array}{c}
\nu_{j,2}\\
\vdots\\
\nu_{j,N}
\end{array}\right)\label{eq:nu-vector}
\end{equation}
The detailed circuit representation of the $\boldsymbol{\nu}$-type
transformer at the $j^{th}$ stage is given in Fig. \eqref{fig:nu-type-transformer-circuit}.
The current on the left branch is related to the currents on the right
branches by

\begin{equation}
I_{\boldsymbol{\nu}_{j}}^{(L)}=\boldsymbol{\nu}_{j}^{T}\left(\begin{array}{c}
I_{\boldsymbol{\nu}_{j},2}^{(R)}\\
\vdots\\
I_{\boldsymbol{\nu}_{j},N}^{(R)}
\end{array}\right)
\end{equation}
and the voltages across the right branches of the $\boldsymbol{\nu}$-type
transformer are related to the voltage across its left branch by the
following formula

\begin{equation}
\left(\begin{array}{c}
V_{\boldsymbol{\nu}_{j},2}^{(R)}\\
\vdots\\
V_{\boldsymbol{\nu}_{j},N}^{(R)}
\end{array}\right)=\boldsymbol{\nu}_{j}V_{\boldsymbol{\nu}_{j}}^{(L)}
\end{equation}

\begin{figure}
\begin{centering}
\includegraphics[scale=1.2]{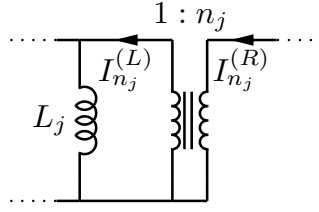} 
\par\end{centering}

\caption{\label{fig:n-type-transformer}Conventions for the direction of currents
for the $n$-type transformer of $j^{th}$ stage in the multiport
Brune circuit in Fig. \eqref{fig:Multiport-Brune-stage}. The inductive
branch $L_{j}$ is shown for orientation purposes with Fig. \eqref{fig:Multiport-Brune-stage}.}
\end{figure}

The detailed circuit diagram with current direction conventions for
the $n$-type transformer is shown in Fig. \eqref{fig:n-type-transformer}.
For this type of transformer the relations between currents and voltages
are given by

\begin{eqnarray}
I_{n_{j}}^{(L)} & = & n_{j}I_{n_{j}}^{(R)}\\
V_{n_{j}}^{(R)} & = & n_{j}V_{n_{j}}^{(L)}
\end{eqnarray}

Now we can begin our analysis. We will proceed from the last(rightmost)
stage to the first(leftmost) in Fig. \eqref{fig:Multiport-Brune-Circuit}
by relating the currents of consecutive stages.

\begin{figure}
\begin{centering}
\includegraphics{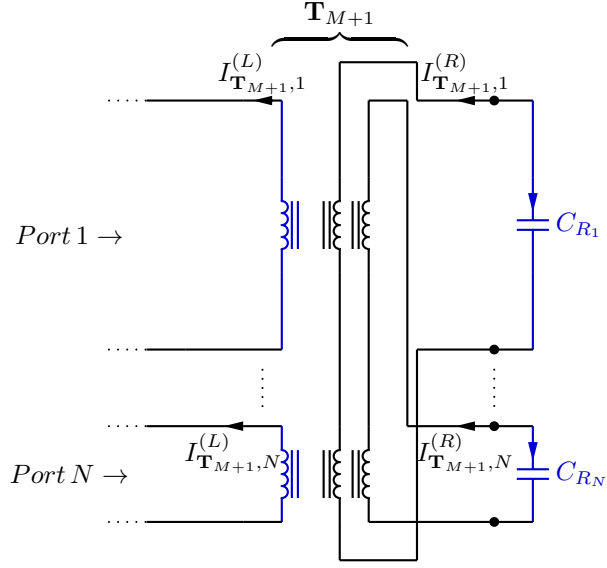} 
\par\end{centering}

\caption{\label{fig:Last-stage-multiport-Brune}Last stage of the multiport
Brune circuit. Tree branches are shown in black and chord branches
are shown in blue. Shunt resistors $R_{j}$'s in the last stage are
replaced by capacitors $C_{R_{j}}$'s for $1\leq j\leq N$ . Their
contribution to the dissipation is computed in Appendix \eqref{sub:Treatment-of-resistors-using-BKD-appendix}.}
\end{figure}

Starting at the last stage we have (see Fig. \eqref{fig:Last-stage-multiport-Brune})

\begin{equation}
\mathbf{I}_{\mathbf{T}_{M+1}}^{\left(R\right)}=-\mathbf{I}_{C_{R}}\label{eq:multiport-Brune-last-stage-current-relations}
\end{equation}
where $\mathbf{I}_{C_{R}}$ are the currents through the capacitors(substituted
for the shunt resistors) of the last stage.

The currents of inter-stage transformers are given by Eq. \eqref{eq:interstage-T-current-relation}

\begin{equation}
\mathbf{I}_{\mathbf{T}_{j}}^{\left(L\right)}=\mathbf{T}_{j}\mathbf{I}_{\mathbf{T}_{j}}^{\left(R\right)}\label{eq:multiport-Brune-Belevitch-current-relations}
\end{equation}
for $1\leq j\leq M+1$, where $\mathbf{T}_{j}$ is the $\left(N\times N\right)$
Belevitch transformer matrix of the $j^{th}$ stage. The currents
of consecutive inter-stage transformers are related by

\begin{equation}
\mathbf{I}_{\mathbf{T}_{j}}^{\left(R\right)}=-\mathbf{e}_{1}I_{C_{j}}+\mathbf{A}_{j}\mathbf{I}_{\mathbf{T}_{j+1}}^{\left(L\right)}\label{eq:consecutive-Belevitch-current-relations}
\end{equation}
for $1\leq j\leq M$, where

\begin{equation}
\mathbf{A}_{j}=\left(\begin{array}{cccc}
1 & -\nu_{j,2} & \cdots & -\nu_{j,N}\\
 & 1 & \boldsymbol{0}\\
 & \boldsymbol{0} & \ddots\\
 &  &  & 1
\end{array}\right)\label{eq:Aj-matrix}
\end{equation}
$\mathbf{A}_{j}$ is a $\left(N\times N\right)$ matrix and $\mathbf{e}_{1}$
is the unit vector $\mathbf{e}_{1}=\left(1,0,\ldots,0\right)^{T}$
of length $N$. We note that in the case of a degenerate stage as
in Fig. \eqref{fig:Multiport-capacitive-degenerate-stage} for the
$j^{th}$ stage we have $\nu_{j,k}=0$ for $2\leq k\leq N$ , hence
$\mathbf{A}_{j}$ is the identity matrix in such a case.

The current $I_{L_{j}}$ through the inductor $L_{j}$ at the $j^{th}$
multiport Brune stage can be written

\begin{equation}
I_{L_{j}}=-I_{C_{j}}+\boldsymbol{\upsilon}_{j}\mathbf{I}_{\mathbf{T}_{j+1}}^{\left(L\right)}\label{eq:multiport-Brune-inductor-current}
\end{equation}
where $\mathbf{\boldsymbol{\upsilon}}_{j}$ is a row vector of length
$N$ :

\begin{equation}
\boldsymbol{\upsilon}_{j}=\left(\begin{array}{cccc}
\left(1-n_{j}\right) & -\nu_{j2} & \cdots & -\nu_{j,N}\end{array}\right)\label{eq:v-vector}
\end{equation}

Using the Eqs. \eqref{eq:multiport-Brune-last-stage-current-relations},
\eqref{eq:multiport-Brune-Belevitch-current-relations}, \eqref{eq:consecutive-Belevitch-current-relations}
and \eqref{eq:multiport-Brune-inductor-current} we can iterate over
the index $j$ backwards starting at $j=M+1$:

\begin{eqnarray*}
\mathbf{I}_{\mathbf{T}_{M+1}}^{(L)} & = & \mathbf{T}_{M+1}\mathbf{I}_{\mathbf{T}_{M+1}}^{(R)}=-\mathbf{T}_{M+1}\mathbf{I}_{C_{R}}\\
I_{L_{M}} & = & -I_{C_{M}}+\boldsymbol{\upsilon}_{M}\mathbf{I}_{\mathbf{T}_{M+1}}^{\left(L\right)}=-I_{C_{M}}-\boldsymbol{\upsilon}_{M}\mathbf{T}_{M+1}\mathbf{I}_{C_{R}}\\
\mathbf{I}_{\mathbf{T}_{M}}^{(R)} & = & -\mathbf{e}_{1}I_{C_{M}}+\mathbf{A}_{M}\mathbf{I}_{\mathbf{T}_{M+1}}^{\left(L\right)}=-\mathbf{e}_{1}I_{C_{M}}-\mathbf{A}_{M}\mathbf{T}_{M+1}\mathbf{I}_{C_{R}}\\
\mathbf{I}_{\mathbf{T}_{M}}^{(L)} & = & \mathbf{T}_{M}\mathbf{I}_{\mathbf{T}_{M}}^{(R)}=-\mathbf{T}_{M}\mathbf{e}_{1}I_{C_{M}}-\mathbf{T}_{M}\mathbf{A}_{M}\mathbf{T}_{M+1}\mathbf{I}_{C_{R}}
\end{eqnarray*}

\begin{eqnarray}
I_{L_{M-1}} & = & -I_{C_{M-1}}+\boldsymbol{\upsilon}_{M-1}\mathbf{I}_{\mathbf{T}_{M}}^{\left(L\right)}=-I_{C_{M-1}}-\boldsymbol{\upsilon}_{M-1}\mathbf{T}_{M}\mathbf{e}_{1}I_{C_{M}}-\boldsymbol{\upsilon}_{M-1}\mathbf{T}_{M}\mathbf{A}_{M}\mathbf{T}_{M+1}\mathbf{I}_{C_{R}}\nonumber \\
\mathbf{I}_{\mathbf{T}_{M-1}}^{(R)} & = & -\mathbf{e}_{1}I_{C_{M-1}}+\mathbf{A}_{M-1}\mathbf{I}_{\mathbf{T}_{M}}^{\left(L\right)}=-\mathbf{e}_{1}I_{C_{M-1}}-\mathbf{A}_{M-1}\mathbf{T}_{M}\mathbf{e}_{1}I_{C_{M}}-\mathbf{A}_{M-1}\mathbf{T}_{M}\mathbf{A}_{M}\mathbf{T}_{M+1}\mathbf{I}_{C_{R}}\nonumber \\
\mathbf{I}_{\mathbf{T}_{M-1}}^{(L)} & = & \mathbf{T}_{M-1}\mathbf{I}_{\mathbf{T}_{M-1}}^{(R)}\nonumber \\
 & = & -\mathbf{T}_{M-1}\mathbf{e}_{1}I_{C_{M-1}}-\mathbf{T}_{M-1}\mathbf{A}_{M-1}\mathbf{T}_{M}\mathbf{e}_{1}I_{C_{M}}-\mathbf{T}_{M-1}\mathbf{A}_{M-1}\mathbf{T}_{M}\mathbf{A}_{M}\mathbf{T}_{M+1}\mathbf{I}_{C_{R}}\nonumber \\
I_{L_{M-2}} & = & -I_{C_{M-2}}+\boldsymbol{\upsilon}_{M-2}\mathbf{I}_{\mathbf{T}_{M-1}}^{\left(L\right)}\nonumber \\
 & = & -I_{C_{M-2}}-\boldsymbol{\upsilon}_{M-2}\mathbf{T}_{M-1}\mathbf{e}_{1}I_{C_{M-1}}-\boldsymbol{\upsilon}_{M-2}\mathbf{T}_{M-1}\mathbf{A}_{M-1}\mathbf{T}_{M}\mathbf{e}_{1}I_{C_{M}}+\nonumber \\
 &  & -\boldsymbol{\upsilon}_{M-2}\mathbf{T}_{M-1}\mathbf{A}_{M-1}\mathbf{T}_{M}\mathbf{A}_{M}\mathbf{T}_{M+1}\mathbf{I}_{C_{R}}\nonumber \\
\vdots & \vdots & \vdots\label{eq:multiport-iteration-for-currents-last-line}
\end{eqnarray}

Hence we conclude that one can write

\begin{equation}
\mathbf{I}_{L}=-\mathbf{F}_{LC}^{eff}\mathbf{I}_{C}
\end{equation}
with for $1\leq j\leq M$ :

\begin{equation}
\begin{cases}
\mathbf{F}_{LC}^{eff}\left(j,k\right)=0 & for\; k<j\\
\mathbf{F}_{LC}^{eff}\left(j,k\right)=1 & for\; k=j\\
\mathbf{F}_{LC}^{eff}\left(j,k\right)=\boldsymbol{\upsilon}_{j}\mathbf{T}_{j+1}\mathbf{e}_{1} & for\; k=j+1,\; and\; j<M\\
\mathbf{F}_{LC}^{eff}\left(j,k\right)=\boldsymbol{\upsilon}_{j}\mathbf{T}_{j+1}\mathbf{A}_{j+1}\ldots\mathbf{T}_{k-1}\mathbf{A}_{k-1}\mathbf{T}_{k}\mathbf{e}_{1} & for\; j+2\leq k\leq M\\
\mathbf{F}_{LC}^{eff}\left(j,k\right)=\boldsymbol{\upsilon}_{j}\mathbf{T}_{j+1}\mathbf{A}_{j+1}\ldots\mathbf{T}_{M}\mathbf{A}_{M}\mathbf{T}_{M+1}\mathbf{e}_{k-M} & for\; j<M\; and\; M+1\leq k\leq M+N\\
\mathbf{F}_{LC}^{eff}\left(j,k\right)=\boldsymbol{\upsilon}_{M}\mathbf{T}_{M+1}\mathbf{e}_{k-M} & for\; j=M\; and\; M+1\leq k\leq M+N
\end{cases}\label{eq:eff-FLC-submatrix-current}
\end{equation}
where $\mathbf{e}_{k}$ is the unit vector of length $N$ non-zero
only at its $k^{th}$ entry such that $\mathbf{e}_{k}\left(j\right)=0$
for $j\neq k$ and $\mathbf{e}_{k}\left(k\right)=1$. We assumed the
following ordering for the capacitors

\begin{equation}
\left\{ C_{1},\ldots,C_{M},C_{R_{1}},\ldots,C_{R_{N}}\right\} 
\end{equation}

To compute $\mathbf{F}_{ZC}^{eff}$ we note the following

\begin{eqnarray}
I_{r_{j}} & = & I_{\mathbf{T}_{j},1}^{(R)}\\
 & = & \mathbf{e}_{1}^{T}\mathbf{I}_{\mathbf{T}_{j}}^{(R)}\label{eq:multiport-Brune-resistor-current}
\end{eqnarray}
for $1\leq j\leq M$.

Referring back to the iteration in Eqs. \eqref{eq:multiport-iteration-for-currents-last-line}
and using Eq. \eqref{eq:multiport-Brune-resistor-current} we can
write

\begin{equation}
\mathbf{I}_{Z}=-\mathbf{F}_{ZC}^{eff}\mathbf{I}_{C}
\end{equation}
with for $1\leq j\leq M$ :

\begin{equation}
\begin{cases}
\mathbf{F}_{ZC}^{eff}\left(j,k\right)=0 & for\; k<j\\
\mathbf{F}_{ZC}^{eff}\left(j,k\right)=1 & for\; k=j\\
\mathbf{F}_{ZC}^{eff}\left(j,k\right)=\mathbf{e}_{1}^{T}\mathbf{A}_{j}\mathbf{T}_{j+1}\ldots\mathbf{T}_{k-1}\mathbf{A}_{k-1}\mathbf{T}_{k}\mathbf{e}_{1} & for\; j+1\leq k\leq M\\
\mathbf{F}_{ZC}^{eff}\left(j,k\right)=\mathbf{e}_{1}^{T}\mathbf{A}_{j}\mathbf{T}_{j+1}\ldots\mathbf{T}_{M}\mathbf{A}_{M}\mathbf{T}_{M+1}\mathbf{e}_{k-M} & for\; M+1\leq k\leq M+N
\end{cases}\label{eq:eff-FZC-submatrix-current}
\end{equation}

\begin{figure}
\begin{centering}
\includegraphics{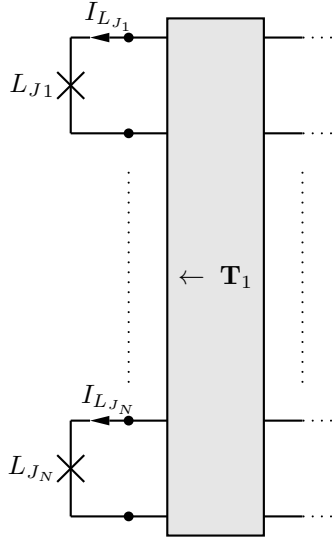} 
\par\end{centering}

\caption{\label{fig:JJ-shunting-all-ports}To compute $\mathbf{F}_{JC}^{eff}$,
for simplicity, we assume that all the ports of the multiport Brune
circuit are shunted by Josephson junctions $L_{J_{1}},\ldots,L_{J_{N}}$.}
\end{figure}

To compute $\mathbf{F}_{JC}^{eff}$ we will assume for simplicity
that all the ports are shunted by Josephson junctions as shown in
Fig. \eqref{fig:JJ-shunting-all-ports} so that all the junctions
belong to the spanning tree. Later we will allow for resistors and
voltage sources shunting the ports of the multiport Brune network.
We note the following

\begin{equation}
\mathbf{I}_{J}=\mathbf{I}_{\mathbf{T}_{1}}^{(L)}\label{eq:JJ-currents-multiport}
\end{equation}
Hence referring back to the iteration in Eq. \eqref{eq:multiport-iteration-for-currents-last-line}
we deduce using Eq. \eqref{eq:JJ-currents-multiport}

\begin{eqnarray}
\mathbf{I}_{J} & = & \mathbf{I}_{\mathbf{T}_{1}}^{(L)}\nonumber \\
 & = & -\mathbf{T}_{1}\mathbf{e}_{1}I_{C_{1}}-\mathbf{T}_{1}\mathbf{A}_{1}\mathbf{T}_{2}\mathbf{e}_{1}I_{C_{2}}\ldots-\mathbf{T}_{1}\mathbf{A}_{1}\ldots\mathbf{T}_{M-1}\mathbf{A}_{M-1}\mathbf{T}_{M}\mathbf{e}_{1}I_{C_{M}}+\nonumber \\
 &  & -\mathbf{T}_{1}\mathbf{A}_{1}\ldots\mathbf{T}_{M}\mathbf{A}_{M}\mathbf{T}_{M+1}\mathbf{I}_{C_{R}}
\end{eqnarray}
Hence for $1\leq j\leq N$

\begin{equation}
\begin{cases}
\mathbf{F}_{JC}^{eff}\left(j,k\right)=\mathbf{e}_{j}^{T}\mathbf{T}_{1}\mathbf{e}_{1} & for\; k=1\\
\mathbf{F}_{JC}^{eff}\left(j,k\right)=\mathbf{e}_{j}^{T}\mathbf{T}_{1}\mathbf{A}_{1}\ldots\mathbf{T}_{k-1}\mathbf{A}_{k-1}\mathbf{T}_{k}\mathbf{e}_{1} & for\;1<k\leq M\\
\mathbf{F}_{JC}^{eff}\left(j,k\right)=\mathbf{e}_{j}^{T}\mathbf{T}_{1}\mathbf{A}_{1}\ldots\mathbf{T}_{M}\mathbf{A}_{M}\mathbf{T}_{M+1}\mathbf{e}_{k-M} & for\; M+1\leq k\leq M+N
\end{cases}\label{eq:eff-FJC-submatrix-current}
\end{equation}
where $\mathbf{F}_{JC}^{eff}$ is defined by

\begin{equation}
\mathbf{I}_{J}=-\mathbf{F}_{JC}^{eff}\mathbf{I}_{C}
\end{equation}

Now that we derived effective loop matrices we will follow the Appendix
\eqref{sub:Burkard's-method} and \eqref{sub:Treatment-of-resistors-using-BKD-appendix}
to compute Hamiltonian matrices and to do dissipation analysis due
to the resistors in the multiport Brune circuit. We will repeat here
some of the definitions - which will be used with the effective loop
matrices- of the Appendix \eqref{sub:Burkard's-method} for convenience.

With the effective loop matrix

\begin{equation}
\mathcal{F}_{C}^{eff}=\left(\begin{array}{c}
\mathbf{F}_{JC}^{eff}\\
\mathbf{F}_{LC}^{eff}
\end{array}\right)\label{eq:multiport-FCeff}
\end{equation}
we can compute the capacitance matrix $\mathcal{C}$ defined in Eq.
\eqref{eq:curly-C-matrix-Burkard-definition}

\begin{equation}
\mathcal{C}=\left(\begin{array}{cc}
\mathbf{C}_{J} & \boldsymbol{0}\\
\boldsymbol{0} & \boldsymbol{0}
\end{array}\right)+\mathcal{F}_{C}^{eff}\mathbf{C}\left(\mathcal{F}_{C}^{eff}\right)^{T}\label{eq:multiport-Brune-curly-C}
\end{equation}
where we have the following partitioning to identify the submatrices
$\mathbf{C}_{0}$ and $\mathbf{C}_{R}$ corresponding to ordinary
capacitances and capacitances to be replaced by shunt resistors in
the last stage of the multiport Brune circuit, respectively

\begin{equation}
\mathbf{C}=\left(\begin{array}{cc}
\mathbf{C}_{0} & \boldsymbol{0}\\
\boldsymbol{0} & \mathbf{C}_{R}
\end{array}\right)\label{eq:multiport-C-matrix}
\end{equation}
with

\begin{equation}
\mathbf{C}_{0}=\left(\begin{array}{ccc}
C_{1} &  & \boldsymbol{0}\\
 & \ddots\\
\boldsymbol{0} &  & C_{M}
\end{array}\right)
\end{equation}
and

\begin{equation}
\mathbf{C}_{R}=\left(\begin{array}{ccc}
C_{R_{1}} &  & \boldsymbol{0}\\
 & \ddots\\
\boldsymbol{0} &  & C_{R_{N}}
\end{array}\right)
\end{equation}
and the matrix $\mathbf{C}_{J}$ in Eq. \eqref{eq:multiport-Brune-curly-C}
for the Josephson junction capacitances is

\begin{equation}
\mathbf{C}_{J}=\left(\begin{array}{ccc}
C_{J_{1}} &  & \boldsymbol{0}\\
 & \ddots\\
\boldsymbol{0} &  & C_{J_{N}}
\end{array}\right)
\end{equation}

Partitioning also $\mathcal{F}_{C}^{eff}$ according to the partitioning
in Eq. \eqref{eq:multiport-C-matrix} as in Eq. \eqref{eq:multiport-FC-partitioning-appendix}

\begin{equation}
\mathcal{F}_{C}^{eff}=\left(\begin{array}{cc}
\mathcal{F}_{C_{0}}^{eff} & \mathcal{F}_{C_{R}}^{eff}\end{array}\right)\label{eq:curly-FC-partitioning-multiport-Brune}
\end{equation}
we can decompose $\mathcal{C}$ in Eq. \eqref{eq:multiport-Brune-curly-C}
as in Eq. \eqref{eq:curly-C-decomposition}

\begin{eqnarray}
\mathcal{C} & = & \left(\begin{array}{cc}
\mathbf{C}_{J} & \boldsymbol{0}\\
\boldsymbol{0} & \boldsymbol{0}
\end{array}\right)+\mathcal{F}_{C}^{eff}\mathbf{C}\left(\mathcal{F}_{C}^{eff}\right)^{T}\\
 & = & \mathcal{C}_{0}+\mathcal{C}_{R}
\end{eqnarray}
where we defined as we did in Eqs. \eqref{eq:curly-C0-definition}
and \eqref{eq:curly-CR-definition}

\begin{eqnarray}
\mathcal{C}_{0} & = & \left(\begin{array}{cc}
\mathbf{C}_{J} & \boldsymbol{0}\\
\boldsymbol{0} & \boldsymbol{0}
\end{array}\right)+\mathcal{F}_{C_{0}}^{eff}\mathbf{C}_{0}\left(\mathcal{F}_{C_{0}}^{eff}\right)^{T}\label{eq:curly-C0-definition-multiport-Brune}\\
\mathcal{C}_{R} & = & \mathcal{F}_{C_{R}}^{eff}\mathbf{C}_{R}\left(\mathcal{F}_{C_{R}}^{eff}\right)^{T}\label{eq:curly-CR-definition-multiport-Brune}
\end{eqnarray}
$\mathcal{C}_{0}$ is the capacitance matrix that appears in the system
Hamiltonian in Eq. \eqref{eq:Hamiltonian-multiport} below for the
multiport Brune circuit whereas $\mathcal{C}_{R}$ is a dissipative
term which we will treat later below by making the substitution $\mathbf{C}_{R}\leftarrow\left(i\omega\right)^{-1}\mathbf{R}^{-1}$
as described in Eq. \eqref{eq:shunt-resistance-substitution} for
the shunt resistors in the last stage.

$\mathbf{L}_{t}^{-1}$ and $\mathcal{G}$ matrices defined in Eqs.
\eqref{eq:Lt-inv-definition}, \eqref{eq:curly-G-definition} are
simply

\begin{equation}
\mathbf{L}_{t}^{-1}=\left(\begin{array}{ccc}
1/L_{1} &  & \boldsymbol{0}\\
 & \ddots\\
\boldsymbol{0} &  & 1/L_{M}
\end{array}\right)
\end{equation}

\begin{equation}
\mathcal{G}=\left(\begin{array}{c}
\boldsymbol{0}\\
\mathbf{I}_{L}
\end{array}\right)
\end{equation}
since there are no chord inductors in the circuit. Hence we have by
Eq. \eqref{eq:M0-definition}

\begin{eqnarray}
\mathbf{M}_{0} & = & \mathcal{G}\mathbf{L}_{t}^{-1}\mathcal{G}^{T}\\
 & = & \left(\begin{array}{ccc}
\boldsymbol{0}_{J} &  & \boldsymbol{0}\\
 & 1/L_{1}\\
 & \ddots\\
\boldsymbol{0} &  & 1/L_{M}
\end{array}\right)\label{eq:M0-multiport-Brune}
\end{eqnarray}
where $\boldsymbol{0}_{J}$ is a $N_{J}\times N_{J}$ zero matrix.
Here since we assumed that all ports are shunted by Josephson junctions
$N_{J}=N$.

Using the Eq. \eqref{eq:Burkard-Hamiltonian-extended} we have the
following Hamiltonian for the multiport Brune circuit

\begin{equation}
\mathcal{H_{S}}=\frac{1}{2}\boldsymbol{Q}^{T}\mathcal{C}_{0}^{-1}\boldsymbol{Q}+U\left(\boldsymbol{\Phi}\right)\label{eq:Hamiltonian-multiport}
\end{equation}
where

\begin{equation}
U\left(\boldsymbol{\Phi}\right)=-\left(\frac{\Phi_{0}}{2\pi}\right)^{2}\mathbf{L}_{J}^{-1}\cos\left(\boldsymbol{\mathbf{\varphi}}_{J}\right)+\frac{1}{2}\boldsymbol{\Phi}^{T}\mathbf{M_{0}}\boldsymbol{\Phi}\label{eq:Potential-energy-function-1-1}
\end{equation}

To treat the dissipation we first note that the diagonal tree impedance
matrix

\begin{equation}
\mathbf{Z}=\left(\begin{array}{ccc}
r_{1} &  & \boldsymbol{0}\\
 & \ddots\\
\boldsymbol{0} &  & r_{M}
\end{array}\right)\label{eq:Z-tree-impedance-matrix}
\end{equation}
consists of in-series resistances $r_{1},\ldots,r_{M}$ of the multiport
Brune circuit in Fig. \eqref{fig:Multiport-Brune-Circuit}. However,
as we noted in Appendix \eqref{sub:Treatment-of-resistors-using-BKD-appendix}
we will treat each in-series resistor $r_{j}$ separately for $1\leq j\leq M$.
So instead of using the full $\mathbf{F}_{ZC}^{eff}$ matrix defined
in Eq. \eqref{eq:eff-FZC-submatrix-current} we will use its rows
$\mathbf{F}_{r_{j},C_{0}}^{eff}$'s corresponding to the in-series
resistors $r_{j}$'s for $1\leq j\leq M$, in our dissipation treatment
below.

We compute the coupling $\bar{\mathbf{m}}_{j}$ of the bath due to
the resistor $r_{j}$ for $1\leq j\leq M$ to the circuit degrees
of freedom using Eq. \eqref{eq:m0-vector} with the effective row
vector $\mathbf{F}_{r_{j},C_{0}}^{eff}$ (since we treat resistors
one at a time $\mathbf{F}_{r_{j},C_{0}}^{eff}$ is the $j^{th}$ row
of the matrix $\mathbf{F}_{ZC_{0}}^{eff}$ defined in Eq. \eqref{eq:FZC-partitioning}
corresponding to the resistor $r_{j}$)

\begin{eqnarray}
\bar{\mathbf{m}}_{j} & = & \mathcal{F}_{C_{0}}^{eff}\mathbf{C}_{0}\left(\mathbf{F}_{r_{j},C_{0}}^{eff}\right)^{T}\label{eq:mj-multiport-Brune}\\
 & = & \left(\begin{array}{c}
\mathbf{\bar{m}}_{j,J}\\
\mathbf{\bar{m}}_{j,L}
\end{array}\right)\label{eq:rj-dissipation-analysis-multiport-1}
\end{eqnarray}
and using Eq. \eqref{eq:CZrj-with-FrjC0}

\begin{equation}
\mathbf{\bar{C}}_{Z,r_{j}}\left(\omega\right)=-i\omega r_{j}\left[1+i\omega r_{j}\mathbf{F}_{r_{j},C_{0}}^{eff}\mathbf{C}_{0}\left(\mathbf{F}_{r_{j},C_{0}}^{eff}\right)^{T}\right]^{-1}\label{eq:rj-dissipation-analysis-multiport-2}
\end{equation}
which is related to the kernel as derived in Eq. \eqref{eq:Kernel-appendix}
(we note that $\mathbf{\bar{C}}_{Z,r_{j}}\left(\omega\right)$ is
a scalar)

\begin{equation}
K_{j}\left(\omega\right)=-\omega^{2}\bar{\mathbf{C}}_{Z,r_{j}}\left(\omega\right)\label{eq:rj-dissipation-analysis-multiport-3}
\end{equation}
and the spectrum of the bath due to the resistor $r_{j}$ is given
by Eq. \eqref{eq:spectral-density-of-the-bath}

\begin{eqnarray}
J_{j}\left(\omega\right) & = & Im\left[K_{j}\left(\omega\right)\right]\label{eq:rj-dissipation-analysis-multiport-4}\\
 & = & \frac{\omega^{3}r_{j}}{1+\omega^{2}r_{j}^{2}\left[\mathbf{F}_{r_{j},C_{0}}^{eff}\mathbf{C}_{0}\left(\mathbf{F}_{r_{j},C_{0}}^{eff}\right)^{T}\right]^{2}}\label{eq:rj-dissipation-analysis-multiport-5}
\end{eqnarray}

The contribution of the resistor $r_{j}$ to the relaxation rate is
given by the formula in Eq. \eqref{eq:T1-rate-for-rj}

\begin{equation}
\frac{1}{T_{1,r_{j}}}=\frac{4}{\hbar}\left|\left\langle 0\left|\mathbf{\bar{m}}_{j}\cdot\mathbf{\Phi}\right|1\right\rangle \right|^{2}J_{j}\left(\omega_{01}\right)\coth\left(\frac{\hbar\omega_{01}}{2k_{B}T}\right)\label{eq:T1-rate-due-to-rj-mp-Brune}
\end{equation}
where $\omega_{01}$ is the qubit frequency. \textcolor{black}{In
this multiqubit case, this formula can be used to get to the relaxation
rate of any qubit, by choosing the appropriate initial and final quantum
states. For example, the relaxation rate of qubit 1 may be calculated
by taking $\ket{0}$ to be the global ground state in which each qubit
is 0, and $\ket{1}$ to be the state in which qubit 1 is in its upper
eigenstate, while the other qubits remain in their ground state; we
would write this more fully as the state $\ket{1000...}$. Taking
this to be the state $\ket{0100...}$ will give $T_{1}$ for the second
qubit, and so forth. In fact, this formula can be used to get the
relaxation rate between any two multi-qubit eigenstates. A further
refinement of these prescriptions may be needed if the eigenstates
of the multiqubit system are entangled.}

To compute the dissipative contribution of the resistors $\left\{ R_{1},\ldots,R_{N}\right\} $
shunting the last stage we make the substitution $\mathbf{C}_{R}\leftarrow\left(i\omega\right)^{-1}\mathbf{R}^{-1}$
done in Eq. \eqref{eq:shunt-resistance-substitution} to get the resistance
matrix defined in Eq. \eqref{eq:curly-R-appendix}

\begin{equation}
\mathcal{R}^{-1}=\mathcal{F}_{C_{R}}^{eff}\mathbf{R}^{-1}\left(\mathcal{F}_{C_{R}}^{eff}\right)^{T}
\end{equation}
where $\mathbf{R}$ is defined as

\begin{equation}
\mathbf{R}=\left(\begin{array}{ccc}
R_{1} &  & \boldsymbol{0}\\
 & \ddots\\
\boldsymbol{0} &  & R_{N}
\end{array}\right)
\end{equation}

We again treat the shunt resistors $\left\{ R_{1},\ldots,R_{N}\right\} $
one at a time, that is, to compute the contribution of the resistor
$R_{j}$ to the relaxation rate we set $R_{k}=\infty$ for $1\leq k\leq N$
and $k\neq j$ and we short circuit in series resistors by setting
$r_{k}=0$ for $1\leq k\leq M$. The coupling vector $\bar{\mathbf{m}}_{R_{j}}$
which couples the bath due to $R_{j}$ to the circuit degrees of freedom
is given in Eq. \eqref{eq:mRj} as

\begin{equation}
\bar{\mathbf{m}}_{R_{j}}=\mathcal{F}_{R_{j},C_{R}}^{eff}\label{eq:curly-FRj-CR-eff}
\end{equation}
where $\mathcal{F}_{R_{j},C_{R}}^{eff}$ is the $j^{th}$ column of
the matrix $\mathcal{F}_{C_{R}}^{eff}$ and the dissipation kernel
due to $R_{j}$ is given in Eq. \eqref{eq:KRj} as

\begin{equation}
K_{R_{j}}\left(\omega\right)=\frac{i\omega}{R_{j}}
\end{equation}
and the spectral density $J_{R_{j}}$ of the bath due to $R_{j}$
is given by Eq. \eqref{eq:spectral-density-of-the-bath-shunt-resistors}

\begin{eqnarray}
J_{R_{j}}\left(\omega\right) & = & \mathrm{Im}\left[K_{R_{j}}\left(\omega\right)\right]\\
 & = & \frac{\omega}{R_{j}}\label{eq:JRj-mp-Brune}
\end{eqnarray}
Hence by Eq. \eqref{eq:T1-rate-for-Rj} we compute the contribution
of the resistor $R_{j}$ to the relaxation rate as

\begin{equation}
\frac{1}{T_{1,R_{j}}}=\frac{4}{\hbar}\left|\left\langle 0\left|\mathbf{\bar{m}}_{R_{j}}\cdot\mathbf{\Phi}\right|1\right\rangle \right|^{2}J_{R_{j}}\left(\omega_{01}\right)\coth\left(\frac{\hbar\omega_{01}}{2k_{B}T}\right)\label{eq:T1-rate-due-to-Rj-mp-Brune}
\end{equation}

\subsection{\label{sub:Resistors-shunting-the-ports}Resistors Shunting the Ports}

\begin{figure}
\begin{centering}
\includegraphics{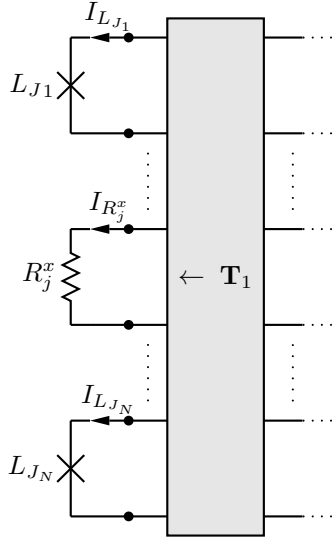} 
\par\end{centering}

\caption{\label{fig:multiport-shunted-by-JJ's-and-resistor}The multiport Brune
circuit shunted by a resistor at its $j^{th}$ port and by Josephson
junctions at the remaining ports.}
\end{figure}

One can also think of shunting some of the ports in Fig. \eqref{fig:JJ-shunting-all-ports}
by resistors. For simplicity let's assume that we shunt only the $j^{th}$
port of the multiport Brune circuit by a resistor $R_{j}^{x}$ and
the rest of the ports by Josephson junctions as shown in Fig. \eqref{fig:multiport-shunted-by-JJ's-and-resistor}.
The branch of the resistor $R_{j}^{x}$ belongs to the spanning tree
hence its treatment will be similar to the treatment of the in-series
resistors $r_{j}$ in the previous section. We will derive an effective
loop row-vector $\mathbf{F}_{R_{j}^{x},C}^{eff}$ corresponding to
$R_{j}^{x}$. Before doing this we note that the $j^{th}$ row of
the effective loop matrix $\mathbf{F}_{JC}^{eff}$ derived in Eq.
\eqref{eq:eff-FJC-submatrix-current} above needs to be dropped since
we replaced the $j^{th}$ Josephson junction in Fig. \eqref{fig:JJ-shunting-all-ports}
by the resistor $R_{j}^{x}$.

We first note the following

\begin{equation}
I_{R_{j}^{x}}=I_{\mathbf{T}_{1},j}^{(L)}
\end{equation}
Hence from the iteration in Eq. \eqref{eq:multiport-iteration-for-currents-last-line}
we conclude

\begin{eqnarray}
I_{R_{j}^{x}} & = & \mathbf{e}_{j}^{T}\mathbf{I}_{\mathbf{T}_{1}}^{(L)}\nonumber \\
 & = & -\mathbf{e}_{j}^{T}\mathbf{T}_{1}\mathbf{e}_{1}I_{C_{1}}-\mathbf{e}_{j}^{T}\mathbf{T}_{1}\mathbf{A}_{1}\mathbf{T}_{2}\mathbf{e}_{1}I_{C_{2}}\ldots-\mathbf{e}_{j}^{T}\mathbf{T}_{1}\mathbf{A}_{1}\ldots\mathbf{T}_{M-1}\mathbf{A}_{M-1}\mathbf{T}_{M}\mathbf{e}_{1}I_{C_{M}}+\nonumber \\
 &  & -\mathbf{e}_{j}^{T}\mathbf{T}_{1}\mathbf{A}_{1}\ldots\mathbf{T}_{M}\mathbf{A}_{M}\mathbf{T}_{M+1}\mathbf{I}_{C_{R}}
\end{eqnarray}
From which we conclude that

\[
I_{R_{j}^{x}}=-\mathbf{F}_{R_{j}^{x},C}^{eff}\mathbf{I}_{C}
\]
where $\mathbf{F}_{R_{j}^{x},C}^{eff}$ is a row vector of length
$\left(M+N\right)$ with

\begin{equation}
\begin{cases}
\mathbf{F}_{R_{j}^{x},C}^{eff}\left(k\right)=\mathbf{e}_{j}^{T}\mathbf{T}_{1}\mathbf{e}_{1} & for\; k=1\\
\mathbf{F}_{R_{j}^{x},C}^{eff}\left(k\right)=\mathbf{e}_{j}^{T}\mathbf{T}_{1}\mathbf{A}_{1}\ldots\mathbf{T}_{k-1}\mathbf{A}_{k-1}\mathbf{T}_{k}\mathbf{e}_{1} & for\;1<k\leq M\\
\mathbf{F}_{R_{j}^{x},C}^{eff}\left(k\right)=\mathbf{e}_{j}^{T}\mathbf{T}_{1}\mathbf{A}_{1}\ldots\mathbf{T}_{M}\mathbf{A}_{M}\mathbf{T}_{M+1}\mathbf{e}_{k-M} & for\; M+1\leq k\leq M+N
\end{cases}\label{eq:eff-FRxjC-submatrix-current}
\end{equation}

So one then needs to append $\mathbf{F}_{R_{j}^{x},C}^{eff}$ to $\mathbf{F}_{ZC}^{eff}$
defined in Eq. \eqref{eq:eff-FZC-submatrix-current} as its last row.
Also $R_{j}^{x}$ will appear at the last diagonal entry of the tree
impedance matrix $\mathbf{Z}$ defined in Eq. \eqref{eq:Z-tree-impedance-matrix}

\begin{equation}
\mathbf{Z}=\left(\begin{array}{cccc}
r_{1} &  &  & \boldsymbol{0}\\
 & \ddots\\
 &  & r_{M}\\
\boldsymbol{0} &  &  & R_{j}^{x}
\end{array}\right)
\end{equation}

To compute the contribution of $R_{j}^{x}$ to the dissipation rate
we follow the same procedure done for the resistors $r_{j}$, $1\leq j\leq M$
in Eqs. \eqref{eq:mj-multiport-Brune}-\eqref{eq:T1-rate-due-to-rj-mp-Brune}
above and compute the coupling $\bar{\mathbf{m}}_{R_{j}^{x}}$ of
the bath due to the resistor $R_{j}^{x}$ to the circuit degrees of
freedom using Eq. \eqref{eq:m0-vector} with the effective matrix
$\mathbf{F}_{R_{j}^{x},C_{0}}^{eff}$

\begin{eqnarray}
\bar{\mathbf{m}}_{R_{j}^{x}} & = & \mathcal{F}_{C_{0}}^{eff}\mathbf{C}_{0}\left(\mathbf{F}_{R_{j}^{x},C_{0}}^{eff}\right)^{T}\\
 & = & \left(\begin{array}{c}
\mathbf{\bar{m}}_{R_{j}^{x},J}\\
\mathbf{\bar{m}}_{R_{j}^{x},L}
\end{array}\right)
\end{eqnarray}
where we assume the following partitioning for $\mathbf{F}_{R_{j}^{x},C}^{eff}$

\begin{equation}
\mathbf{F}_{R_{j}^{x},C}^{eff}=\left(\begin{array}{cc}
\mathbf{F}_{R_{j}^{x},C_{0}}^{eff} & \mathbf{F}_{R_{j}^{x},C_{R}}^{eff}\end{array}\right)
\end{equation}
where $\mathbf{F}_{R_{j}^{x},C_{0}}^{eff}$ and $\mathbf{F}_{R_{j}^{x},C_{R}}^{eff}$
are row vectors of length $M$ and $N$, respectively.

Using Eq. \eqref{eq:CZrj-with-FrjC0}

\begin{equation}
\mathbf{\bar{C}}_{Z,R_{j}^{x}}\left(\omega\right)=-i\omega R_{j}^{x}\left[1+i\omega R_{j}^{x}\mathbf{F}_{R_{j}^{x},C_{0}}^{eff}\mathbf{C}_{0}\left(\mathbf{F}_{R_{j}^{x},C_{0}}^{eff}\right)^{T}\right]^{-1}
\end{equation}
which is related to the kernel as derived in Eq. \eqref{eq:Kernel-appendix}

\begin{equation}
K_{R_{j}^{x}}\left(\omega\right)=-\omega^{2}\bar{\mathbf{C}}_{Z,R_{j}^{x}}\left(\omega\right)
\end{equation}
and the spectrum of the bath due to the resistor $R_{j}^{x}$ is given
by Eq. \eqref{eq:spectral-density-of-the-bath}

\begin{eqnarray}
J_{R_{j}^{x}}\left(\omega\right) & = & Im\left[K_{R_{j}^{x}}\left(\omega\right)\right]\\
 & = & \frac{\omega^{3}R_{j}^{x}}{1+\omega^{2}\left(R_{j}^{x}\right)^{2}\left[\mathbf{F}_{R_{j}^{x},C_{0}}^{eff}\mathbf{C}_{0}\left(\mathbf{F}_{R_{j}^{x},C_{0}}^{eff}\right)^{T}\right]^{2}}
\end{eqnarray}
And by Eq. \eqref{eq:T1-rate-for-rj} the contribution of $R_{j}^{x}$
to the loss rate

\[
\frac{1}{T_{1,R_{j}^{x}}}=\frac{4}{\hbar}\left|\left\langle 0\left|\mathbf{\bar{m}}_{R_{j}^{x}}\cdot\mathbf{\Phi}\right|1\right\rangle \right|^{2}J_{R_{j}^{x}}\left(\omega_{01}\right)\coth\left(\frac{\hbar\omega_{01}}{2k_{B}T}\right)
\]

If we have several resistors shunting the multiport Brune circuit
the analysis is straightforward. One then needs to drop the rows in
the $\mathbf{F}_{JC}^{eff}$ matrix corresponding to the ports shunted
by resistors and repeat the dissipation analysis above for each $R_{j}^{x}$.
A full $\mathbf{F}_{ZC}^{eff}$ is obtained by appending the rows
$\mathbf{F}_{R_{j}^{x},C}^{eff}$ corresponding to each $R_{j}^{x}$
which are appended to $\mathbf{Z}$.

\subsection{\label{sub:Voltage-sources-shunting-the-ports}Voltage Sources Shunting
the Ports}

\begin{figure}
\begin{centering}
\includegraphics{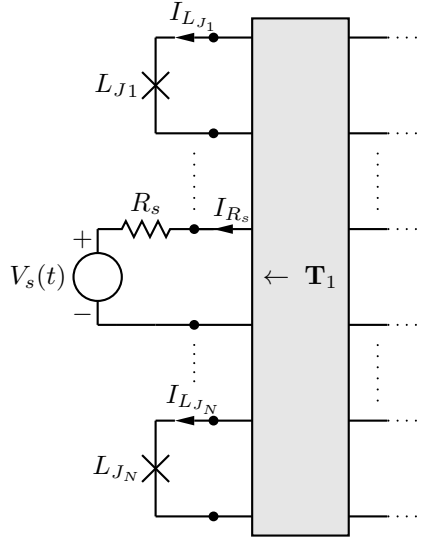} 
\par\end{centering}

\caption{\label{fig:Voltage-shunting-multiport-Brune}Voltage source $V_{s}\left(t\right)$
in series with a resistance $R_{s}$ shunting the $j^{th}$ port of
the multiport Brune circuit.}
\end{figure}

Now we consider shunting one of the ports by a possibly time-dependent
voltage source $V_{s}\left(t\right)$ as shown in Fig. \eqref{fig:Voltage-shunting-multiport-Brune}.
We also assume that the voltage source has an in series resistance
$R_{s}$. Assuming also that $V_{s}$ and $R_{s}$ shunt the $j^{th}$
port the treatment of the resistor $R_{s}$ follows the same analysis
we did above for the resistor $R_{j}^{x}$ in the previous section.
To treat the voltage source we first note the following

\begin{equation}
I_{V_{s}}=I_{R_{s}}=I_{\mathbf{T}_{1},j}^{(L)}
\end{equation}
Hence with the same loop analysis we did for the resistor $R_{j}^{x}$
above we can derive an effective loop matrix $\mathbf{F}_{V_{s},C}^{eff}$
such that

\begin{equation}
I_{V_{s}}=-\mathbf{F}_{V_{s},C}^{eff}\mathbf{I}_{C}
\end{equation}
where $\mathbf{F}_{V_{s},C}^{eff}=\mathbf{F}_{R_{j}^{x},C}^{eff}$
with $\mathbf{F}_{R_{j}^{x},C}^{eff}$ being defined in Eq. \eqref{eq:eff-FRxjC-submatrix-current}.

Now $\mathbf{F}_{V_{s},C}^{eff}$ will appear as one of the rows of
the $\mathbf{F}_{VC}$ matrix. One can then use $\mathbf{F}_{VC}$
to first compute $\bar{\mathbf{m}}_{V}$ defined in Eq. \eqref{eq:Burkard-mv-vector-definition}
and then the coupling vectors $\mathbf{C}_{V}$ and $\mathcal{C}_{V}$
defined in Eqs. \eqref{eq:Burkard-bold-CV} and \eqref{eq:Burkard-CV},
respectively, in Appendix \eqref{sub:Burkard's-method}. Making the
substitution $\mathbf{C}_{R}\leftarrow\left(i\omega\right)^{-1}\mathbf{R}^{-1}$
in Eq. \eqref{eq:shunt-resistance-substitution} and following the
analysis at the end of Appendix \eqref{sub:Burkard's-method} we get
the following Hamiltonian derived in Eq. \eqref{eq:Burkard-Hamiltonian-extended}
for the multiport Brune circuit shunted by the voltage source at one
of its ports

\begin{equation}
\mathcal{H}_{S}=\frac{1}{2}\left(\mathbf{Q}-\left(\mathbf{C}_{V,0}+\mathcal{C}_{V,R}\left(t\right)\right)*\mathbf{V}_{V}\left(t\right)\right)^{T}\mathcal{C}_{0}^{-1}\left(\mathbf{Q}-\left(\mathbf{C}_{V,0}+\mathcal{C}_{V,R}\left(t\right)\right)*\mathbf{V}_{V}\left(t\right)\right)+U\left(\mathbf{\Phi}\right)
\end{equation}
where the voltage source vector $\mathbf{V}_{V}\left(t\right)$ has
$V_{s}\left(t\right)$ at its corresponding entry. The matrices $\mathbf{C}_{V,0}$
and $\mathcal{C}_{V,R}\left(t\right)$ are defined, respectively,
in Eqs. \eqref{eq:CV0-definition} and \eqref{eq:curly-CVR-definition}
in Appendix \eqref{sub:Analysis-of-voltage-source-couplings} which
we repeat here for convenience

\begin{equation}
\mathbf{C}_{V,0}=\mathcal{F}_{C_{0}}^{eff}\mathbf{C}_{0}\left(\mathbf{F}_{VC_{0}}^{eff}\right)^{T}
\end{equation}

\begin{equation}
\mathcal{C}_{V,R}\left(\omega\right)=\mathbf{C}_{V,R}\left(\omega\right)+\mathcal{C}_{V}\left(\omega\right)\label{eq:curly-CVR-definition-main-text}
\end{equation}
where in the definition of the frequency-independent coupling matrix
$\mathbf{C}_{V,0}$ we assumed the following partitioning for $\mathbf{F}_{VC}^{eff}$
as in Eq. \eqref{eq:FVC-partitioning}

\begin{equation}
\mathbf{F}_{VC}^{eff}=\left(\begin{array}{cc}
\mathbf{F}_{VC_{0}}^{eff} & \mathbf{F}_{VC_{R}}^{eff}\end{array}\right)
\end{equation}
For the definitions of the matrices $\mathbf{C}_{V,R}\left(\omega\right)$
and $\mathcal{C}_{V}\left(\omega\right)$ appearing in Eq. \eqref{eq:curly-CVR-definition-main-text}
we refer the reader to Eqs. \eqref{eq:CVR-definition} and \eqref{eq:curly-CV-update},
respectively, in Appendix \eqref{sub:Analysis-of-voltage-source-couplings}.
We recall that effective loop matrices should be used in those definitions.

\textcolor{black}{We would like to note here that the analysis above
is done only to obtain the time-dependent Hamiltonian in the case
of a AC voltage drive $V_{s}\left(t\right)$; we will only use our
formalism to compute $T_{1}$ times only for unexcited systems, i.e.
when there is no time-dependent voltage applied.}

\section{Examples}

Here we illustrate the method described in the previous section with
a generic 2-port 1-stage Brune circuit.

\subsection{\label{sub:2-port-1-stage-generic-cct-example}2-port 1-stage Generic
Brune Circuit}

\begin{figure}
\begin{centering}
\includegraphics{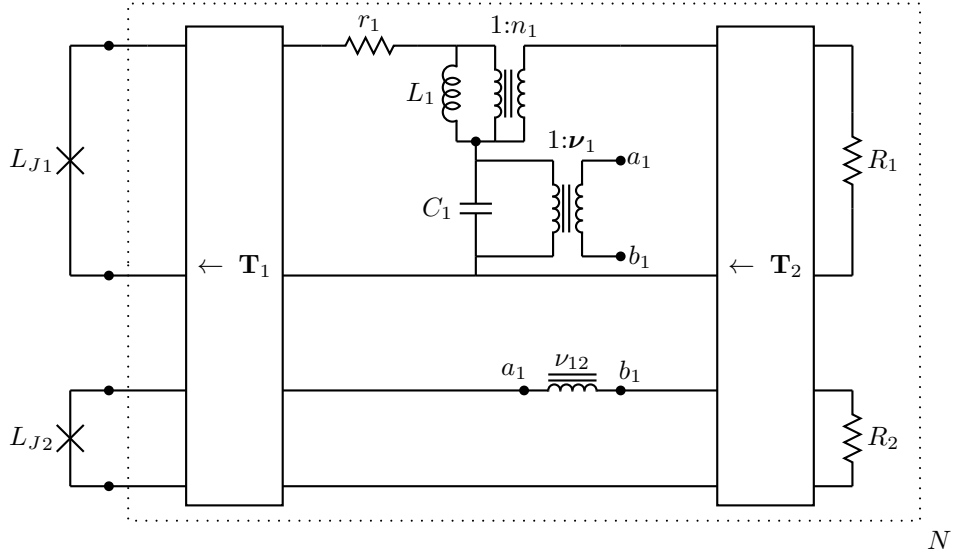} 
\par\end{centering}

\caption{\label{fig:2-port-1-stage-multiport-Brune}2-port 1-stage multiport
Brune circuit. The part of the circuit labeled $N$ contained in the
dotted box is the 2-port Brune circuit shunted by two Josephson junctions
at each of its ports.}
\end{figure}

We consider a generic 2-port 1-stage Brune circuit shown in Fig. \eqref{fig:2-port-1-stage-multiport-Brune}.
The 2-port Brune circuit is shown in the dotted box. This 2-port Brune
circuit is shunted by two Josephson junctions $L_{J_{1}}$ and $L_{J_{2}}$
at each of its ports. There is only one reactive stage which corresponds
to a single pole at a finite frequency in the partial fraction expansion
for the impedance fit in Eq. \eqref{eq:Vector-Fit}.

\begin{figure}
\begin{centering}
\includegraphics{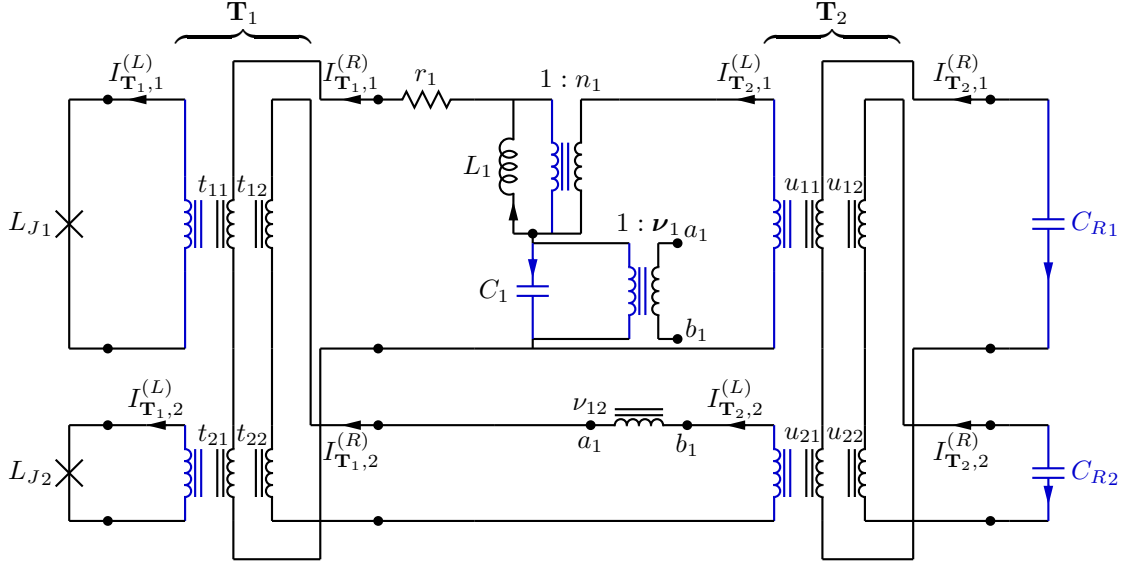} 
\par\end{centering}

\caption{\label{fig:Modified-2-port-Brune-circuit}Modified 2-port Brune circuit.
Chord branches are shown in blue, tree branches are shown in black.}
\end{figure}

First we need to replace the resistors $R_{1}$ and $R_{2}$ in the
last stage by capacitors $C_{R_{1}}$ and $C_{R_{2}}$ as shown in
Fig. \eqref{fig:Modified-2-port-Brune-circuit}. We will later make
the substitutions $C_{R_{1}}\leftarrow\frac{1}{i\omega R_{1}}$ and
$C_{R_{2}}\leftarrow\frac{1}{i\omega R_{2}}$ for the dissipation
analysis.

There are three capacitors in the circuit in Fig. \eqref{fig:Modified-2-port-Brune-circuit}.
All the capacitors are in chord branches. The currents through those
capacitors are given by the vector

\begin{equation}
\mathbf{I}_{C}=\left(\begin{array}{c}
I_{C_{1}}\\
I_{C_{R_{1}}}\\
I_{C_{R_{2}}}
\end{array}\right)
\end{equation}
As described in the previous section we will eliminate transformer
branches and write the currents through tree branches as functions
of capacitor currents. This way we will compute effective loop matrices
$\mathbf{F}_{JC}^{eff}$, $\mathbf{F}_{LC}^{eff}$ and $\mathbf{F}_{ZC}^{eff}$
.

As we see in Fig. \eqref{fig:Modified-2-port-Brune-circuit} the two
Belevitch transformers $\mathbf{T}_{1}$ and $\mathbf{T}_{2}$ are
given by the turns ratios matrices

\begin{equation}
\mathbf{T}_{1}=\left(\begin{array}{cc}
t_{11} & t_{12}\\
t_{21} & t_{22}
\end{array}\right)
\end{equation}

\begin{equation}
\mathbf{T}_{2}=\left(\begin{array}{cc}
u_{11} & u_{12}\\
u_{21} & u_{22}
\end{array}\right)
\end{equation}

We can now write

\begin{eqnarray}
\mathbf{I}_{\mathbf{T}_{2}}^{(L)} & = & \mathbf{T}_{2}\mathbf{I}_{\mathbf{T}_{2}}^{(R)}=-\mathbf{T}_{2}\mathbf{I}_{C_{R}}\\
 & = & -\left(\begin{array}{cc}
u_{11} & u_{12}\\
u_{21} & u_{22}
\end{array}\right)\left(\begin{array}{c}
I_{C_{R_{1}}}\\
I_{C_{R_{2}}}
\end{array}\right)
\end{eqnarray}
where we used $\mathbf{I}_{\mathbf{T}_{2}}^{(R)}=-\mathbf{I}_{C_{R}}$.
The current $I_{L_{1}}$ through the inductor $L_{1}$ is given by

\begin{eqnarray}
I_{L_{1}} & = & -I_{C_{1}}+\boldsymbol{\upsilon}_{1}\mathbf{I}_{\mathbf{T}_{2}}^{(L)}\\
 & = & -I_{C_{1}}-\left(\left(1-n_{1}\right),-\nu_{12}\right)\left(\begin{array}{cc}
u_{11} & u_{12}\\
u_{21} & u_{22}
\end{array}\right)\left(\begin{array}{c}
I_{C_{R_{1}}}\\
I_{C_{R_{2}}}
\end{array}\right)\\
 & = & -I_{C_{1}}-\left(\left(1-n_{1}\right)u_{11}-u_{21}\nu_{12},\,\left(1-n_{1}\right)u_{12}-u_{22}\nu_{12}\right)\left(\begin{array}{c}
I_{C_{R_{1}}}\\
I_{C_{R_{2}}}
\end{array}\right)\nonumber 
\end{eqnarray}
from which we identify

\begin{equation}
\mathbf{F}_{LC}^{eff}=\left(1,\,\left(1-n_{1}\right)u_{11}-u_{21}\nu_{12},\,\left(1-n_{1}\right)u_{12}-u_{22}\nu_{12}\right)\label{eq:FLC-eff-mp-ex}
\end{equation}
where $\mathbf{F}_{LC}^{eff}$ is defined by the relation

\begin{equation}
I_{L_{1}}=-\mathbf{F}_{LC}^{eff}\mathbf{I}_{C}
\end{equation}

We now move to the transformer $\mathbf{T}_{1}$ and write

\begin{eqnarray}
\mathbf{I}_{\mathbf{T}_{1}}^{(R)} & = & -I_{C_{1}}\mathbf{e}_{1}+\mathbf{A}_{1}\mathbf{I}_{\mathbf{T}_{2}}^{(L)}\\
 & = & -I_{C_{1}}\left(\begin{array}{c}
1\\
0
\end{array}\right)+\left(\begin{array}{cc}
1 & -\nu_{12}\\
0 & 1
\end{array}\right)\mathbf{I}_{\mathbf{T}_{2}}^{(L)}\\
 & = & -I_{C_{1}}\left(\begin{array}{c}
1\\
0
\end{array}\right)-\left(\begin{array}{cc}
1 & -\nu_{12}\\
0 & 1
\end{array}\right)\left(\begin{array}{cc}
u_{11} & u_{12}\\
u_{21} & u_{22}
\end{array}\right)\left(\begin{array}{c}
I_{C_{R_{1}}}\\
I_{C_{R_{2}}}
\end{array}\right)\label{eq:IT1R}
\end{eqnarray}
and

\begin{eqnarray}
\mathbf{I}_{\mathbf{T}_{1}}^{(L)} & = & \mathbf{T}_{1}\mathbf{I}_{\mathbf{T}_{1}}^{(R)}\label{eq:T1-currents-2-port-example-start}\\
 & = & -I_{C_{1}}\mathbf{T}_{1}\mathbf{e}_{1}+\mathbf{T}_{1}\mathbf{A}_{1}\mathbf{I}_{\mathbf{T}_{2}}^{(L)}\\
 & = & -I_{C_{1}}\left(\begin{array}{c}
t_{11}\\
t_{21}
\end{array}\right)-\left(\begin{array}{cc}
t_{11} & t_{12}\\
t_{21} & t_{22}
\end{array}\right)\left(\begin{array}{cc}
1 & -\nu_{12}\\
0 & 1
\end{array}\right)\left(\begin{array}{cc}
u_{11} & u_{12}\\
u_{21} & u_{22}
\end{array}\right)\left(\begin{array}{c}
I_{C_{R_{1}}}\\
I_{C_{R_{2}}}
\end{array}\right)\nonumber \\
 & = & -I_{C_{1}}\left(\begin{array}{c}
t_{11}\\
t_{21}
\end{array}\right)+\label{eq:T1-currents-2-port-example-end}\\
 &  & -\left(\begin{array}{cc}
t_{11}u_{11}+u_{21}\left(t_{12}-t_{11}\nu_{12}\right) & t_{11}u_{12}+u_{22}\left(t_{12}-t_{11}\nu_{12}\right)\\
t_{21}u_{11}+u_{21}\left(t_{22}-t_{21}\nu_{12}\right) & t_{21}u_{12}+u_{22}\left(t_{22}-t_{21}\nu_{12}\right)
\end{array}\right)\left(\begin{array}{c}
I_{C_{R_{1}}}\\
I_{C_{R_{2}}}
\end{array}\right)\nonumber 
\end{eqnarray}

Using the relation $\mathbf{I}_{J}=\mathbf{I}_{\mathbf{T}_{1}}^{(L)}$
and the Eqs. \eqref{eq:T1-currents-2-port-example-start}-\eqref{eq:T1-currents-2-port-example-end}
above we identify

\begin{equation}
\mathbf{F}_{JC}^{eff}=\left(\begin{array}{ccc}
t_{11} & t_{11}u_{11}+u_{21}\left(t_{12}-t_{11}\nu_{12}\right) & t_{11}u_{12}+u_{22}\left(t_{12}-t_{11}\nu_{12}\right)\\
t_{21} & t_{21}u_{11}+u_{21}\left(t_{22}-t_{21}\nu_{12}\right) & t_{21}u_{12}+u_{22}\left(t_{22}-t_{21}\nu_{12}\right)
\end{array}\right)\label{eq:FJC-eff-mp-ex}
\end{equation}
where $\mathbf{F}_{JC}^{eff}$ is defined by the relation $\mathbf{I}_{J}=-\mathbf{F}_{JC}^{eff}\mathbf{I}_{C}$.

Using Eqs. \eqref{eq:FLC-eff-mp-ex}, \eqref{eq:FJC-eff-mp-ex} and
making the partitioning in Eq. \eqref{eq:curly-FC-partitioning-multiport-Brune}
we get

\begin{equation}
\mathcal{F}_{C_{0}}^{eff}=\left(\begin{array}{c}
t_{11}\\
t_{21}\\
1
\end{array}\right)\label{eq:FC0-eff-mp-ex}
\end{equation}
and hence by Eq. \eqref{eq:curly-C0-definition-multiport-Brune}

\begin{eqnarray}
\mathcal{C}_{0} & = & \left(\begin{array}{cc}
\mathbf{C}_{J} & \boldsymbol{0}\\
\boldsymbol{0} & \boldsymbol{0}
\end{array}\right)+\mathcal{F}_{C_{0}}^{eff}\mathbf{C}_{0}\left(\mathcal{F}_{C_{0}}^{eff}\right)^{T}\\
 & = & \left(\begin{array}{ccc}
C_{J_{1}}+t_{11}^{2}C_{1} & t_{11}t_{21}C_{1} & t_{11}C_{1}\\
t_{21}t_{11}C_{1} & C_{J_{2}}+t_{21}^{2}C_{1} & t_{21}C_{1}\\
t_{11}C_{1} & t_{21}C_{1} & C_{1}
\end{array}\right)
\end{eqnarray}
where we noted that $\mathbf{C}_{0}=C_{1}$ and assumed that $\mathbf{C}_{J}=\left(\begin{array}{cc}
C_{J_{1}} & 0\\
0 & C_{J_{2}}
\end{array}\right)$. We also note that a non-zero $\mathbf{C}_{J}$ is necessary here
to have a non-singular $\mathcal{C}_{0}$ matrix.

$\mathbf{M}_{0}$ is given by Eq. \eqref{eq:M0-multiport-Brune} as

\begin{equation}
\mathbf{M}_{0}=\left(\begin{array}{ccc}
0 & 0 & 0\\
0 & 0 & 0\\
0 & 0 & \frac{1}{L_{1}}
\end{array}\right)
\end{equation}
\textcolor{black}{Hence the Hamiltonian is given by Eq. \eqref{eq:Hamiltonian-multiport}
as}

\textcolor{black}{{} 
\begin{equation}
\mathcal{H_{S}}=\frac{1}{2}\boldsymbol{Q}^{T}\mathcal{C}_{0}^{-1}\boldsymbol{Q}+U\left(\boldsymbol{\Phi}\right)
\end{equation}
with}

\textcolor{black}{{} 
\begin{equation}
U\left(\boldsymbol{\Phi}\right)=-\left(\frac{\Phi_{0}}{2\pi}\right)^{2}\mathbf{L}_{J}^{-1}\cos\left(\boldsymbol{\mathbf{\varphi}}_{J}\right)+\frac{1}{2}\boldsymbol{\Phi}^{T}\mathbf{M_{0}}\boldsymbol{\Phi}
\end{equation}
and}

\textcolor{black}{{} 
\begin{equation}
\mathbf{L}_{J}=\left(\begin{array}{cc}
L_{J_{1}} & 0\\
0 & L_{J_{2}}
\end{array}\right)
\end{equation}
}

Noting also $I_{r_{1}}=I_{\mathbf{T}_{1},1}^{(R)}$ we identify by
Eq. \eqref{eq:IT1R}

\begin{equation}
\mathbf{F}_{ZC}^{eff}=\left(\begin{array}{ccc}
1 & u_{11}-u_{21}\nu_{12} & u_{12}-u_{22}\nu_{12}\end{array}\right)
\end{equation}
$\mathbf{F}_{ZC}^{eff}$ is defined by $I_{r_{1}}=-\mathbf{F}_{ZC}^{eff}\mathbf{I}_{C}$.
We note

\begin{equation}
\mathbf{F}_{r_{1},C_{0}}^{eff}=1\label{eq:Fr1C0-eff-mp-ex}
\end{equation}
and using Eq. \eqref{eq:mj-multiport-Brune} and Eqs. \eqref{eq:FC0-eff-mp-ex},
\eqref{eq:Fr1C0-eff-mp-ex} we compute

\begin{eqnarray}
\bar{\mathbf{m}}_{1} & = & \mathcal{F}_{C_{0}}^{eff}\mathbf{C}_{0}\left(\mathbf{F}_{r_{1},C_{0}}^{eff}\right)^{T}\\
 & = & \left(\begin{array}{c}
t_{11}\\
t_{21}\\
1
\end{array}\right)C_{1}
\end{eqnarray}

Using Eq. \eqref{eq:rj-dissipation-analysis-multiport-2} and \eqref{eq:Fr1C0-eff-mp-ex}
we compute

\begin{eqnarray}
\mathbf{\bar{C}}_{Z,r_{1}}\left(\omega\right) & = & -i\omega r_{1}\left[1+i\omega r_{1}\mathbf{F}_{r_{1},C_{0}}^{eff}\mathbf{C}_{0}\left(\mathbf{F}_{r_{1},C_{0}}^{eff}\right)^{T}\right]^{-1}\\
 & = & -i\omega r_{1}\left[1+i\omega r_{1}C_{1}\right]^{-1}
\end{eqnarray}
and by Eq. \eqref{eq:rj-dissipation-analysis-multiport-3} we have

\begin{eqnarray}
K_{1}\left(\omega\right) & = & -\omega^{2}\bar{\mathbf{C}}_{Z,r_{1}}\left(\omega\right)\\
 & = & i\omega^{3}r_{1}\left[1+i\omega r_{1}C_{1}\right]^{-1}
\end{eqnarray}
hence by Eq. \eqref{eq:rj-dissipation-analysis-multiport-5} we get
the spectral density of the bath due to the resistor $r_{1}$ as

\begin{eqnarray}
J_{1}\left(\omega\right) & = & Im\left[K_{1}\left(\omega\right)\right]\\
 & = & \frac{r_{1}\omega^{3}}{1+r_{1}^{2}C_{1}^{2}\omega^{2}}
\end{eqnarray}

\textcolor{black}{The}\textcolor{cyan}{{} }contribution of $r_{1}$
to the loss rate is computed then by using the formula in Eq. \eqref{eq:T1-rate-due-to-rj-mp-Brune}.

To do the dissipation analysis for the last two shunt resistors $R_{1}$
and $R_{2}$ we first note the following, using Eqs. \eqref{eq:FLC-eff-mp-ex},
\eqref{eq:FJC-eff-mp-ex} and making the partitioning in Eq. \eqref{eq:curly-FC-partitioning-multiport-Brune}

\begin{equation}
\mathcal{F}_{C_{R}}^{eff}=\left(\begin{array}{cc}
t_{11}u_{11}+u_{21}\left(t_{12}-t_{11}\nu_{12}\right) & t_{11}u_{12}+u_{22}\left(t_{12}-t_{11}\nu_{12}\right)\\
t_{21}u_{11}+u_{21}\left(t_{22}-t_{21}\nu_{12}\right) & t_{21}u_{12}+u_{22}\left(t_{22}-t_{21}\nu_{12}\right)\\
\left(1-n_{1}\right)u_{11}-u_{21}\nu_{12} & \left(1-n_{1}\right)u_{12}-u_{22}\nu_{12}
\end{array}\right)
\end{equation}
In Eq. \eqref{eq:curly-FRj-CR-eff} $\bar{\mathbf{m}}_{R_{1}}$ is
defined as the first column of $\mathcal{F}_{C_{R}}^{eff}$ and $\bar{\mathbf{m}}_{R_{2}}$
is defined as the second column of $\mathcal{F}_{C_{R}}^{eff}$ so
that

\begin{equation}
\bar{\mathbf{m}}_{R_{1}}=\left(\begin{array}{c}
t_{11}u_{11}+u_{21}\left(t_{12}-t_{11}\nu_{12}\right)\\
t_{21}u_{11}+u_{21}\left(t_{22}-t_{21}\nu_{12}\right)\\
\left(1-n_{1}\right)u_{11}-u_{21}\nu_{12}
\end{array}\right)
\end{equation}
and

\begin{equation}
\bar{\mathbf{m}}_{R_{2}}=\left(\begin{array}{c}
t_{11}u_{12}+u_{22}\left(t_{12}-t_{11}\nu_{12}\right)\\
t_{21}u_{12}+u_{22}\left(t_{22}-t_{21}\nu_{12}\right)\\
\left(1-n_{1}\right)u_{12}-u_{22}\nu_{12}
\end{array}\right)
\end{equation}

The spectral densities $J_{R_{1}}$ and $J_{R_{2}}$ of the baths
due to the resistors $R_{1}$ and $R_{2}$, respectively, are given
by the formula in Eq. \eqref{eq:JRj-mp-Brune} as

\begin{equation}
J_{R_{1}}\left(\omega\right)=\frac{\omega}{R_{1}}
\end{equation}
and

\begin{equation}
J_{R_{2}}\left(\omega\right)=\frac{\omega}{R_{2}}
\end{equation}

Contributions of the shunt resistors $R_{1}$ and $R_{2}$ to the
loss rate are then computed using the formula in Eq. \eqref{eq:T1-rate-due-to-Rj-mp-Brune}.

\subsection{\label{sub:3D-Transmon-example}3-port Data for the 3D-Transmon}

In this section we do a multiport study of the numerical example that
has been studied with the one-port classical Brune analysis in \citep{brune-quantization-paper}.
Below we repeat the description of the system for completeness.

To show the application of the synthesis method we have just described,
we analyse a dataset produced to analyse a 3D transmon similar to
the one reported in an experiment at IBM \citep{RigettiCu}. Our modeling
is performed using the finite-element electromagnetics simulator HFSS
\citep{HFSS}. Since the systems we want to model admit very small
loss \citep{3D,Reagor}, they are very close to the border which separates
passive systems from active ones. Therefore it is necessary to take
care that the simulation resolution is high enough to ensure the passivity
of the simulated impedance. Otherwise the fitted impedance $\mathbf{Z}\left(s\right)$
does not satisfy the $PR$ conditions in Section \eqref{sub:Positive-Real-Property-in-SS}
meaning that there is no passive physical network corresponding to
$\mathbf{Z}\left(s\right)$.

\begin{figure}
\begin{centering}
\includegraphics[scale=0.5]{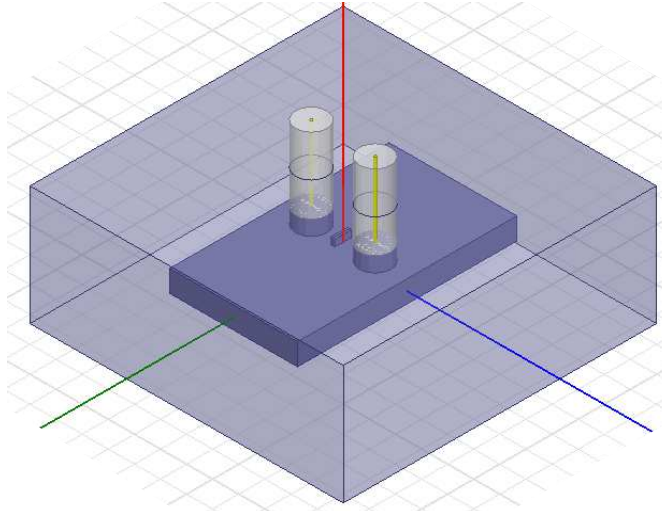} 
\par\end{centering}

\caption{\label{fig:Cavity-Fig1}Geometry of the 3D transmon qubit simulated
in HFSS. Light blue is perfect conductor and dark blue is the vacuum.
The qubit port terminals are defined on a dielectric substrate located
at the position of the red line. Two coaxial ports are positioned
symmetrically on each side of the substrate. The cavity dimensions
are $(height,\: length,\: width)=(4.2mm,\:24.5mm,\:42mm)$.}
\end{figure}

\begin{figure}
\begin{centering}
\includegraphics[scale=0.3]{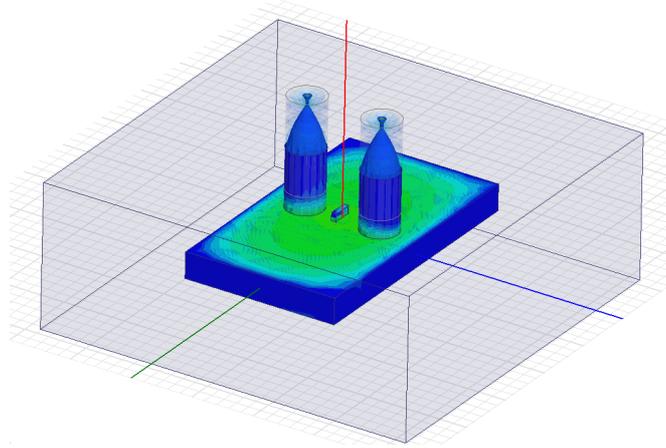} 
\par\end{centering}

\caption{\label{fig:Cavity-Fig2}Fundamental mode (the $TE101$ mode) of the
cavity with frequency $f_{TE101}=6.875GHz$. Green color indicates
electric field regions of higher magnitude compared to blue regions. }
\end{figure}

The simulated device is a 3D transmon, inserted with appropriate antenna
structures into the middle of a rectangular superconducting (aluminium)
box cavity, which is standard in several labs presently for high-coherence
qubit experiments. Fig. \eqref{fig:Cavity-Fig1} shows a perspective
rendering of the device, and Fig. \eqref{fig:Cavity-Fig2} shows an
intensity map of the fundamental mode of the cavity. The simulation
includes two coaxial ports entering the body of the cavity symmetrically
on either side of the qubit. HFSS is used to calculate the device's
three-port $S$ matrix over a wide frequency range, from $3.0$ to
$15.0$ GHz. The three ports are those defined by the two coaxial
connectors and the qubit terminal pair. That is, the metal defining
the Josephson junction itself is absent from the simulation, so that
its very small capacitance and (nonlinear) inductance can be added
back later as a discrete element as in Fig. \eqref{fig:Brune-circuit-state-space}.
The conversion from the $S$ matrix to $\mathbf{Z}_{sim}$ is calculated
using standard formulas \citep{Newcomb,Pozar}. \textcolor{black}{Here
we don't shunt the coaxial ports by $50\Omega$ terminations as we
did in \citep{brune-quantization-paper}. Rather, we keep the simulated
impedance matrix $\mathbf{Z}_{sim}$ with 3 ports.}

\textcolor{black}{To obtain the{} impedance matrix $\mathbf{Z}(s)$
as in Eq. \eqref{eq:Vector-Fit} fitted to the numerical impedance
matrix $\mathbf{Z}_{sim}$, we use the MATLAB package Vector Fitting
(VF) \citep{Vector Fitting}. Applying the compacting technique described
in \citep{Compacting} we get a state-space description for $\mathbf{Z}\left(s\right)$
of dimension $21$ (the dimension before compacting being $51$).
Applying the multiport Brune algorithm described in Section \eqref{sec:Multiport-Brune-Algorithm}
we obtain the circuit parameter values listed in Table \eqref{tab:circuit-parameter-values-Abraham-3-port-data}
with the number of stages $M=12$. There are three degenerate stages,
namely stages $1,11,12$. For the multiport Brune circuit with parameters
listed in Table \eqref{tab:circuit-parameter-values-Abraham-3-port-data}
the first port is the qubit port. Here are the extracted Belevitch
transformer matrices for each stage:}

\begin{table}
\begin{centering}
\begin{tabular}{|c|c|c|c|c|c|c|}
\hline 
$j$  & $r_{j}\left(\Omega\right)$  & $L_{j}\left(nH\right)$  & $C_{j}\left(nF\right)$  & $t_{j}=1/n_{j}$  & $\nu_{j,2}$  & $\nu_{j,3}$\tabularnewline
\hline 
\hline 
$1*$  & $0.0923$  & $0$  & $1.1953\times10^{-4}$  & $0$  & $0$  & $0$\tabularnewline
\hline 
$2$  & $0.0471$  & $7.1890\times10^{2}$  & $2.4523\times10^{-7}$  & $0.9478$  & $-0.0008$  & $-0.0259$\tabularnewline
\hline 
$3$  & $0.0973$  & $2.7674$  & $7.7198\times10^{-4}$  & $0.0986$  & $-0.0002$  & $0.2050$\tabularnewline
\hline 
$4$  & $0.1063$  & $2.7113$  & $7.8675\times10^{-4}$  & $0.0971$  & $0.0003$  & $-0.0020$\tabularnewline
\hline 
$5$  & $0.2136$  & $3.0283\times10^{3}$  & $1.7701\times10^{-7}$  & $0.9915$  & $0.0037$  & $0.0018$\tabularnewline
\hline 
$6$  & $20.7896$  & $2.7344\times10^{2}$  & $1.0464\times10^{-6}$  & $0.7657$  & $0.0002$  & $0.3708$\tabularnewline
\hline 
$7$  & $21.4619$  & $2.7500\times10^{2}$  & $1.0416\times10^{-6}$  & $0.7508$  & $0$  & $-0.0222$\tabularnewline
\hline 
$8$  & $26.6330$  & $2.4557\times10^{4}$  & $6.2335\times10^{-9}$  & $0.9959$  & $9.65\times10^{-5}$  & $-2.311\times10^{-4}$\tabularnewline
\hline 
$9$  & $4.7957$  & $4.9851\times10^{2}$  & $2.0961\times10^{-7}$  & $0.8408$  & $0.0002$  & $0.0122$\tabularnewline
\hline 
$10$  & $30.5600$  & $4.6115\times10^{2}$  & $2.2697\times10^{-7}$  & $0.8409$  & $0.0007$  & $-0.0623$\tabularnewline
\hline 
$11*$  & $84.5207$  & $0$  & $2.4178\times10^{-7}$  & $0$  & $0$  & $0$\tabularnewline
\hline 
$12*$  & $88.4419$  & $0$  & $2.2673\times10^{-7}$  & $0$  & $0$  & $0$\tabularnewline
\hline 
\end{tabular}
\par\end{centering}

\begin{centering}
$R_{1}=1.0837\times10^{7}$, $R_{2}=1.1306\times10^{7}$, $R_{3}=7.7537\times10^{7}$ 
\par\end{centering}

\caption{\label{tab:circuit-parameter-values-Abraham-3-port-data}3-port Brune
circuit parameter values for the dataset corresponding to the setup
in Figs. \eqref{fig:Cavity-Fig1} and \eqref{fig:Cavity-Fig2}. Stages
marked with $*$ are capactive degenerate stages. We note that in
degenerate stages there are no $n$-type and $\boldsymbol{\nu}$-type
transformer couplings. \textcolor{black}{The first port is the qubit
port.}}
\end{table}

\begin{eqnarray}
\mathbf{T}_{1} & = & \left(\begin{array}{ccc}
-1.0000 & -0.0001 & -0.0010\\
0.0008 & -0.7148 & -0.6993\\
0.0007 & 0.6993 & -0.7148
\end{array}\right)
\end{eqnarray}

\begin{equation}
\mathbf{T}_{2}=\left(\begin{array}{ccc}
0.8933 & -0.0132 & -0.4493\\
-0.0132 & -0.9999 & 0.0032\\
-0.4493 & 0.0030 & -0.8934
\end{array}\right)
\end{equation}

\begin{equation}
\mathbf{T}_{3}=\left(\begin{array}{ccc}
0.4315 & 0.0060 & -0.9021\\
0.0127 & 0.9998 & 0.0127\\
0.9020 & -0.0169 & 0.4314
\end{array}\right)
\end{equation}

\begin{equation}
\mathbf{T}_{4}=\left(\begin{array}{ccc}
0.0000 & -1.0000 & 0.0030\\
1.0000 & 0.0000 & 0.0000\\
0.0000 & 0.0030 & 1.0000
\end{array}\right)
\end{equation}

\begin{equation}
\mathbf{T}_{5}=\left(\begin{array}{ccc}
0.0000 & 0.4254 & -0.9050\\
0.0403 & -0.9043 & -0.4250\\
-0.9992 & -0.0365 & -0.0171
\end{array}\right)
\end{equation}

\begin{equation}
\mathbf{T}_{6}=\left(\begin{array}{ccc}
-0.0416 & 0.0024 & -0.9991\\
0.9299 & 0.3659 & -0.0378\\
0.3655 & -0.9306 & -0.0174
\end{array}\right)
\end{equation}

\begin{equation}
\mathbf{T}_{7}=\left(\begin{array}{ccc}
-0.0006 & -0.9994 & -0.0342\\
1.0000 & -0.0006 & -0.0000\\
0.0000 & -0.0342 & 0.9994
\end{array}\right)
\end{equation}

\begin{equation}
\mathbf{T}_{8}=\left(\begin{array}{ccc}
0.9975 & 0.0341 & -0.0615\\
-0.0308 & 0.9981 & 0.0538\\
0.0632 & -0.0517 & 0.9967
\end{array}\right)
\end{equation}

\begin{equation}
\mathbf{T}_{9}=\left(\begin{array}{ccc}
-0.9976 & 0.0011 & -0.0685\\
0.0032 & 0.9995 & -0.0299\\
0.0685 & -0.0301 & -0.9972
\end{array}\right)
\end{equation}

\begin{equation}
\mathbf{T}_{10}=\left(\begin{array}{ccc}
-0.0011 & -0.9999 & -0.0109\\
1.0000 & -0.0011 & 0.0003\\
-0.0003 & -0.0109 & 0.9999
\end{array}\right)
\end{equation}

\begin{equation}
\mathbf{T}_{11}=\left(\begin{array}{ccc}
-0.9775 & -0.1067 & -0.1820\\
-0.1088 & 0.9941 & 0.0015\\
0.1808 & 0.0212 & -0.9833
\end{array}\right)
\end{equation}

\begin{equation}
\mathbf{T}_{12}=\left(\begin{array}{ccc}
-0.0081 & 0.9876 & -0.1566\\
-1.0000 & -0.0080 & 0.0013\\
0 & 0.1566 & 0.9877
\end{array}\right)
\end{equation}

\begin{equation}
\mathbf{T}_{13}=\left(\begin{array}{ccc}
-0.0978 & 0.9951 & -0.0116\\
0.9952 & 0.0978 & -0.0001\\
0.0011 & -0.0116 & -0.9999
\end{array}\right)
\end{equation}

\textcolor{black}{We note that all the Belevitch transformer matrices
above are orthogonal - as expected - up to numerical noise. We would
like to also mention that $\mathbf{T}_{j}$ being proportional to
a permutation matrix for some $j$, $1\leq j\leq M$ would re-order
the ports without introducing a coupling between them. Note that all
the $T_{j}$ matrices, except for the first one, are roughly approximated
by permutation matrices; this can taken to mean that there is only
one ``channel'' whereby there is significant coupling among the
three ports due to the first Belevitch transformer $\mathbf{T}_{1}$.}

\begin{figure}[H]
\begin{centering}
\includegraphics[scale=0.85]{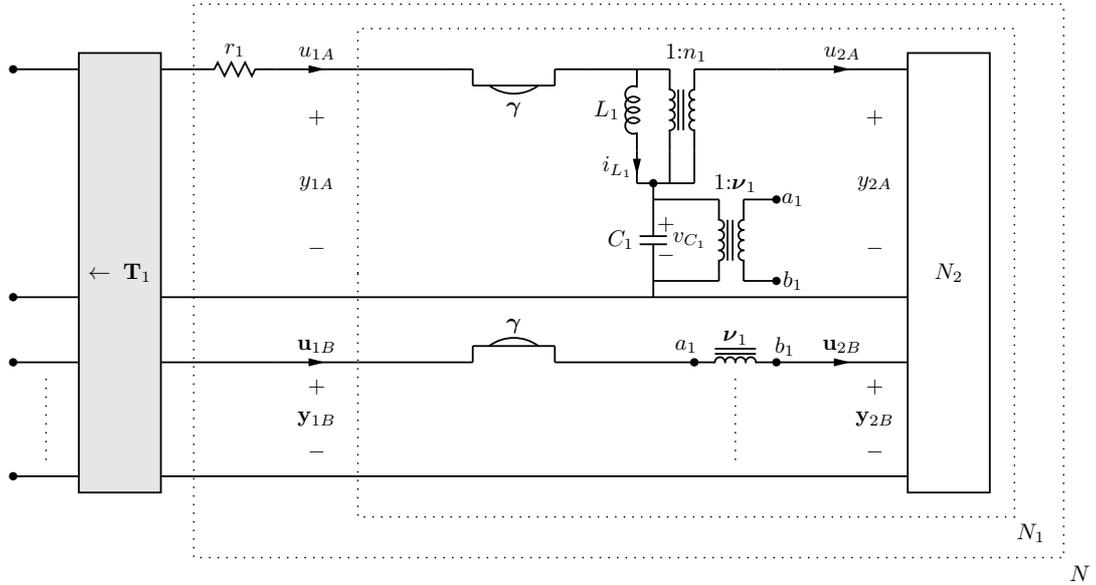} 
\par\end{centering}

\caption{\label{fig:multiport-Brune-stage-non-reciprocal}Multiport Brune stage
circuit in the case of a non-reciprocal response. We observe the appearance
of a multiport gyrator coupling the first port to the remaining ports
with a gyration vector $\boldsymbol{\gamma}$ as shown in \citep{Anderson-Moylan-1975}.}
\end{figure}

\section{Non-reciprocal Brune Stage}

\begin{figure}[H]
\begin{centering}
\includegraphics[scale=0.9]{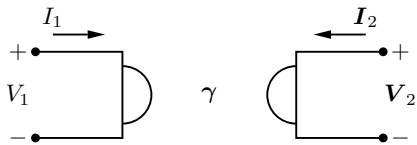} 
\par\end{centering}

\caption{\label{fig:Multiport-gyrator}The multiport gyrator appearing in the
non-reciprocal Brune stage in Fig. \eqref{fig:multiport-Brune-stage-non-reciprocal}.
The gyration ratios are given by the vector $\boldsymbol{\gamma}$.}
\end{figure}

The multiport Brune's Algorithm described in Section \eqref{sec:Multiport-Brune's-method}
produces reciprocal stages for a reciprocal impedance response $\mathbf{Z}\left(s\right)$.
If the response is non-reciprocal,\textcolor{cyan}{{} }\textcolor{black}{i.e.
if the blackbox in Fig. \eqref{fig:Multiport-Blackbox-Impedance.}
contains circulators \cite{Newcomb} for example,}\textcolor{cyan}{{}
}the multiport Brune circuit extracted at each stage is slightly modified
as shown in Fig. \eqref{fig:multiport-Brune-stage-non-reciprocal},
\citep{Anderson-Moylan-1975}. In Fig. \eqref{fig:multiport-Brune-stage-non-reciprocal}
we see that a multiport gyrator with a gyration vector $\boldsymbol{\gamma}$
is extracted right after the resistance $r_{1}$. The circuit symbol
for this multiport gyrator is shown in Fig. \eqref{fig:Multiport-gyrator}.
It has a single port on the left and $\left(N-1\right)$ ports on
the right with the following constitutive relations, \citep{Anderson-Moylan-1975}:

\begin{equation}
\left(\begin{array}{c}
V_{1}\\
\mathbf{V}_{2}
\end{array}\right)=\left(\begin{array}{cc}
0 & -\boldsymbol{\gamma}^{T}\\
\boldsymbol{\gamma} & 0
\end{array}\right)\left(\begin{array}{c}
I_{1}\\
\mathbf{I}_{2}
\end{array}\right)
\end{equation}
where $V_{1}$, $I_{1}$ are the voltage and the current of the left
port and $\mathbf{V}_{2}$, $\mathbf{I}_{2}$ are vectors of length
$\left(N-1\right)$ holding the voltages and currents of the ports
on the right, respectively.

The network $N_{1}$ has the same time-evolution description as in
Eq. \eqref{eq:multiport-Brune-state-space-eqs-N1-time-evolution}.
However the input-output relation is slightly different with the appearance
of the gyration vector $\boldsymbol{\gamma}$ in the $\mathbf{D}_{1}$
matrix:

\begin{equation}
\mathbf{D}_{1}=\left(\begin{array}{cc}
\mathbf{D}_{2AA}/n_{1}^{2} & -\boldsymbol{\gamma}^{T}+\mathbf{D}_{2AB}/n_{1}\\
\boldsymbol{\gamma+}\mathbf{D}_{2BA}/n_{1} & \mathbf{D}_{2BB}
\end{array}\right)
\end{equation}
That is $\boldsymbol{\gamma}$ is extracted by taking the anti-symmetric
part of $\mathbf{D}_{1}$.

We note that the quantization of the multiport Brune circuit with
gyrators - in the most general non-reciprocal case - is not solved
by the formalism reported here, and remains an open problem.

\section{Conclusion}

\textcolor{black}{As the scaling up of superconducting quantum processors
continue, it is clear that one will need to deal with microwave circuits
of increasing complexity. In fact, a route to scaleup is clearly taking
shape: it is increasingly accepted that a clear step towards realizing
a fault-tolerant quantum computing is by using the surface code \citep{Kitaev-toric-code}.
The surface code is a quantum error-correcting code with a relatively
high threshold (estimated threshold error rate around $1\%$) and
with realistic requirements for, e.g., local couplings beween qubits.
The implementation of the surface code will require maintaining the
excellent fidelity of qubit operations, made possible by high Q-factor
resonators and other components, in systems of increasing scale and
complexity \cite{DiVincenzo-skew-lattice}. There will clearly be
a need for highly accurate characterization of these multiqubit systems,
so that their optimization can proceed according to a rational plan.
Already in previous work, we have seen first attempts at such design
efforts, with the focus on the problems with spurious modes that inevitably
appear in complex microwave structures. In \citep{T1-Houck} the significant
contribution of off-resonant modes in the spontaneous emission rates
of qubits was already demonstrated. This has led to a method for controlling
the spontaneous emission rates of qubits by incorporating microwave
filters in the superconducting quantum processors. There has been
very recent research activity in this direction \citep{Bronn-Purcell-Filter,Eyob-Purcell-Filter}:
a new push is being made to design ``Purcell filters'' for the suppression
of spontaneous emission of the qubits while keeping the qubit-resonator
coupling at a strength required for measurement. Indeed, having multiple
modes is not always a nuisance. There are indications that one can
engineer the multi-mode structure of microwave circuits to tailor
qubit interactions via the help of filters for the realization of
high-fidelity two-qubits gates \citep{high-constrast}. }

\textcolor{black}{We are hopeful that the multiport method that we
have provided here will be useful in the further efforts to achieve
a more accurate analysis and design of these filter structures mentioned
above, and of all other aspects of the structures needed to achieve
surface-code action. Our work provides a complete algorithm for going
from a proposed electric structure to its Hamiltonian description.
It should be pointed out, however, that the present work leaves incomplete
the one further step of this design process, in which the Schr{ö}dinger
equation for this Hamiltonian is solved to obtain energy eigenvalues
and eigenfunctions that are vital in, e.g, assessing the action of
gating pulses, or of predicting qubit decoherence rates. Since our
analysis endeavors to be exact, we obtain Hamiltonians with many degrees
of freedom, with potential functions with a large hierarchy of stiffnesses
\cite{Brito}. Such a high dimensional problem can be almost impossible
to solve by straightforward discretization techniques. However, the
multi-scale nature of the problem can be used to systematically identify
``fast\textquotedbl{} and ``slow\textquotedbl{} degrees of freedom
in the problem \cite{Brito}; the principles of the Born-Oppenheimer
approximation can be applied to have a very efficient approximate
treatment of the fast variables, so that the dynamics of many fewer
slow variables, which can typically be directly associated with the
qubits or the principal resonant modes, can be determined accurately,
leading to a physically appealing and computationally accurate description
of the quantum physics of the system. We look forward to adapting
the present mathematical analyses so that they are fully integrated
into a full attack on the problem of scaling quantum hardware.}

\section{Appendices}

\subsection{Review of Lumped Element Circuit Quantization Methods}

In this \textcolor{black}{appendix} we will review two formalisms
\citep{BKD,Burkard} developed for the quantization of the lumped
element superconducting circuits. We won't attempt to do a full review
of each formalism. We will rather content ourselves with describing
how to combine parts of each formalism for the purpose of quantizing
Brune circuits. In the following we will refer to \citep{BKD} as
``BKD'' and to \citep{Burkard} as ``Burkard''; ``KCL'' stands
for Kirchhoff's current law and ``KVL'' stands for Kirchhoff's voltage
law.

Both methods derive classical equations of motions for the lumped
element circuits and identify canonical variables. They both start
with a graph theoretical analysis by choosing a spanning tree in the
circuit to write KCL and KVL relations involving current and voltage
variables in the circuit. We will follow the graph analysis done in
Burkard. This will allow us to write an equation of motion and to
identify the canonical degrees of freedom of the circuit. We will
need to make slight modifications to the theory to be able to treat
Brune circuits. Once we have the equation of motion, we will interpret
it as an equation of motion of BKD to do a dissipation analysis and
compute relaxation rates.

\subsubsection{\label{sub:Burkard's-method}Derivation of the Equation of Motion
by the Burkard's Method}

The first step in Burkard is to find a spanning tree containing all
the Josephson junctions, voltage sources and impedances in the circuit.
One can also put linear inductors in the tree. However there should
be no capacitors in the tree so that all capacitors are in the chord
branches. Linear inductors are also allowed to be in the chord branches.
Burkard assumes that there is no loop containing only Josephson junctions,
voltage sources and impedances which is physically justified since
in reality each loop would have a finite self-inductance.

With the choice of such a spanning tree we can partition the current
and voltage vectors as follows

\begin{equation}
\mathbf{I}_{\mathrm{tr}}=\left(\mathbf{I}_{J},\,\mathbf{I}_{L},\,\mathbf{I}_{V},\,\mathbf{I}_{Z}\right)
\end{equation}

\begin{equation}
\mathbf{V}_{\mathrm{tr}}=\left(\mathbf{V}_{J},\,\mathbf{V}_{L},\,\mathbf{V}_{V},\,\mathbf{V}_{Z}\right)
\end{equation}

\begin{equation}
\mathbf{I}_{\mathrm{ch}}=\left(\mathbf{I}_{C_{J}},\,\mathbf{I}_{C},\,\mathbf{I}_{K}\right)
\end{equation}

\begin{equation}
\mathbf{V}_{\mathrm{ch}}=\left(\mathbf{V}_{C_{J}},\,\mathbf{V}_{C},\,\mathbf{V}_{K}\right)
\end{equation}
where $\mathbf{I}_{\mathrm{tr}}$ and $\mathbf{I}_{\mathrm{ch}}$
are the vectors holding the currents through the tree and chord branches
and $\mathbf{V}_{\mathrm{tr}}$ and $\mathbf{V}_{\mathrm{ch}}$ are
vectors holding the voltages across the tree and chord branches, respectively.
Labels $J$, $L$, $K$, $V$, $Z$, $C_{J}$, $C$ correspond to
Josephson junctions, tree inductors, chord inductors, voltage sources,
impedances, Josephson junction capacitances and ordinary capacitances,
respectively. In terms of those loop variables Kirchhoff's laws can
be written as

\begin{equation}
\mathbf{F}\mathbf{I}_{\mathrm{ch}}=-\mathbf{I}_{\mathrm{tr}}\label{eq:KCL-appendix}
\end{equation}

\begin{equation}
\mathbf{F}^{T}\mathbf{V}_{\mathrm{tr}}=\mathbf{V}_{\mathrm{ch}}-\mathbf{\dot{\Phi}}_{x}\label{eq:KVL-appendix}
\end{equation}
where we have introduced the loop matrix $\mathbf{F}$ defined in
Eq. (3) of Burkard. $\mathbf{\Phi}_{x}=\left(\Phi_{1},\ldots,\Phi_{F}\right)$
is the flux bias vector holding the external fluxes threading $F$
fundamental loops of the circuit, each fundamental loop being defined
by a chord branch. The loop matrix $\mathbf{F}$ can be partitioned
as

\begin{equation}
\mathbf{F}=\left(\begin{array}{ccc}
\mathbf{I} & \mathbf{F}_{JC} & \mathbf{F}_{JK}\\
\boldsymbol{0} & \mathbf{F}_{LC} & \mathbf{F}_{LK}\\
\boldsymbol{0} & \mathbf{F}_{VC} & \mathbf{F}_{VK}\\
\boldsymbol{0} & \mathbf{F}_{ZC} & \mathbf{F}_{ZK}
\end{array}\right)
\end{equation}
The first column is due to Josephson junction capacitances $C_{J}$'s
shunting only the Josephson junctions.

Burkard further assumes that the voltage sources and the impedances
are not inductively shunted in the sense that

\begin{equation}
\mathbf{F}_{VK}=\mathbf{F}_{ZK}=0
\end{equation}

Then by writing KCL for Josephson junctions and tree inductors and
KVL for chord capacitors Burkard derives the first-order equation
of motion (Eq. (19) in Burkard - we fixed a sign typo)

\begin{equation}
\mathcal{C}\mathbf{\dot{\Phi}}=\mathbf{Q}-\mathbf{C}_{V}\mathbf{V}_{V}+\mathcal{F}_{C}\mathbf{C}_{Z}*\mathbf{V}_{C}\label{eq:Burkard-eq-of-motion-(19)}
\end{equation}
where the vector $\mathbf{\Phi}$ holds the flux degrees of freedom
corresponding to the fluxes across the Josephson junctions (J) and
tree inductor branches (L)

\begin{equation}
\mathbf{\Phi}=\left(\begin{array}{c}
\mathbf{\Phi}_{J}\\
\mathbf{\Phi}_{L}
\end{array}\right)
\end{equation}
with $\mathbf{\Phi}_{J}=\Phi_{0}\boldsymbol{\varphi}_{J}/2\pi$, $\boldsymbol{\varphi}_{J}$
being the vector of phases across the Josephson junctions. The canonical
charge variables are given by the vector $\mathbf{Q}$

\begin{equation}
\mathbf{Q}=-\left(\begin{array}{c}
\mathbf{Q}_{J}\\
\mathbf{Q}_{L}
\end{array}\right)-\mathcal{F}_{K}\mathbf{Q}_{K}
\end{equation}
with

\begin{equation}
\mathcal{F}_{K}=\left(\begin{array}{c}
\mathbf{F}_{JK}\\
\mathbf{F}_{LK}
\end{array}\right)
\end{equation}

\[
\mathcal{F}_{C}=\left(\begin{array}{c}
\mathbf{F}_{JC}\\
\mathbf{F}_{LC}
\end{array}\right)
\]
and the capacitance matrices in Eq. \eqref{eq:Burkard-eq-of-motion-(19)}
are given by

\begin{equation}
\mathcal{C}=\left(\begin{array}{cc}
\mathbf{C}_{J} & \boldsymbol{0}\\
\boldsymbol{0} & \boldsymbol{0}
\end{array}\right)+\mathcal{F}_{C}\mathbf{C}\mathcal{F}_{C}^{T}\label{eq:curly-C-matrix-Burkard-definition}
\end{equation}

\begin{equation}
\mathbf{C}_{V}=\mathcal{F}_{C}\mathbf{C}\mathbf{F}_{VC}^{T}\label{eq:Burkard-bold-CV}
\end{equation}

\begin{equation}
\mathbf{C}_{Z}\left(\omega\right)=\left(i\omega\right)\mathbf{C}\mathbf{F}_{ZC}^{T}\mathbf{Z}\left(\omega\right)\mathbf{F}_{ZC}\mathbf{C}
\end{equation}
where $\mathbf{C}$ is the diagonal matrix of ordinary capacitances
in the circuit except Josephson junctions' capacitances such that

\begin{equation}
\mathbf{Q}_{C}=\mathbf{C}\mathbf{V}_{C}
\end{equation}

The last term $\mathcal{F}_{C}\mathbf{C}_{Z}*\mathbf{V}_{C}$ in Eq.
\eqref{eq:Burkard-eq-of-motion-(19)} is the dissipative term. This
term can also be written in frequency domain as

\begin{equation}
\mathcal{F}_{C}\mathbf{C}_{Z}\left(\omega\right)\mathbf{V}_{C}=-\mathcal{F}_{C}\mathbf{C}\mathbf{F}_{ZC}^{T}\mathbf{Z}\left(\omega\right)\mathbf{I}_{Z}\label{eq:Burkard-dissipative-term}
\end{equation}
$\mathbf{I}_{Z}$ can be written as function of flux degrees of freedom
$\mathbf{\Phi}$ and voltage sources $\mathbf{V}_{V}$ as

\begin{eqnarray}
\mathbf{I}_{Z} & = & -\left(i\omega\right)\left[\mathbf{I}+\left(i\omega\right)\mathbf{F}_{ZC}\mathbf{C}\mathbf{F}_{ZC}^{T}\mathbf{Z}\left(\omega\right)\right]^{-1}\mathbf{F}_{ZC}\mathbf{C}\mathcal{F}_{C}^{T}\mathbf{\dot{\Phi}}\nonumber \\
 &  & -\left(i\omega\right)\left[\mathbf{I}+\left(i\omega\right)\mathbf{F}_{ZC}\mathbf{C}\mathbf{F}_{ZC}^{T}\mathbf{Z}\left(\omega\right)\right]^{-1}\mathbf{F}_{ZC}\mathbf{C}\mathbf{F}_{VC}^{T}\mathbf{V}_{V}\label{eq:IZ-solution}
\end{eqnarray}

If we define

\begin{eqnarray}
\mathbf{\bar{m}} & = & \mathcal{F}_{C}\mathbf{C}\mathbf{F}_{ZC}^{T}\label{eq:Burkard-m-vector-definition}\\
\mathbf{\bar{m}}_{V} & = & \mathbf{F}_{VC}\mathbf{C}\mathbf{F}_{ZC}^{T}\label{eq:Burkard-mv-vector-definition}
\end{eqnarray}
and as in Eq. (28) of Burkard (with a sign change)

\begin{equation}
\mathbf{\bar{C}}_{Z}\left(\omega\right)=-\left(i\omega\right)\mathbf{Z}\left(\omega\right)\left[\mathbf{I}+\left(i\omega\right)\mathbf{F}_{ZC}\mathbf{C}\mathbf{F}_{ZC}^{T}\mathbf{Z}\left(\omega\right)\right]^{-1}\label{eq:Burkard-CZ-definition}
\end{equation}
We can rewrite Eq. \eqref{eq:IZ-solution} in terms of the newly defined
quantities in Eqs. \eqref{eq:Burkard-m-vector-definition}, \eqref{eq:Burkard-mv-vector-definition}
and \eqref{eq:Burkard-CZ-definition} as

\begin{equation}
\mathbf{I}_{Z}=\mathbf{Z}^{-1}\left(\omega\right)\mathbf{\bar{C}}_{Z}\left(\omega\right)\mathbf{\bar{m}}^{T}\mathbf{\dot{\Phi}}+\mathbf{Z}^{-1}\left(\omega\right)\mathbf{\bar{C}}_{Z}\left(\omega\right)\mathbf{\bar{m}}_{V}^{T}\mathbf{V}_{V}\label{eq:IZ-solution-shorter}
\end{equation}

Replacing the solution for $\mathbf{I}_{Z}$ in Eq. \eqref{eq:IZ-solution-shorter}
in Eq. \eqref{eq:Burkard-dissipative-term} we get

\begin{equation}
\mathcal{F}_{C}\mathbf{C}_{Z}\left(\omega\right)\mathbf{V}_{C}=-\mathbf{\bar{m}}\mathbf{\bar{C}}_{Z}\left(\omega\right)\mathbf{\bar{m}}^{T}\mathbf{\dot{\Phi}}-\mathbf{\bar{m}}\mathbf{\bar{C}}_{Z}\left(\omega\right)\mathbf{\bar{m}}_{V}^{T}\mathbf{V}_{V}\label{eq:Burkard-dissipative-term-solution}
\end{equation}

Defining also

\begin{equation}
\mathcal{C}_{Z}\left(\omega\right)=\mathbf{\bar{m}}\mathbf{\bar{C}}_{Z}\left(\omega\right)\mathbf{\bar{m}}^{T}\label{eq:Burkard-CZ}
\end{equation}
and

\begin{equation}
\mathcal{C}_{V}\left(\omega\right)=\mathbf{\bar{m}}\mathbf{\bar{C}}_{Z}\left(\omega\right)\mathbf{\bar{m}}_{V}^{T}\label{eq:Burkard-CV}
\end{equation}

We can write the dissipative term in Eq. \eqref{eq:Burkard-dissipative-term-solution}
in terms of the quantities in Eqs. \eqref{eq:Burkard-CZ} and \eqref{eq:Burkard-CV}
as

\begin{equation}
\mathcal{F}_{C}\mathbf{C}_{Z}\left(\omega\right)\mathbf{V}_{C}=-\mathcal{C}_{Z}\left(\omega\right)\mathbf{\dot{\Phi}}-\mathcal{C}_{V}\left(\omega\right)\mathbf{V}_{V}\label{eq:Burkard-dissipative-term-solution-shorter}
\end{equation}
Note that we have extracted an additional non-dissipative term $\mathcal{C}_{V}\left(\omega\right)\mathbf{V}_{V}$
in the equation above.

Plugging the dissipative term in Eq. \eqref{eq:Burkard-dissipative-term-solution-shorter}
back in the equation of motion in Eq. \eqref{eq:Burkard-eq-of-motion-(19)}
we get in time domain

\begin{equation}
\left(\mathcal{C}+\mathcal{C}_{Z}\left(t\right)\right)*\mathbf{\dot{\Phi}}=\mathbf{Q}-\left(\mathbf{C}_{V}+\mathcal{C}_{V}\left(t\right)\right)*\mathbf{V}_{V}\left(t\right)\label{eq:Burkard-equation-of-motion-general-Vv}
\end{equation}

Note that the Eq. \eqref{eq:Burkard-equation-of-motion-general-Vv}
is more general than Burkard's equation of motion in Eq. (25) of \citep{Burkard}
since we allow a general voltage source term $\mathbf{V}_{V}\left(t\right)$
with possible frequency components both at DC and AC. The vector $\mathcal{C}_{V}\left(t\right)$
is due to AC components of $\mathbf{V}_{V}\left(t\right)$. In the
case of a DC voltage source term $\mathbf{V}_{V}$, $\mathcal{C}_{V}\left(\omega\right)\mathbf{V}_{V}\left(\omega\right)=0$
for $\omega\neq0$ and $\mathcal{C}_{V}\left(\omega\right)=0$ for
$\omega=0$ so we recover Burkard's equation of motion. Eq. \eqref{eq:Burkard-equation-of-motion-general-Vv}
will lead to the more general Hamiltonian in Eq. \eqref{eq:Burkard-Hamiltonian}
below.

Taking the time derivative of both sides in Eq. \eqref{eq:Burkard-equation-of-motion-general-Vv}
we obtain

\begin{equation}
\left(\mathcal{C}+\mathcal{C}_{Z}\right)*\mathbf{\ddot{\Phi}}=\mathbf{\dot{Q}}-\dot{\mathcal{C}}_{V}*\mathbf{V}_{V}\label{eq:Burkard-equation-of-motion-2nd-order}
\end{equation}

\textcolor{black}{Using the identity $\left(\mathcal{C}+\mathcal{C}_{Z}\right)*\mathbf{\ddot{\Phi}}=\left(\dot{\mathcal{C}}+\mathcal{\dot{C}}_{Z}\right)*\mathbf{\dot{\Phi}}$
it is interesting to note here that since the vector $\mathbf{\dot{\Phi}}$
is equal to the vector of voltages across the Josephson junction branches
and tree inductors, the factor $\left(\mathcal{\dot{C}}+\mathcal{\dot{C}}_{Z}\right)$
is an admittance matrix and $\left(\mathcal{\dot{C}}+\mathcal{\dot{C}}_{Z}\right)*\mathbf{\dot{\Phi}}=\mathbf{\dot{Q}}-\dot{\mathcal{C}}_{V}*\mathbf{V}_{V}$
is the KCL for the tree branches.}

Taking the dissipative term $\mathcal{C}_{Z}*\ddot{\mathbf{\Phi}}$
in Eq. \eqref{eq:Burkard-equation-of-motion-2nd-order} to the right-hand
side by noting the identity

\begin{equation}
\mathcal{C}_{Z}*\mathbf{\ddot{\Phi}}=\ddot{\mathcal{C}}_{Z}*\mathbf{\Phi}
\end{equation}
And using the Eq. (29) $\mathbf{\dot{Q}}=-\frac{\partial U}{\partial\mathbf{\Phi}}$
of Burkard with the potential $U$

\begin{equation}
U\left(\mathbf{\Phi}\right)=-\mathbf{L}_{J}^{-1}\mathbf{cos}\mathbf{\boldsymbol{\varphi}}+\frac{1}{2}\mathbf{\Phi}^{T}\mathbf{M}_{0}\mathbf{\Phi}+\mathbf{\Phi}^{T}\mathbf{N}\mathbf{\Phi}_{x}\label{eq:Burkard-potential}
\end{equation}
where

\begin{equation}
\mathbf{M}_{0}=\mathcal{G}\mathbf{L}_{t}^{-1}\mathcal{G}^{T}\label{eq:M0-definition}
\end{equation}

\begin{equation}
\mathbf{N}=\mathcal{G}\mathbf{L}_{t}^{-1}\left(\begin{array}{cc}
0 & \mathbf{I}_{K}\end{array}\right)^{T}
\end{equation}
with $\mathbf{L}_{t}^{-1}$ being the inverse inductance matrix such
that

\begin{equation}
\left(\begin{array}{c}
\mathbf{I}_{L}\\
\mathbf{I}_{K}
\end{array}\right)=\mathbf{L}_{t}^{-1}\left(\begin{array}{c}
\mathbf{\Phi}_{L}\\
\mathbf{\Phi}_{K}
\end{array}\right)\label{eq:Lt-inv-definition}
\end{equation}
and

\begin{equation}
\mathcal{G}=\left(\begin{array}{cc}
\boldsymbol{0} & -\mathbf{F}_{JK}\\
\mathbf{I}_{L} & -\mathbf{F}_{LK}
\end{array}\right)\label{eq:curly-G-definition}
\end{equation}
we get

\begin{equation}
\mathcal{C}*\ddot{\mathbf{\Phi}}=-\mathbf{I}_{c}\mathbf{sin}\boldsymbol{\varphi}-\mathbf{M}_{0}\mathbf{\Phi}-\mathcal{\ddot{C}}_{Z}*\mathbf{\Phi}-\dot{\mathcal{C}}_{V}*\mathbf{V}_{V}-\mathbf{N}\mathbf{\Phi}_{x}\label{eq:Burkard-equation-of-motion-BKD-type}
\end{equation}
where $\mathbf{I}_{c}=\frac{\Phi_{0}}{2\pi}\mathbf{L}_{J}^{-1}$ is
the diagonal matrix of critical currents.

\textcolor{black}{We would like mention here that the dissipation
analysis that follows is to be applied only to unexcited systems,
i.e. to systems where the voltage sources $\mathbf{V}_{V}\left(t\right)$
have only a DC component.}

\subsubsection{\label{sub:Treatment-of-resistors-using-BKD-appendix}Treatment of
Resistors and Relaxation Rate Calculations Using the BKD Method}

Before writing a Hamiltonian for the equation of motion in Eq. \eqref{eq:Burkard-equation-of-motion-BKD-type}
we have one more step to do. We again need to stretch Burkard theory
slightly to include the resistors $\left\{ R_{1},\ldots,R_{N}\right\} $
shunting the last stage of the multiport Brune circuit in our analysis.
The trick we used for that purpose as described in the main text for
the one-port Brune circuit was to replace those resistors by capacitors
and to make the substitution $C_{j}\leftarrow\frac{1}{i\omega R_{j}}$
for $1\leq j\leq N$ after we obtained the equation of motion. To
do a similar analysis for the multiport Brune circuit, here we extend
Burkard's method to include chord resistors.

We start by partitioning the matrix $\mathbf{C}$

\begin{equation}
\mathbf{C}=\left(\begin{array}{cc}
\mathbf{C}_{0} & \boldsymbol{0}\\
\boldsymbol{0} & \mathbf{C}_{R}
\end{array}\right)\label{eq:multiport-C-partitioning-appendix}
\end{equation}
where $\mathbf{C}_{0}$ and $\mathbf{C}_{R}$ are diagonal matrices
of size $M\times M$ and $N\times N$, respectively. Diagonal entries
of $\mathbf{C}_{0}$ hold the values of ordinary capacitors in the
circuit whereas $\mathbf{C}_{R}$ is an auxiliary matrix which will
be substituted later as

\begin{equation}
\mathbf{C}_{R}\leftarrow\left(i\omega\right)^{-1}\mathbf{R}^{-1}\label{eq:shunt-resistance-substitution}
\end{equation}
where $\mathbf{R}$ is the $N\times N$ diagonal matrix holding the
values of the resistances shunting the last stage

\begin{equation}
\mathbf{R}=\left(\begin{array}{ccc}
R_{1} &  & \boldsymbol{0}\\
 & \ddots\\
\boldsymbol{0} &  & R_{N}
\end{array}\right)
\end{equation}

We also partition $\mathcal{F}_{C}$ respecting the partitioning of
$\mathbf{C}$ in Eq. \eqref{eq:multiport-C-partitioning-appendix}
as follows

\begin{equation}
\mathcal{F}_{C}=\left(\begin{array}{cc}
\mathcal{F}_{C_{0}} & \mathcal{F}_{C_{R}}\end{array}\right)\label{eq:multiport-FC-partitioning-appendix}
\end{equation}
where $\mathcal{F}_{C_{0}}$, $\mathcal{F}_{C_{R}}$ are submatrices
with $M$ and $N$ columns, respectively.

Now we can decompose the capacitance matrix $\mathcal{C}$ in Eq.
\eqref{eq:curly-C-matrix-Burkard-definition} using Eqs. \eqref{eq:multiport-C-partitioning-appendix}
and \eqref{eq:multiport-FC-partitioning-appendix} as

\begin{eqnarray}
\mathcal{C} & = & \left(\begin{array}{cc}
\mathbf{C}_{J} & \boldsymbol{0}\\
\boldsymbol{0} & \boldsymbol{0}
\end{array}\right)+\mathcal{F}_{C}\mathbf{C}\mathcal{F}_{C}^{T}\\
 & = & \left(\begin{array}{cc}
\mathbf{C}_{J} & \boldsymbol{0}\\
\boldsymbol{0} & \boldsymbol{0}
\end{array}\right)+\mathcal{F}_{C_{0}}\mathbf{C}_{0}\mathcal{F}_{C_{0}}^{T}+\mathcal{F}_{C_{R}}\mathbf{C}_{R}\mathcal{F}_{C_{R}}^{T}\\
 & = & \mathcal{C}_{0}+\mathcal{C}_{R}\label{eq:curly-C-decomposition}
\end{eqnarray}
where we defined

\begin{eqnarray}
\mathcal{C}_{0} & = & \left(\begin{array}{cc}
\mathbf{C}_{J} & \boldsymbol{0}\\
\boldsymbol{0} & \boldsymbol{0}
\end{array}\right)+\mathcal{F}_{C_{0}}\mathbf{C}_{0}\mathcal{F}_{C_{0}}^{T}\label{eq:curly-C0-definition}\\
\mathcal{C}_{R} & = & \mathcal{F}_{C_{R}}\mathbf{C}_{R}\mathcal{F}_{C_{R}}^{T}\label{eq:curly-CR-definition}
\end{eqnarray}

Making the substitution in Eq. \eqref{eq:shunt-resistance-substitution}
we get the following dissipative term on left-hand side of the equation
of motion in Eq. \eqref{eq:Burkard-equation-of-motion-BKD-type}

\begin{eqnarray}
\mathcal{C}_{R}*\mathbf{\ddot{\mathbf{\Phi}}} & = & \dot{\mathcal{C}}_{R}*\mathbf{\dot{\mathbf{\Phi}}}\\
 & = & \left(\mathcal{F}_{C_{R}}\mathbf{R}^{-1}\mathcal{F}_{C_{R}}^{T}\right)*\mathbf{\dot{\mathbf{\Phi}}}\\
 & = & \mathcal{R}^{-1}\mathbf{\dot{\mathbf{\Phi}}}\label{eq:dissipative-term-due-to-R}
\end{eqnarray}
where in the first line we used differentiation property of the convolution
operator, in the second line we used the Eqs. \eqref{eq:shunt-resistance-substitution}
and \eqref{eq:curly-CR-definition} and in the third line we defined

\begin{equation}
\mathcal{R}^{-1}=\mathcal{F}_{C_{R}}\mathbf{R}^{-1}\mathcal{F}_{C_{R}}^{T}\label{eq:curly-R-appendix}
\end{equation}
which is frequency independent, justifying dropping the convolution
operator in Eq. \eqref{eq:dissipative-term-due-to-R}. Now taking
the dissipative term in Eq. \eqref{eq:dissipative-term-due-to-R}
due to the last shunt resistors $\mathbf{R}$ to the right-hand side
of the equation of motion in Eq. \eqref{eq:Burkard-equation-of-motion-BKD-type}
and noting also the decomposition in Eq. \eqref{eq:curly-C-decomposition},
we re-write the equation of motion \eqref{eq:Burkard-equation-of-motion-BKD-type}

\begin{equation}
\mathcal{C}_{0}*\mathbf{\ddot{\mathbf{\Phi}}}=-\mathbf{I}_{c}\mathbf{sin}\boldsymbol{\varphi}-\mathcal{R}^{-1}\mathbf{\dot{\mathbf{\Phi}}}-\mathbf{M}_{0}\mathbf{\Phi}-\mathcal{\ddot{C}}_{Z}*\mathbf{\Phi}-\dot{\mathcal{C}}_{V}*\mathbf{V}_{V}-\mathbf{N}\mathbf{\Phi}_{x}\label{eq:Burkard-equation-of-motion-BKD-type-2}
\end{equation}

Following Burkard we can write the following Hamiltonian for the lossless
part of the dynamics described by the Eq. \eqref{eq:Burkard-equation-of-motion-BKD-type-2}

\begin{equation}
\mathcal{H}_{S}=\frac{1}{2}\left(\mathbf{Q}-\left(\mathbf{C}_{V}+\mathcal{C}_{V}\right)*\mathbf{V}_{V}\left(t\right)\right)^{T}\mathcal{C}_{0}^{-1}\left(\mathbf{Q}-\left(\mathbf{C}_{V}+\mathcal{C}_{V}\right)*\mathbf{V}_{V}\left(t\right)\right)+U\left(\mathbf{\Phi}\right)\label{eq:Burkard-Hamiltonian}
\end{equation}
where the potential function $U\left(\mathbf{\Phi}\right)$ is defined
in Eq. \eqref{eq:Burkard-potential}. A point to note here is that
although $\mathbf{C}_{V}$ defined in Eq. \eqref{eq:Burkard-bold-CV}
is frequency independent, after the substitution in Eq. \eqref{eq:shunt-resistance-substitution}
it will acquire frequency dependence which we will analyze further
down below in Section \eqref{sub:Analysis-of-voltage-source-couplings}.

Comparing again the equation motion in Eq. (61) of BKD with the Eq.
\eqref{eq:Burkard-equation-of-motion-BKD-type-2} above we identify

\begin{equation}
\mathbf{R}_{BKD}^{-1}\leftrightarrow\mathcal{R}^{-1}
\end{equation}
where $\mathbf{R}_{BKD}$ is now the resistivity matrix defined in
Eq. (25) of BKD. We note however that $\mathbf{R}_{BKD}^{-1}$ is
diagonal whereas $\mathcal{R}^{-1}$ is in general non-diagonal.

To compute the contribution of the shunt resistor $R_{j}$ alone to
the relaxation rate we set $R_{k}=\infty$ for $1\leq k\leq N$ and
$k\neq j$ to write the term $\mathcal{R}^{-1}\mathbf{\dot{\mathbf{\Phi}}}$
in the equation of motion in Eq. \eqref{eq:Burkard-equation-of-motion-BKD-type-2}
in the frequency domain as

\begin{equation}
\left.\left(i\omega\right)\mathcal{R}^{-1}\mathbf{\mathbf{\Phi}}\right|_{R_{k}\rightarrow\infty,k\neq j}=\left(i\omega\right)R_{j}^{-1}\mathcal{F}_{R_{j},C_{R}}\mathcal{F}_{R_{j},C_{R}}^{T}\mathbf{\Phi}\label{eq:FRjCR-definition-appendix}
\end{equation}
where $\mathcal{F}_{R_{j},C_{R}}$ is the $j^{th}$ column of the
matrix $\mathcal{F}_{C_{R}}$ corresponding to the resistor $R_{j}$
(or the capacitor $C_{R_{j}}$). Comparing now

\begin{equation}
\mathbf{M}_{d}\left(\omega\right)\leftrightarrow\left(i\omega\right)R_{j}^{-1}\mathcal{F}_{R_{j},C_{R}}\mathcal{F}_{R_{j},C_{R}}^{T}
\end{equation}
where $\mathbf{M}_{d}\left(\omega\right)$ is the dissipation matrix
defined in Eq. (72)-(75) of BKD we identify the coupling vector

\begin{equation}
\bar{\mathbf{m}}_{R_{j}}=\mathcal{F}_{R_{j},C_{R}}\label{eq:mRj}
\end{equation}
and the dissipation kernel is given by

\begin{equation}
K_{R_{j}}\left(\omega\right)=\frac{i\omega}{R_{j}}\label{eq:KRj}
\end{equation}

Using Eq. (93) of BKD we define the spectral density of the bath corresponding
to the resistor $R_{j}$ as (correcting the sign and dropping the
scale factor)

\begin{eqnarray}
J_{R_{j}}\left(\omega\right) & = & \mathrm{Im}\left[K_{R_{j}}\left(\omega\right)\right]\label{eq:spectral-density-of-the-bath-shunt-resistors}\\
 & = & \frac{\omega}{R_{j}}\label{eq:JRj}
\end{eqnarray}
Hence the contribution of $R_{j}$ to the relaxation rate is given
by the formula in Eq. (124) of BKD as

\begin{equation}
\frac{1}{T_{1,R_{j}}}=\frac{4}{\hbar}\left|\left\langle 0\left|\mathbf{\bar{m}}_{R_{j}}\cdot\mathbf{\Phi}\right|1\right\rangle \right|^{2}J_{R_{j}}\left(\omega_{01}\right)\coth\left(\frac{\hbar\omega_{01}}{2k_{B}T}\right)\label{eq:T1-rate-for-Rj}
\end{equation}
where $\omega_{01}$ is the qubit frequency. Here we used flux variable
$\mathbf{\Phi}$ and $\mathbf{\bar{m}}_{R_{j}}$ since we dropped
the scale factor in Eq. \eqref{eq:spectral-density-of-the-bath-shunt-resistors}
(BKD uses phase variable $\mathbf{\boldsymbol{\varphi}}$ and the
normalized vector $\mathbf{m}$).

We need to also consider the effect of last shunt resistors on the
matrices $\mathcal{\ddot{C}}_{Z}$ and $\dot{\mathcal{C}}_{V}$ appearing
in the equation of motion in Eq. \eqref{eq:Burkard-equation-of-motion-BKD-type-2}
above. For this we first make the following partitioning for the matrix
$\mathbf{F}_{ZC}$

\begin{equation}
\mathbf{F}_{ZC}=\left(\begin{array}{cc}
\mathbf{F}_{ZC_{0}} & \mathbf{F}_{ZC_{R}}\end{array}\right)\label{eq:FZC-partitioning}
\end{equation}
where the submatrices $\mathbf{F}_{ZC_{0}}$ and $\mathbf{F}_{ZC_{R}}$
have $M$ and $N$ columns, respectively.

Then using also the partitioning in Eq. \eqref{eq:multiport-C-partitioning-appendix}
and the substitution Eq. \eqref{eq:shunt-resistance-substitution}
we can re-write the matrix $\mathbf{\bar{C}}_{Z}\left(\omega\right)$
defined in Eq. \eqref{eq:Burkard-CZ-definition} as $\mathbf{\bar{C}}_{Z,R}\left(\omega\right)$

\begin{eqnarray}
\mathbf{\bar{C}}_{Z,R}\left(\omega\right) & = & \left.\mathbf{\bar{C}}_{Z}\left(\omega\right)\right\rfloor _{\mathbf{C}_{R}\leftarrow\left(i\omega\right)^{-1}\mathbf{R}^{-1}}\label{eq:CZR-substitution}\\
 & = & -\left(i\omega\right)\mathbf{Z}_{R}\left(\omega\right)\left[\mathbf{I}+\left(i\omega\right)\mathbf{F}_{ZC_{0}}\mathbf{C}_{0}\mathbf{F}_{ZC_{0}}^{T}\mathbf{Z}_{R}\left(\omega\right)\right]^{-1}
\end{eqnarray}
where we have defined

\begin{equation}
\mathbf{Z}_{R}\left(\omega\right)=\mathbf{Z}\left[\mathbf{I}+\mathbf{F}_{ZC_{R}}\mathbf{R}^{-1}\mathbf{F}_{ZC_{R}}^{T}\mathbf{Z}\right]^{-1}\label{eq:ZR-matrix}
\end{equation}
with

\begin{equation}
\mathbf{Z}=\left(\begin{array}{ccc}
r_{1} &  & \boldsymbol{0}\\
 & \ddots\\
\boldsymbol{0} &  & r_{M}
\end{array}\right)\label{eq:Z-matrix-of-in-series-resistances-appendix}
\end{equation}
being the diagonal matrix of in-series resistances $r_{1},\ldots,r_{M}$
in the circuit.

Since we will consider resistors one at a time we will take the limit
$\mathbf{R}\rightarrow\infty$ (which corresponds to open circuiting
the last stage in the Brune circuit) in Eq. \eqref{eq:ZR-matrix}
above to define

\begin{eqnarray}
\mathbf{\bar{C}}_{Z,r}\left(\omega\right) & = & \left.\mathbf{\bar{C}}_{Z,R}\left(\omega\right)\right\rfloor _{\mathbf{R}\rightarrow\infty}\label{eq:CZr-definition}\\
 & = & -\left(i\omega\right)\mathbf{Z}\left[\mathbf{I}+\left(i\omega\right)\mathbf{F}_{ZC_{0}}\mathbf{C}_{0}\mathbf{F}_{ZC_{0}}^{T}\mathbf{Z}\right]^{-1}
\end{eqnarray}
We will see later below that $\mathbf{\bar{C}}_{Z,r}\left(\omega\right)$
is proportional to the dissipation kernel due to in-series resistors.

We need to also update the coupling matrix $\bar{\mathbf{m}}$ defined
in Eqs. \eqref{eq:Burkard-m-vector-definition} above to account for
the effect of shunt resistors on the terms due to in-series resistors.
First we again make a partitioning

\begin{equation}
\mathcal{F}_{C}=\left(\begin{array}{cc}
\mathcal{F}_{C_{0}} & \mathcal{F}_{C_{R}}\end{array}\right)\label{eq:FC-partitioning}
\end{equation}
Then using this partitioning and Eqs. \eqref{eq:Burkard-m-vector-definition},
\eqref{eq:multiport-C-partitioning-appendix}, \eqref{eq:shunt-resistance-substitution}
and \eqref{eq:FZC-partitioning} we can write the decomposition

\begin{equation}
\bar{\mathbf{m}}=\bar{\mathbf{m}}_{0}+\bar{\mathbf{m}}_{R}\left(\omega\right)\label{eq:m-decomposition}
\end{equation}
where we defined

\begin{equation}
\bar{\mathbf{m}}_{0}=\mathcal{F}_{C_{0}}\mathbf{C}_{0}\mathbf{F}_{ZC_{0}}^{T}\label{eq:m0-vector}
\end{equation}

\begin{equation}
\bar{\mathbf{m}}_{R}\left(\omega\right)=\left(i\omega\right)^{-1}\mathcal{F}_{C_{R}}\mathbf{R}^{-1}\mathbf{F}_{ZC_{R}}^{T}\label{eq:mR-vector}
\end{equation}

Here, however, we will only use the zeroth order term in Eq. \eqref{eq:m0-vector}
to define

\begin{equation}
\mathcal{C}_{Z,r}\left(\omega\right)=\mathbf{\bar{m}}_{0}\mathbf{\bar{C}}_{Z,r}\left(\omega\right)\mathbf{\bar{m}}_{0}^{T}\label{eq:Burkard-CZr}
\end{equation}
Note that the frequency dependent factors $\left(i\omega\right)^{-1}$
appearing in higher orders terms due to the coupling matrix $\bar{\mathbf{m}}_{R}\left(\omega\right)$
in Eq. \eqref{eq:mR-vector} above can be absorbed in the corresponding
$\mathbf{\bar{C}}_{Z}\left(\omega\right)$ matrices if one wants to
investigate higher order effects.

Comparing Eq. \eqref{eq:Burkard-equation-of-motion-BKD-type-2} with
the equation of motion of BKD (Eq. (61) in \citep{BKD}) we make the
following identification

\begin{eqnarray}
\mathbf{M}_{d} & \leftrightarrow & \mathcal{\ddot{C}}_{Z,r}\\
\mathbf{M}_{d}\left(\omega\right) & \leftrightarrow & -\omega^{2}\mathcal{C}_{Z,r}\left(\omega\right)\label{eq:BKD-Burkard-dissipation-matrix-mapping}
\end{eqnarray}
where on the left we have the dissipation matrix defined in BKD and
on the right we have the dissipation matrix appearing in Eq. \eqref{eq:Burkard-equation-of-motion-BKD-type-2}.
The capacitance matrix $\mathcal{C}_{0}$ in Eq. \eqref{eq:Burkard-equation-of-motion-BKD-type-2}
maps directly to the capacitance matrix $\mathbf{C}$ in BKD and the
stiffness matrices $\mathbf{M}_{0}$ appearing in Eq. \eqref{eq:Burkard-equation-of-motion-BKD-type-2}
and BKD map to each other.

To do a dissipation analysis we will treat the Eq. \eqref{eq:Burkard-equation-of-motion-BKD-type-2}
as an equation of motion of BKD and we will do a Caldeira-Leggett
analysis using the identification in Eq. \eqref{eq:BKD-Burkard-dissipation-matrix-mapping}.
By Eq. (64) in BKD we have

\begin{equation}
\mathbf{M}_{d}\left(\omega\right)=\bar{\mathbf{m}}\mathbf{\bar{L}}_{Z}^{-1}\left(\omega\right)\bar{\mathbf{m}}^{T}
\end{equation}
And by Eq. \eqref{eq:Burkard-CZr}

\begin{equation}
\mathcal{C}_{Z,r}\left(\omega\right)=\bar{\mathbf{m}}_{0}\bar{\mathbf{C}}_{Z,r}\left(\omega\right)\bar{\mathbf{m}}_{0}^{T}\label{eq:curly-CZr-repetition}
\end{equation}
Hence by using Eq. \eqref{eq:BKD-Burkard-dissipation-matrix-mapping}
we further identify

\begin{equation}
\bar{\mathbf{L}}_{Z}^{-1}\left(\omega\right)\leftrightarrow-\omega^{2}\bar{\mathbf{C}}_{Z,r}\left(\omega\right)\label{eq:BKD-LZ-equiv-CZ}
\end{equation}
Coupling matrices $\bar{\mathbf{m}}$ and $\bar{\mathbf{m}}_{0}$
map to each other directly

\begin{equation}
\bar{\mathbf{m}}\leftrightarrow\bar{\mathbf{m}}_{0}
\end{equation}

To compute the contribution of each resistor to the relaxation rate
$1/T_{1}$ we will treat each resistor seperately. In that case by
Eq. (73) in BKD we have

\begin{equation}
K\left(\omega\right)=\bar{\mathbf{L}}_{Z}^{-1}\left(\omega\right)
\end{equation}
where $K\left(\omega\right)$ is a scalar. Hence by Eq. \eqref{eq:BKD-LZ-equiv-CZ}
we get

\begin{equation}
K_{j}\left(\omega\right)=-\omega^{2}\bar{\mathbf{C}}_{Z,r_{j}}\left(\omega\right)\label{eq:Kernel-appendix}
\end{equation}
for the equation of motion in Eq. \eqref{eq:Burkard-equation-of-motion-BKD-type-2},
where we defined

\begin{eqnarray}
\bar{\mathbf{C}}_{Z,r_{j}}\left(\omega\right) & = & \left.\bar{\mathbf{C}}_{Z,r}\left(\omega\right)\right|_{r_{k}=0,\, for\, k\neq j}\label{eq:CZrj-appendix}\\
 & = & -i\omega r_{j}\left[1+i\omega r_{j}\mathbf{F}_{r_{j},C_{0}}\mathbf{C}_{0}\mathbf{F}_{r_{j},C_{0}}^{T}\right]^{-1}\label{eq:CZrj-with-FrjC0}
\end{eqnarray}
where $\mathbf{F}_{r_{j},C_{0}}$ is the $j^{th}$ row of the matrix
$\mathbf{F}_{ZC_{0}}$ defined in Eq. \eqref{eq:FZC-partitioning}.
We note that $\bar{\mathbf{C}}_{Z,r_{j}}\left(\omega\right)$ is a
scalar. That is we are treating only the in-series resistor $r_{j}$
by short circuiting the other in-series resistors setting $r_{k}=0$
for $k\neq j$. Note that shunt resistors $\mathbf{R}$ are already
open circuited by taking the limit of $\mathbf{R}\rightarrow\infty$
in the definition of $\bar{\mathbf{C}}_{Z,r}\left(\omega\right)$
in Eq. \eqref{eq:CZr-definition}. We also define the coupling vector
$\bar{\mathbf{m}}_{j}$ of the bath due to the resistor $r_{j}$ to
the circuit degrees of freedom by taking the $j^{th}$ column of the
coupling matrix $\bar{\mathbf{m}}_{0}$ in Eq. \eqref{eq:m0-vector}.

Using Eq. (93) of BKD we define the spectral density of the bath corresponding
to the resistor $r_{j}$ as (again correcting the sign and dropping
the scale factor)

\begin{equation}
J_{j}\left(\omega\right)=\mathrm{Im}\left[K_{j}\left(\omega\right)\right]\label{eq:spectral-density-of-the-bath}
\end{equation}

We can now write the contribution of the resistor $r_{j}$ to the
relaxation rate using the formula in Eq. (124) of BKD as

\begin{equation}
\frac{1}{T_{1,r_{j}}}=\frac{4}{\hbar}\left|\left\langle 0\left|\mathbf{\bar{m}}_{j}\cdot\mathbf{\Phi}\right|1\right\rangle \right|^{2}J_{j}\left(\omega_{01}\right)\coth\left(\frac{\hbar\omega_{01}}{2k_{B}T}\right)\label{eq:T1-rate-for-rj}
\end{equation}
where $\omega_{01}$ is the qubit frequency. Here we again used flux
variable $\mathbf{\Phi}$ and $\mathbf{\bar{m}}_{j}$ since we dropped
the scale factor in Eq. \eqref{eq:spectral-density-of-the-bath}.

\subsubsection{\label{sub:Analysis-of-voltage-source-couplings}Analysis of Voltage
Source Couplings}

Now we will update the $\mathbf{C}_{V}$ and the $\mathcal{C}_{V}$
matrices to account for the effect of the resistors $\mathbf{R}$
shunting the last Brune stage. Making first the partitioning

\begin{equation}
\mathbf{F}_{VC}=\left(\begin{array}{cc}
\mathbf{F}_{VC_{0}} & \mathbf{F}_{VC_{R}}\end{array}\right)\label{eq:FVC-partitioning}
\end{equation}
Then using the definition for $\mathbf{C}_{V}$ in Eq. \eqref{eq:Burkard-bold-CV}
together with Eqs. \eqref{eq:multiport-C-partitioning-appendix},
\eqref{eq:shunt-resistance-substitution}, \eqref{eq:multiport-FC-partitioning-appendix}
and \eqref{eq:FVC-partitioning} we can decompose $\mathbf{C}_{V}$
as

\begin{equation}
\mathbf{C}_{V}=\mathbf{C}_{V,0}+\mathbf{C}_{V,R}\left(\omega\right)
\end{equation}
where we have defined

\begin{equation}
\mathbf{C}_{V,0}=\mathcal{F}_{C_{0}}\mathbf{C}_{0}\mathbf{F}_{VC_{0}}^{T}\label{eq:CV0-definition}
\end{equation}
and

\begin{equation}
\mathbf{C}_{V,R}\left(\omega\right)=\left(i\omega\right)^{-1}\mathcal{F}_{C_{R}}\mathbf{R}^{-1}\mathbf{F}_{VC_{R}}^{T}\label{eq:CVR-definition}
\end{equation}
We note the frequency dependence of $\mathbf{C}_{V,R}$.

A similar analysis can be done also for the $\bar{\mathbf{m}}_{V}$
matrix defined in Eq. \eqref{eq:Burkard-mv-vector-definition}. Then
again using the partitioning in Eq. \eqref{eq:FVC-partitioning} and
Eqs. \eqref{eq:Burkard-mv-vector-definition}, \eqref{eq:multiport-C-partitioning-appendix},
\eqref{eq:shunt-resistance-substitution} and \eqref{eq:FZC-partitioning}
we can write the decomposition

\begin{equation}
\bar{\mathbf{m}}_{V}=\bar{\mathbf{m}}_{V,0}+\bar{\mathbf{m}}_{V,R}\left(\omega\right)\label{eq:mV-decomposition}
\end{equation}
where we defined

\begin{equation}
\bar{\mathbf{m}}_{V,0}=\mathbf{F}_{VC_{0}}\mathbf{C}_{0}\mathbf{F}_{ZC_{0}}^{T}\label{eq:mV0-vector}
\end{equation}

\begin{equation}
\bar{\mathbf{m}}_{V,R}\left(\omega\right)=\left(i\omega\right)^{-1}\mathbf{F}_{VC_{R}}\mathbf{R}^{-1}\mathbf{F}_{ZC_{R}}^{T}\label{eq:mVR-vector}
\end{equation}

Now using the definition for $\mathcal{C}_{V}$ in Eq. \eqref{eq:Burkard-CV}
together with the decompositions Eq. \eqref{eq:m-decomposition},
\eqref{eq:mV-decomposition} and the updated $\mathbf{\bar{C}}_{Z}\left(\omega\right)$
defined as $\mathbf{\bar{C}}_{Z,R}\left(\omega\right)$ in Eq. \eqref{eq:CZR-substitution}
we can re-write $\mathcal{C}_{V}$ as

\begin{eqnarray}
\mathcal{C}_{V}\left(\omega\right) & = & \mathbf{\bar{m}}\left.\mathbf{\bar{C}}_{Z}\left(\omega\right)\right\rfloor _{\mathbf{C}_{R}\leftarrow\left(i\omega\right)^{-1}\mathbf{R}^{-1}}\mathbf{\bar{m}}_{V}^{T}\\
 & = & \left(\bar{\mathbf{m}}_{0}+\bar{\mathbf{m}}_{R}\left(\omega\right)\right)\mathbf{\bar{C}}_{Z,R}\left(\omega\right)\left(\bar{\mathbf{m}}_{V,0}+\bar{\mathbf{m}}_{V,R}\left(\omega\right)\right)^{T}\label{eq:curly-CV-update}
\end{eqnarray}
Unlike the dissipation matrix $\mathcal{C}_{Z}$ we will keep the
full expression for the term $\mathcal{C}_{V}\left(\omega\right)$
since it is not a dissipative term. However we will combine the frequency
dependent term $\mathbf{C}_{V,R}\left(\omega\right)$ defined in Eq.
\eqref{eq:CVR-definition} with $\mathcal{C}_{V}\left(\omega\right)$
in Eq. \eqref{eq:curly-CV-update} to define

\begin{equation}
\mathcal{C}_{V,R}\left(\omega\right)=\mathbf{C}_{V,R}\left(\omega\right)+\mathcal{C}_{V}\left(\omega\right)\label{eq:curly-CVR-definition}
\end{equation}

Using the matrices defined in Eq. \eqref{eq:Burkard-CZr} and \eqref{eq:curly-CVR-definition}
above we write again the equation of motion in Eq. \eqref{eq:Burkard-equation-of-motion-BKD-type-2}
as

\begin{equation}
\mathcal{C}_{0}*\ddot{\mathbf{\Phi}}=-\mathbf{I}_{c}\mathbf{sin}\boldsymbol{\varphi}-\mathcal{R}^{-1}\dot{\mathbf{\Phi}}-\mathbf{M}_{0}\mathbf{\Phi}-\mathcal{\ddot{C}}_{Z,r}*\mathbf{\Phi}-\dot{\mathcal{C}}_{V,R}*\mathbf{V}_{V}-\mathbf{N}\mathbf{\Phi}_{x}\label{eq:Burkard-equation-of-motion-BKD-type-3}
\end{equation}

The Hamiltonian of the system described by the equation of motion
in Eq. \eqref{eq:Burkard-equation-of-motion-BKD-type-3} is given
by

\begin{equation}
\mathcal{H}_{S}=\frac{1}{2}\left[\mathbf{Q}-\left(\mathbf{C}_{V,0}+\mathcal{C}_{V,R}\left(t\right)\right)*\mathbf{V}_{V}\left(t\right)\right]^{T}\mathcal{C}_{0}^{-1}\left[\mathbf{Q}-\left(\mathbf{C}_{V,0}+\mathcal{C}_{V,R}\left(t\right)\right)*\mathbf{V}_{V}\left(t\right)\right]+U\left(\mathbf{\Phi}\right)\label{eq:Burkard-Hamiltonian-extended}
\end{equation}
where the potential function $U\left(\mathbf{\Phi}\right)$ is defined
in Eq. \eqref{eq:Burkard-potential}. This time-dependent Hamiltonian
is the extension of the Hamiltonian given in Eq. (36) of Burkard to
a time-dependent voltage source vector $\mathbf{V}_{V}\left(t\right)$.

\subsection{Derivation of the Effective Kirchhoff's Voltage Law}

\subsubsection{\label{sub:Effective-Kirchhoff's-voltage-law-one-port}Effective
Kirchhoff's Voltage Law for the One-Port State-Space Brune Circuit}

Here we present the effective Kirchhoff analysis for the Kirchhoff's
voltage law for the one-port state-space Brune circuit in Fig. \eqref{fig:Modified-state-space-Brune-circuit}.
The treatment here will be along similar lines with the analysis we
did in Section \eqref{sub:Quantization-state-space-Brune-circuit-1-port}
to get the effective Kirchhoff's current law.

We start by writing the Kirchhoff's voltage law for the circuit in
Fig. \eqref{fig:Modified-state-space-Brune-circuit}

\[
\mathbf{F}^{\mathrm{T}}\mathbf{V}_{\mathrm{tr}}=\mathbf{V}_{\mathrm{ch}}
\]
where the voltages are partitioned as

\begin{equation}
\mathbf{V}_{\mathrm{tr}}=\left(V_{J},\mathbf{V}_{L},\mathbf{V}_{Z},\mathbf{V}_{T}^{\left(tr\right)}\right)\label{eq:tree-voltage-vector-1-port-ss-appendix}
\end{equation}

\begin{equation}
\mathbf{V}_{\mathrm{ch}}=\left(\mathbf{V}_{C},\mathbf{V}_{T}^{\left(ch\right)}\right)\label{eq:chord-voltage-vector-1-port-ss-appendix}
\end{equation}
and the loop matrix matrix $\mathbf{F}^{\mathrm{T}}$ is partitioned
by Eq. \eqref{eq:F-matrix-one-port-ss} as

\[
\mathbf{F}^{T}=\left(\begin{array}{cccc}
\mathbf{F}_{JC}^{T} & \mathbf{F}_{LC}^{T} & \mathbf{F}_{ZC}^{T} & \mathbf{F}_{TC}^{T}\\
\mathbf{F}_{JT}^{T} & \mathbf{F}_{LT}^{T} & \mathbf{F}_{ZT}^{T} & \mathbf{F}_{TT}^{T}
\end{array}\right)
\]

We will show that we can get an effective Kirchhoff's voltage law

\[
\left(\mathbf{F}^{\mathrm{T}}\right)^{eff}\mathbf{V}_{\mathrm{tr}}^{eff}=\mathbf{V}_{\mathrm{ch}}^{eff}
\]
for some effective loop matrix $\left(\mathbf{F}^{\mathrm{T}}\right)^{eff}$
partitioned as

\[
\left(\mathbf{F}^{\mathrm{T}}\right)^{eff}=\left(\begin{array}{ccc}
\left(\mathbf{F}_{JC}^{T}\right)^{eff} & \left(\mathbf{F}_{LC}^{T}\right)^{eff} & \left(\mathbf{F}_{ZC}^{T}\right)^{eff}\end{array}\right)
\]
by eliminating the transformer branches' voltage variables

\begin{eqnarray}
\mathbf{V}_{\mathrm{tr}}^{eff} & = & \left(V_{J},\mathbf{V}_{L},\mathbf{V}_{Z}\right)\\
\mathbf{V}_{\mathrm{ch}}^{eff} & = & \mathbf{V}_{C}
\end{eqnarray}

First we note the following in the circuit in Fig. \eqref{fig:Modified-state-space-Brune-circuit}

\begin{equation}
\mathbf{V}_{T}^{(ch)}=-\mathbf{V}_{L}
\end{equation}
Hence by the transformer voltage relations we have

\begin{eqnarray}
\mathbf{V}_{T}^{(tr)} & = & \mathbf{N}\mathbf{V}_{T}^{(ch)}\\
 & =- & \mathbf{N}\mathbf{V}_{L}\label{eq:tree-transformer-voltages-one-port-ss}
\end{eqnarray}
where $\mathbf{N}$ is the diagonal turns ratio matrix defined in
Eq. \eqref{eq:turns-ratio-matrix-one-port-ss}.

Now writing voltages of the chord capacitors as a function of voltages
of tree branches in the circuit in Fig. \eqref{fig:Modified-state-space-Brune-circuit}
we get

\begin{equation}
\mathbf{V}_{C}=\mathbf{F}_{JC}^{T}V_{J}+\mathbf{F}_{LC}^{T}\mathbf{V}_{L}+\mathbf{F}_{ZC}^{T}\mathbf{V}_{Z}+\mathbf{F}_{TC}^{T}\mathbf{V}_{T}^{\left(tr\right)}\label{eq:capacitor-voltage-one-port-ss}
\end{equation}
with

\begin{eqnarray}
\mathbf{F}_{JC}^{T} & = & \left(\begin{array}{c}
1\\
\vdots\\
1
\end{array}\right)\\
\mathbf{F}_{LC}^{T} & = & \left(\begin{array}{cccc}
1\\
1 & 1 & \boldsymbol{0}\\
\vdots & \vdots & \ddots\\
1 & 1 & \cdots & 1\\
1 & 1 & \cdots & 1
\end{array}\right)\\
\mathbf{F}_{ZC}^{T} & = & \left(\begin{array}{cccc}
1\\
1 & 1 & \boldsymbol{0}\\
\vdots & \vdots & \ddots\\
1 & 1 & \cdots & 1\\
1 & 1 & \cdots & 1
\end{array}\right)\\
\mathbf{F}_{TC}^{T} & = & \left(\begin{array}{ccc}
0 &  & \boldsymbol{0}\\
1 & \ddots\\
\vdots & \ddots & 0\\
1 & \cdots & 1
\end{array}\right)
\end{eqnarray}
where $\mathbf{F}_{JC}^{T}$ is a vector of length $\left(M+1\right)$
and $\mathbf{F}_{LC}^{T}$, $\mathbf{F}_{ZC}^{T}$, $\mathbf{F}_{TC}^{T}$
are matrices of size $\left(M+1\right)\times M$.

Using Eqs. \eqref{eq:tree-transformer-voltages-one-port-ss} and \eqref{eq:capacitor-voltage-one-port-ss}
we get

\begin{equation}
\mathbf{V}_{C}=\mathbf{F}_{JC}^{T}V_{J}+(\mathbf{F}_{LC}^{T}-\mathbf{F}_{TC}^{T}\mathbf{N})\mathbf{V}_{L}+\mathbf{F}_{ZC}^{T}\mathbf{V}_{Z}
\end{equation}
from which we conclude

\begin{eqnarray}
\left(\mathbf{F}_{JC}^{T}\right)^{eff} & = & \mathbf{F}_{JC}^{T}\label{eq:effective-Kirchhoff-voltage-submatrices-1-port-ss-1st}\\
\left(\mathbf{F}_{LC}^{T}\right)^{eff} & = & \mathbf{F}_{LC}^{T}-\mathbf{F}_{TC}^{T}\mathbf{N}\\
\left(\mathbf{F}_{ZC}^{T}\right)^{eff} & = & \mathbf{F}_{ZC}^{T}\label{eq:effective-Kirchhoff-voltage-submatrices-1-port-ss-last}
\end{eqnarray}
and the effective Kirchhoff's voltage law

\begin{equation}
\left(\mathbf{F}^{T}\right)^{eff}\mathbf{V}_{\mathrm{tr}}^{eff}=\mathbf{V}_{\mathrm{ch}}^{eff}
\end{equation}

Comparing above Eqs. \eqref{eq:effective-Kirchhoff-voltage-submatrices-1-port-ss-1st}-\eqref{eq:effective-Kirchhoff-voltage-submatrices-1-port-ss-last}
with $\eqref{eq:effective-FLC-one-port-ss}$, \eqref{eq:effective-FJC-one-port-ss}
and \eqref{eq:effective-FZC-one-port-ss} of Section \eqref{sub:Quantization-state-space-Brune-circuit-1-port}
we conclude

\begin{eqnarray}
\left(\mathbf{F}_{JC}^{T}\right)^{eff} & = & \left(\mathbf{F}_{JC}^{eff}\right)^{T}\\
\left(\mathbf{F}_{LC}^{T}\right)^{eff} & = & \left(\mathbf{F}_{LC}^{eff}\right)^{T}\\
\left(\mathbf{F}_{ZC}^{T}\right)^{eff} & = & \left(\mathbf{F}_{ZC}^{eff}\right)^{T}
\end{eqnarray}
Hence

\begin{equation}
\left(\mathbf{F}^{T}\right)^{eff}=\left(\mathbf{F}^{eff}\right)^{T}
\end{equation}

\subsubsection{\label{sub:Effective-Kirchhoff's-voltage-law-multiport}Effective
Kirchhoff's Voltage Law for the Multiport Brune Circuit}

In this section we will derive an effective Kirchhoff's voltage law
for multiport Brune circuit in Fig. \eqref{fig:Multiport-Brune-Circuit}.
We write Kirchhoff's voltage law

\begin{equation}
\mathbf{F}^{\mathrm{T}}\mathbf{V}_{\mathrm{tr}}=\mathbf{V}_{\mathrm{ch}}\label{eq:KVL-app}
\end{equation}
with partioning in Eqs. \eqref{eq:tree-voltage-vector-multiport}
and \eqref{eq:chord-voltage-vector-multiport} for voltage vectors
$\mathbf{V}_{\mathrm{tr}}$ and $\mathbf{V}_{\mathrm{ch}}$ respectively.
We note that the relation in Eq. \eqref{eq:KVL-app} derives from
a graph theoretical analysis \citep{Burkard} of the multiport Brune
network.

We will further partition voltage vectors $\mathbf{V}_{T}^{(tr)}$
and $\mathbf{V}_{T}^{(ch)}$ for the transformer branches as follows

\begin{eqnarray}
\mathbf{V}_{T}^{\left(tr\right)} & = & \left(\mathbf{V}_{n}^{(R)},\mathbf{V}_{\mathbf{T}}^{(R)},\mathbf{V}_{\boldsymbol{\nu}}^{(R)}\right)\\
\mathbf{V}_{T}^{\left(ch\right)} & = & \left(\mathbf{V}_{n}^{(L)},\mathbf{V}_{\mathbf{T}}^{(L)},\mathbf{V}_{\boldsymbol{\nu}}^{(L)}\right)
\end{eqnarray}
where

\begin{eqnarray}
\mathbf{V}_{n}^{(R)} & = & \left(V_{n_{1}}^{(R)},\ldots,V_{n_{M}}^{(R)}\right)\\
\mathbf{V}_{\mathbf{T}}^{(R)} & = & \left(\mathbf{V}_{\mathbf{T}_{1}}^{(R)},\ldots,\mathbf{V}_{\mathbf{T}_{M+1}}^{(R)}\right)\\
\mathbf{V}_{\boldsymbol{\nu}}^{(R)} & = & \left(\mathbf{V}_{\boldsymbol{\nu}_{1}}^{(R)},\ldots,\mathbf{V}_{\boldsymbol{\nu}_{M}}^{(R)}\right)
\end{eqnarray}
and

\begin{eqnarray}
\mathbf{V}_{n}^{(L)} & = & \left(V_{n_{1}}^{(L)},\ldots,V_{n_{M}}^{(L)}\right)\\
\mathbf{V}_{\mathbf{T}}^{(L)} & = & \left(\mathbf{V}_{\mathbf{T}_{1}}^{(L)},\ldots,\mathbf{V}_{\mathbf{T}_{M+1}}^{(L)}\right)\\
\mathbf{V}_{\boldsymbol{\nu}}^{(L)} & = & \left(V_{\boldsymbol{\nu}_{1}}^{(L)},\ldots,V_{\boldsymbol{\nu}_{M}}^{(L)}\right)
\end{eqnarray}
with

\begin{eqnarray}
\mathbf{V}_{\mathbf{T}_{j}}^{(L)(R)} & = & \left(\begin{array}{c}
V_{\mathbf{T}_{j},1}^{(L)(R)}\\
\vdots\\
V_{\mathbf{T}_{j},N}^{(L)(R)}
\end{array}\right)\\
\mathbf{V}_{\boldsymbol{\nu}_{j}}^{(R)} & = & \left(\begin{array}{c}
V_{\boldsymbol{\nu}_{j},2}^{(R)}\\
\vdots\\
V_{\boldsymbol{\nu}_{j},N}^{(R)}
\end{array}\right)
\end{eqnarray}
where $\mathbf{V}_{\mathbf{T}_{j}}^{(L)(R)}$ are vectors of length
$N$ for $1\leq j\leq M+1$ and $\mathbf{V}_{\boldsymbol{\nu}_{j}}^{(R)}$
are vectors of length $\left(N-1\right)$ for $1\leq j\leq M$.

We will move in the opposite direction to the direction we have chosen
in Section \eqref{sec:Quantization-of-the-multiport-Brune-circuit}
through the multiport Brune circuit while deriving effective loop
matrices. That is we will start at the leftmost part of the multiport
Brune circuit in Fig. \eqref{fig:Multiport-Brune-Circuit} and move
to the right. We again assume that shunt resistors $R_{j}$'s in the
last stage are replaced by capacitors $C_{R_{j}}$'s for $1\leq j\leq N$
as shown in Fig. \eqref{fig:Last-stage-multiport-Brune} and that
all of the ports of the multiport Brune circuit are shunted by Josephson
junctions as shown in Fig. \eqref{fig:JJ-shunting-all-ports}; we
get the relation

\begin{equation}
\mathbf{V}_{\mathbf{T}_{1}}^{\left(L\right)}=\mathbf{V}_{J}\label{eq:multiport-Brune-JJ-voltage-relation-1st-stage}
\end{equation}
where $\mathbf{V}_{J}$ is the vector holding the voltages across
the Josephson junctions shunting the ports of the multiport Brune
circuit. We again assume the following ordering for the capacitors

\begin{equation}
\left\{ C_{1},\ldots,C_{M},C_{R_{1}},\ldots,C_{R_{N}}\right\} 
\end{equation}

The voltages of inter-stage transformers are given by Eq. \eqref{eq:interstage-T-voltage-relation}

\begin{equation}
\mathbf{V}_{\mathbf{T}_{j}}^{\left(R\right)}=\mathbf{T}_{j}^{T}\mathbf{V}_{\mathbf{T}_{j}}^{\left(L\right)}\label{eq:multiport-Brune-Belevitch-voltage-relations}
\end{equation}
for $1\leq j\leq M+1$, where $\mathbf{T}_{j}$ is the $\left(N\times N\right)$
Belevitch transformer matrix of the $j^{th}$ stage. The voltages
of consecutive inter-stage transformers are related by

\begin{equation}
\mathbf{V}_{\mathbf{T}_{j+1}}^{\left(L\right)}=\mathbf{A}_{j}^{T}\mathbf{e}_{1}V_{r_{j}}+\mathbf{\boldsymbol{\upsilon}}_{j}^{T}V_{L_{j}}+\mathbf{A}_{j}^{T}\mathbf{V}_{\mathbf{T}_{j}}^{\left(R\right)}\label{eq:consecutive-Belevitch-voltage-relations}
\end{equation}
for $1\leq j\leq M$, where $\mathbf{e}_{1}=\left(\begin{array}{cccc}
1 & 0 & \ldots & 0\end{array}\right)^{T}$ is the unit vector, $\mathbf{A}_{j}$ is the $\left(N\times N\right)$
matrix defined in Eq. \eqref{eq:Aj-matrix} and $\boldsymbol{\upsilon}_{j}$
is the row vector defined in Eq. \eqref{eq:v-vector}.

We can write the voltage $V_{C_{j}}$ across the capacitor $C_{j}$
at the $j^{th}$ stage as

\begin{equation}
V_{C_{j}}=V_{L_{j}}+V_{r_{j}}+\mathbf{e}_{1}^{T}\mathbf{V}_{\mathbf{T}_{j}}^{(R)}\label{eq:VCj-relation-multiport}
\end{equation}

We are now going to iterate over the index $j$ starting at $j=1$
through the stages in the multiport Brune circuit using the Eqs. \eqref{eq:multiport-Brune-JJ-voltage-relation-1st-stage},
\eqref{eq:multiport-Brune-Belevitch-voltage-relations}, \eqref{eq:consecutive-Belevitch-voltage-relations}
and \eqref{eq:VCj-relation-multiport}

\begin{eqnarray*}
\mathbf{V}_{\mathbf{T}_{1}}^{(R)} & = & \mathbf{T}_{1}^{T}\mathbf{V}_{\mathbf{T}_{1}}^{\left(L\right)}=\mathbf{T}_{1}^{T}\mathbf{V}_{J}\\
\mathbf{V}_{\mathbf{T}_{2}}^{\left(L\right)} & = & \mathbf{A}_{1}^{T}\mathbf{e}_{1}V_{r_{1}}+\mathbf{\boldsymbol{\upsilon}}_{1}^{T}V_{L_{1}}+\mathbf{A}_{1}^{T}\mathbf{V}_{\mathbf{T}_{1}}^{\left(R\right)}=\mathbf{A}_{1}^{T}\mathbf{e}_{1}V_{r_{1}}+\mathbf{\boldsymbol{\upsilon}}_{1}^{T}V_{L_{1}}+\mathbf{A}_{1}^{T}\mathbf{T}_{1}^{T}\mathbf{V}_{J}\\
V_{C_{1}} & = & V_{L_{1}}+V_{r_{1}}+\mathbf{e}_{1}^{T}\mathbf{V}_{\mathbf{T}_{1}}^{(R)}=V_{L_{1}}+V_{r_{1}}+\mathbf{e}_{1}^{T}\mathbf{T}_{1}^{T}\mathbf{V}_{J}\\
\mathbf{V}_{\mathbf{T}_{2}}^{\left(R\right)} & = & \mathbf{T}_{2}^{T}\mathbf{V}_{\mathbf{T}_{2}}^{\left(L\right)}=\mathbf{T}_{2}^{T}\mathbf{A}_{1}^{T}\mathbf{e}_{1}V_{r_{1}}+\mathbf{T}_{2}^{T}\mathbf{\boldsymbol{\upsilon}}_{1}^{T}V_{L_{1}}+\mathbf{T}_{2}^{T}\mathbf{A}_{1}^{T}\mathbf{T}_{1}^{T}\mathbf{V}_{J}\\
\mathbf{V}_{\mathbf{T}_{3}}^{\left(L\right)} & = & \mathbf{A}_{2}^{T}\mathbf{e}_{1}V_{r_{2}}+\mathbf{\boldsymbol{\upsilon}}_{2}^{T}V_{L_{2}}+\mathbf{A}_{2}^{T}\mathbf{V}_{\mathbf{T}_{2}}^{\left(R\right)}\\
 & = & \mathbf{A}_{2}^{T}\mathbf{T}_{2}^{T}\mathbf{A}_{1}^{T}\mathbf{e}_{1}V_{r_{1}}+\mathbf{A}_{2}^{T}\mathbf{e}_{1}V_{r_{2}}+\mathbf{A}_{2}^{T}\mathbf{T}_{2}^{T}\mathbf{\boldsymbol{\upsilon}}_{1}^{T}V_{L_{1}}+\mathbf{\boldsymbol{\upsilon}}_{2}^{T}V_{L_{2}}+\\
 &  & +\mathbf{A}_{2}^{T}\mathbf{T}_{2}^{T}\mathbf{A}_{1}^{T}\mathbf{T}_{1}^{T}\mathbf{V}_{J}\\
V_{C_{2}} & = & V_{L_{2}}+V_{r_{2}}+\mathbf{e}_{1}^{T}\mathbf{V}_{\mathbf{T}_{2}}^{(R)}\\
 & = & \mathbf{e}_{1}^{T}\mathbf{T}_{2}^{T}\mathbf{A}_{1}^{T}\mathbf{e}_{1}V_{r_{1}}+V_{r_{2}}+\mathbf{e}_{1}^{T}\mathbf{T}_{2}^{T}\mathbf{\boldsymbol{\upsilon}}_{1}^{T}V_{L_{1}}+V_{L_{2}}+\\
 &  & +\mathbf{e}_{1}^{T}\mathbf{T}_{2}^{T}\mathbf{A}_{1}^{T}\mathbf{T}_{1}^{T}\mathbf{V}_{J}
\end{eqnarray*}

\begin{eqnarray}
\mathbf{V}_{\mathbf{T}_{3}}^{\left(R\right)} & = & \mathbf{T}_{3}^{T}\mathbf{V}_{\mathbf{T}_{3}}^{\left(L\right)}\nonumber \\
 & = & \mathbf{T}_{3}^{T}\mathbf{A}_{2}^{T}\mathbf{T}_{2}^{T}\mathbf{A}_{1}^{T}\mathbf{e}_{1}V_{r_{1}}+\mathbf{T}_{3}^{T}\mathbf{A}_{2}^{T}\mathbf{e}_{1}V_{r_{2}}+\nonumber \\
 &  & +\mathbf{T}_{3}^{T}\mathbf{A}_{2}^{T}\mathbf{T}_{2}^{T}\mathbf{\boldsymbol{\upsilon}}_{1}^{T}V_{L_{1}}+\mathbf{T}_{3}^{T}\mathbf{\boldsymbol{\upsilon}}_{2}^{T}V_{L_{2}}+\mathbf{T}_{3}^{T}\mathbf{A}_{2}^{T}\mathbf{T}_{2}^{T}\mathbf{A}_{1}^{T}\mathbf{T}_{1}^{T}\mathbf{V}_{J}\nonumber \\
\mathbf{V}_{\mathbf{T}_{4}}^{\left(L\right)} & = & \mathbf{A}_{3}^{T}\mathbf{e}_{1}V_{r_{3}}+\mathbf{\boldsymbol{\upsilon}}_{3}^{T}V_{L_{3}}+\mathbf{A}_{3}^{T}\mathbf{V}_{\mathbf{T}_{3}}^{\left(R\right)}\nonumber \\
 & = & \mathbf{A}_{3}^{T}\mathbf{T}_{3}^{T}\mathbf{A}_{2}^{T}\mathbf{T}_{2}^{T}\mathbf{A}_{1}^{T}\mathbf{e}_{1}V_{r_{1}}+\mathbf{A}_{3}^{T}\mathbf{T}_{3}^{T}\mathbf{A}_{2}^{T}\mathbf{e}_{1}V_{r_{2}}+\mathbf{A}_{3}^{T}\mathbf{e}_{1}V_{r_{3}}+\nonumber \\
 &  & +\mathbf{A}_{3}^{T}\mathbf{T}_{3}^{T}\mathbf{A}_{2}^{T}\mathbf{T}_{2}^{T}\mathbf{\boldsymbol{\upsilon}}_{1}^{T}V_{L_{1}}+\mathbf{A}_{3}^{T}\mathbf{T}_{3}^{T}\mathbf{\boldsymbol{\upsilon}}_{2}^{T}V_{L_{2}}+\mathbf{\boldsymbol{\upsilon}}_{3}^{T}V_{L_{3}}+\mathbf{A}_{3}^{T}\mathbf{T}_{3}^{T}\mathbf{A}_{2}^{T}\mathbf{T}_{2}^{T}\mathbf{A}_{1}^{T}\mathbf{T}_{1}^{T}\mathbf{V}_{J}\nonumber \\
V_{C_{3}} & = & V_{L_{3}}+V_{r_{3}}+\mathbf{e}_{1}^{T}\mathbf{V}_{\mathbf{T}_{3}}^{(R)}\nonumber \\
 & = & \mathbf{e}_{1}^{T}\mathbf{T}_{3}^{T}\mathbf{A}_{2}^{T}\mathbf{T}_{2}^{T}\mathbf{A}_{1}^{T}\mathbf{e}_{1}V_{r_{1}}+\mathbf{e}_{1}^{T}\mathbf{T}_{3}^{T}\mathbf{A}_{2}^{T}\mathbf{e}_{1}V_{r_{2}}+V_{r_{3}}+\nonumber \\
 &  & +\mathbf{e}_{1}^{T}\mathbf{T}_{3}^{T}\mathbf{A}_{2}^{T}\mathbf{T}_{2}^{T}\mathbf{\boldsymbol{\upsilon}}_{1}^{T}V_{L_{1}}+\mathbf{e}_{1}^{T}\mathbf{T}_{3}^{T}\mathbf{\boldsymbol{\upsilon}}_{2}^{T}V_{L_{2}}+V_{L_{3}}+\nonumber \\
 &  & +\mathbf{e}_{1}^{T}\mathbf{T}_{3}^{T}\mathbf{A}_{2}^{T}\mathbf{T}_{2}^{T}\mathbf{A}_{1}^{T}\mathbf{T}_{1}^{T}\mathbf{V}_{J}\label{eq:multiport-iteration-for-voltages-last-line}\\
\vdots & \vdots & \vdots\nonumber 
\end{eqnarray}
Hence from the above relations we define the following for $1\leq j\leq M$
and $1\leq k\leq M$

\begin{equation}
\begin{cases}
\left(\mathbf{F}_{LC}^{T}\right)^{eff}\left(j,k\right)=0 & for\; j<k\leq M\\
\left(\mathbf{F}_{LC}^{T}\right)^{eff}\left(j,k\right)=1 & for\; k=j\\
\left(\mathbf{F}_{LC}^{T}\right)^{eff}\left(j,k\right)=\mathbf{e}_{1}^{T}\mathbf{T}_{k+1}^{T}\mathbf{\boldsymbol{\upsilon}}_{k}^{T} & for\; k=j-1\\
\left(\mathbf{F}_{LC}^{T}\right)^{eff}\left(j,k\right)=\mathbf{e}_{1}^{T}\mathbf{T}_{j}^{T}\mathbf{A}_{j-1}^{T}\mathbf{T}_{j-1}^{T}\ldots\mathbf{A}_{k+1}^{T}\mathbf{T}_{k+1}^{T}\mathbf{\boldsymbol{\upsilon}}_{k}^{T} & for\;1\leq k\leq j-2
\end{cases}\label{eq:eff-submatrix-FLC-voltage}
\end{equation}

\begin{equation}
\begin{cases}
\left(\mathbf{F}_{ZC}^{T}\right)^{eff}\left(j,k\right)=0 & for\; j<k\leq M\\
\left(\mathbf{F}_{ZC}^{T}\right)^{eff}\left(j,k\right)=1 & for\; k=j\\
\left(\mathbf{F}_{ZC}^{T}\right)^{eff}\left(j,k\right)=\mathbf{e}_{1}^{T}\mathbf{T}_{j}^{T}\mathbf{A}_{j-1}^{T}\mathbf{T}_{j-1}^{T}\ldots\mathbf{A}_{k+1}^{T}\mathbf{T}_{k+1}^{T}\mathbf{A}_{k}^{T}\mathbf{e}_{1} & for\;1\leq k\leq j-1
\end{cases}\label{eq:eff-submatrix-FZC-voltage}
\end{equation}
and for $1\leq k\leq N$

\begin{equation}
\begin{cases}
\left(\mathbf{F}_{JC}^{T}\right)^{eff}\left(j,k\right)=\mathbf{e}_{1}^{T}\mathbf{T}_{1}^{T}\mathbf{e}_{k} & for\; j=1\\
\left(\mathbf{F}_{JC}^{T}\right)^{eff}\left(j,k\right)=\mathbf{e}_{1}^{T}\mathbf{T}_{j}^{T}\mathbf{A}_{j-1}^{T}\mathbf{T}_{j-1}^{T}\ldots\mathbf{A}_{1}^{T}\mathbf{T}_{1}^{T}\mathbf{e}_{k} & for\;1<j\leq M
\end{cases}\label{eq:eff-submatrix-FJC-voltage}
\end{equation}

To compute effective loop submatrices for $j>M$ we note from Eq.
\eqref{eq:multiport-iteration-for-voltages-last-line} the following

\begin{eqnarray}
\mathbf{V}_{\mathbf{T}_{M+1}}^{\left(L\right)} & = & \mathbf{A}_{M}^{T}\mathbf{T}_{M}^{T}\ldots\mathbf{A}_{2}^{T}\mathbf{T}_{2}^{T}\mathbf{A}_{1}^{T}\mathbf{e}_{1}V_{r_{1}}+\ldots+\mathbf{A}_{M}^{T}\mathbf{e}_{1}V_{r_{M}}+\nonumber \\
 &  & +\mathbf{A}_{M}^{T}\mathbf{T}_{M}^{T}\ldots\mathbf{A}_{2}^{T}\mathbf{T}_{2}^{T}\mathbf{\boldsymbol{\upsilon}}_{1}^{T}V_{L_{1}}+\ldots+\mathbf{\boldsymbol{\upsilon}}_{M}^{T}V_{L_{M}}+\nonumber \\
 &  & +\mathbf{A}_{M}^{T}\mathbf{T}_{M}^{T}\ldots\mathbf{A}_{1}^{T}\mathbf{T}_{1}^{T}\mathbf{V}_{J}\label{eq:multiport-Brune-left-voltage-last-stage-transformer}
\end{eqnarray}

Writing also the voltage relations for the last stage transformer
using Eq. \eqref{eq:multiport-Brune-Belevitch-voltage-relations}

\begin{equation}
\mathbf{V}_{\mathbf{T}_{M+1}}^{\left(R\right)}=\mathbf{T}_{M+1}^{T}\mathbf{V}_{\mathbf{T}_{M+1}}^{\left(L\right)}
\end{equation}
And noting

\begin{eqnarray}
\mathbf{V}_{C_{R}} & = & \mathbf{V}_{\mathbf{T}_{M+1}}^{\left(R\right)}\\
 & = & \mathbf{T}_{M+1}^{T}\mathbf{V}_{\mathbf{T}_{M+1}}^{\left(L\right)}
\end{eqnarray}
We conclude using Eq. \eqref{eq:multiport-Brune-left-voltage-last-stage-transformer}

\begin{eqnarray}
\mathbf{V}_{C_{R}} & = & \mathbf{T}_{M+1}^{T}\mathbf{A}_{M}^{T}\mathbf{T}_{M}^{T}\ldots\mathbf{A}_{2}^{T}\mathbf{T}_{2}^{T}\mathbf{A}_{1}^{T}\mathbf{e}_{1}V_{r_{1}}\ldots+\mathbf{T}_{M+1}^{T}\mathbf{A}_{M}^{T}\mathbf{e}_{1}V_{r_{M}}+\nonumber \\
 &  & +\mathbf{T}_{M+1}^{T}\mathbf{A}_{M}^{T}\mathbf{T}_{M}^{T}\ldots\mathbf{A}_{2}^{T}\mathbf{T}_{2}^{T}\mathbf{\boldsymbol{\upsilon}}_{1}^{T}V_{L_{1}}\ldots+\mathbf{T}_{M+1}^{T}\mathbf{\boldsymbol{\upsilon}}_{M}^{T}V_{L_{M}}+\nonumber \\
 &  & +\mathbf{T}_{M+1}^{T}\mathbf{A}_{M}^{T}\mathbf{T}_{M}^{T}\ldots\mathbf{A}_{1}^{T}\mathbf{T}_{1}^{T}\mathbf{V}_{J}\label{eq:multiport-Brune-voltage-relations-for-CR}
\end{eqnarray}

From Eq. \eqref{eq:multiport-Brune-voltage-relations-for-CR} above
we define for $1\leq k\leq M$

\begin{equation}
\begin{cases}
\left(\mathbf{F}_{LC}^{T}\right)^{eff}\left(j,k\right)=\mathbf{e}_{j-M}^{T}\mathbf{T}_{M+1}^{T}\mathbf{A}_{M}^{T}\mathbf{T}_{M}^{T}\ldots\mathbf{A}_{k+1}^{T}\mathbf{T}_{k+1}^{T}\mathbf{\boldsymbol{\upsilon}}_{k}^{T} & for\; k<M\; and\; M+1\leq j\leq M+N\\
\left(\mathbf{F}_{LC}^{T}\right)^{eff}\left(j,k\right)=\mathbf{e}_{j-M}^{T}\mathbf{T}_{M+1}^{T}\mathbf{\boldsymbol{\upsilon}}_{M}^{T} & for\; k=M\; and\; M+1\leq j\leq M+N
\end{cases}\label{eq:eff-submatrix-FLC-voltage-CR}
\end{equation}

\begin{equation}
\begin{cases}
\left(\mathbf{F}_{ZC}^{T}\right)^{eff}\left(j,k\right)=\mathbf{e}_{j-M}^{T}\mathbf{T}_{M+1}^{T}\mathbf{A}_{M}^{T}\mathbf{T}_{M}^{T}\ldots\mathbf{A}_{k+1}^{T}\mathbf{T}_{k+1}^{T}\mathbf{A}_{k}^{T}\mathbf{e}_{1} & for\; M+1\leq j\leq M+N\end{cases}\label{eq:eff-submatrix-FZC-voltage-CR}
\end{equation}
where $\mathbf{e}_{j}$ is the unit vector of length $N$ non-zero
only at its $j^{th}$ entry such that $\mathbf{e}_{j}\left(k\right)=0$
for $k\neq j$ and $\mathbf{e}_{j}\left(j\right)=1$.

And for $1\leq k\leq N$

\begin{equation}
\begin{cases}
\left(\mathbf{F}_{JC}^{T}\right)^{eff}\left(j,k\right)=\mathbf{e}_{j-M}^{T}\mathbf{T}_{M+1}^{T}\mathbf{A}_{M}^{T}\mathbf{T}_{M}^{T}\ldots\mathbf{A}_{1}^{T}\mathbf{T}_{1}^{T}\mathbf{e}_{k} & for\; M+1\leq j\leq M+N\end{cases}\label{eq:eff-submatrix-FJC-voltage-CR}
\end{equation}

Hence from Eqs. \eqref{eq:eff-submatrix-FLC-voltage}, \eqref{eq:eff-submatrix-FZC-voltage},
\eqref{eq:eff-submatrix-FJC-voltage} and \eqref{eq:eff-submatrix-FLC-voltage-CR},
\eqref{eq:eff-submatrix-FZC-voltage-CR}, \eqref{eq:eff-submatrix-FJC-voltage-CR}
we conclude

\begin{equation}
\left(\mathbf{F}^{T}\right)^{eff}\mathbf{V}_{\mathrm{tr}}^{eff}=\mathbf{V}_{\mathrm{ch}}^{eff}
\end{equation}
with

\begin{equation}
\left(\mathbf{F}^{T}\right)^{eff}=\left(\begin{array}{ccc}
\left(\mathbf{F}_{JC}^{T}\right)^{eff} & \left(\mathbf{F}_{LC}^{T}\right)^{eff} & \left(\mathbf{F}_{ZC}^{T}\right)^{eff}\end{array}\right)
\end{equation}
and comparing Eqs. \eqref{eq:eff-submatrix-FLC-voltage}, \eqref{eq:eff-submatrix-FZC-voltage},
\eqref{eq:eff-submatrix-FJC-voltage} and \eqref{eq:eff-submatrix-FLC-voltage-CR},
\eqref{eq:eff-submatrix-FZC-voltage-CR}, \eqref{eq:eff-submatrix-FJC-voltage-CR}
to Eqs. \eqref{eq:eff-FLC-submatrix-current}, \eqref{eq:eff-FZC-submatrix-current}
and \eqref{eq:eff-FJC-submatrix-current} of Section \eqref{sec:Quantization-of-the-multiport-Brune-circuit}
we conclude

\begin{equation}
\left(\mathbf{F}^{T}\right)^{eff}=\left(\mathbf{F}^{eff}\right)^{T}
\end{equation}

\newpage{}


\begin{thebibliography}{10}
\bibitem{brune-quantization-paper}Firat Solgun, David W. Abraham,
and David P. DiVincenzo, ``Blackbox quantization of superconducting
circuits using exact impedance synthesis'', Phys. Rev. B 90, 134504,
(2014).

\bibitem{BKD} G. Burkard, R. H. Koch, and D. P. DiVincenzo, ``Multi-level
quantum description of decoherence in superconducting qubits'', Phys.
Rev. B 69, 064503 (2004).

\bibitem{Burkard}G. Burkard, ``Circuit theory for decoherence in
superconducting charge qubits'', Phys. Rev. B 71, 144511 (2005).

\bibitem{blais} Alexandre Blais, Ren-Shou Huang, Andreas Wallraff,
S. M. Girvin, and R. J. Schoelkopf, ``Cavity quantum electrodynamics
for superconducting electrical circuits: an architecture for quantum
computation,'' Phys. Rev. A \textbf{69}, 062320 (2004).

\bibitem{3D} Hanhee Paik, D. I. Schuster, Lev S. Bishop, G. Kirchmair,
G. Catelani, A. P. Sears, B. R. Johnson, M. J. Reagor, L. Frunzio,
L. Glazman, S. M. Girvin, M. H. Devoret, and R. J. Schoelkopf, ``Observation
of high coherence in Josephson junction qubits measured in a three-dimensional
circuit QED architecture,'' Phys. Rev. Lett. \textbf{107}, 240501
(2011).

\bibitem{RigettiCu} Chad Rigetti, Stefano Poletto, Jay M. Gambetta,
B. L. T. Plourde, Jerry M. Chow, A. D. Corcoles, John A. Smolin, Seth
T. Merkel, J. R. Rozen, George A. Keefe, Mary B. Rothwell, Mark B.
Ketchen, and M. Steffen, ``Superconducting qubit in waveguide cavity
with coherence time approaching 0.1ms,'' Phys. Rev. B \textbf{86},
100506(R) (2012).

\bibitem{Blaisunpub} Jerome Bourassa, Jay M. Gambetta, and Alexandre
Blais, ``Multi-mode circuit quantum electrodynamics,'', Abstract
Y29.00005, APS March Meeting, Dallas, 2011.

\bibitem{Nigg} Simon E. Nigg, Hanhee Paik, Brian Vlastakis, Gerhard
Kirchmair, Shyam Shankar, Luigi Frunzio, Michel Devoret, Robert Schoelkopf,
and Steven Girvin, ``Black-box superconducting circuit quantization,''
Phys. Rev. Lett. \textbf{108}, 240502 (2012).

\bibitem{PMC}E. R. Beringer, ``Resonant Cavities as Microwave Circuit
Elements,'' in \emph{Principles of Microwave Circuits}, edited by
C. G. Montgomery, R. H. Dicke, and E. M. Purcell (MIT Radiation Laboratory,
vol. 8, 1945), p. 215, Section 7.4.

\bibitem{Foster} Foster, R. M., ``A reactance theorem'', Bell Systems
Technical Journal, vol.3, no. 2, pp. 259\textendash{}267, November
1924.

\bibitem{Brune}O. Brune, \emph{Synthesis of a finite two-terminal
network whose driving-point impedance is a prescribed function of
frequency}, Doctoral thesis, MIT, 1931.

\bibitem{HFSS}Ansys HFSS (High Frequency Structural Simulator), http://www.ansys.com.

\bibitem{Vector Fitting}B. Gustavsen and A. Semlyen, \textquotedbl{}Rational
approximation of frequency domain responses by vector fitting\textquotedbl{},
IEEE Trans. Power Delivery, vol. 14, no. 3, pp. 1052-1061, July 1999;
http://www.sintef.no/Projectweb/VECTFIT/.

\bibitem{VF Passivity Enforcement}B. Gustavsen and A. Semlyen, \textquotedbl{}Enforcing
passivity for admittance matrices approximated by rational functions\textquotedbl{},
IEEE Trans. Power Systems, vol. 16, no. 1, pp. 97-104, Feb. 2001.

\bibitem{Newcomb}Robert W. Newcomb, \emph{Linear Multiport Synthesis},
McGraw-Hill Book Company, 1966.

\bibitem{Anderson-Vongpanitlerd}Brian D. O. Anderson and Sumeth Vongpanitlerd,
\emph{Network Analysis and Synthesis, A Modern Systems Theory Approach},
Dover Publications, Inc., Mineola, New York, 2006.

\bibitem{Antoulas}Athanasios C. Antoulas, \emph{Approximation of
Large-Scale Dynamical Systems}, SIAM, Jun 25, 2009.

\bibitem{Anderson-Moylan-1975}B. D. O. Anderson and P. J. Moylan,
``The Brune Synthesis in State-Space Terms'', Circuit Theory and
Applications, Vol. 3, 193-199 (1975).

\bibitem{Wilkinson}Wilkinson, James H. (1984). \textquotedbl{}The
perfidious polynomial\textquotedbl{}. In ed. by Gene H. Golub. Studies
in Numerical Analysis. Mathematical Association of America.

\bibitem{Anderson-Riccati-Brune-1999}B. D. O. Anderson, ``Riccati
Equations, Network Theory and Brune Synthesis: Old Solutions for Contemporary
Problems'', Dynamical Systems, Control, Coding, Computer Vision,
Progress in Systems and Control Theory Volume 25, 1999, pp 1-25.

\bibitem{ideal-transformers-thesis}Delson, Jerome King, \emph{Networks
involving ideal transformers}, Dissertation (Ph.D.), California Institute
of Technology, (1953).

\bibitem{Compacting}B. Gustavsen and A. Semlyen, ``A Robust Approach
for System Identification in the Frequency Domain'', IEEE Trans.
Power Delivery, vol. 19, no. 3, pp. 1167-1173, July 2004.

\bibitem{IC-Interconnect-analysis}Mustafa Celik, Lawrence Pileggi,
Altan Odabasioglu, \emph{IC Interconnect Analysis}, Kluwer Academic
Publishers, 2002.

\bibitem{Yakubovic} V. A. Yakubovic, ``The Solution of Certain Matrix
Inequalities in Automatic Control Theory'', Doklady Akademii Nauk,
SSSR, Vol. 143, 1962, pp. 1304-1307.

\bibitem{Kalman-Positive-Real-Lemma-1} R. E. Kalman, ``Lyapunov
Functions for the Problem of Lur'e in Automatic Control'', Proceedings
of the National Academy of Sciences, Vol. 49, No. 2, Feb. 1963, pp.
201-205.

\bibitem{Kalman-Positive-Real-Lemma-2} R. E. Kalman, ``On a New
Characterization of Linear Passive Systems'', Proceedings of the
First Allerton Conference on Circuit and System Theory, University
of Illinois, Nov. 1963, pp. 456-470.

\bibitem{Popov} V. M. Popov, ``Hyperstability and Optimality of
Automatic Systems with Several Control Functions'', Rev. Roumaine
Sci. Tech. - Elektrotechn. et Energ., Vol. 9, No. 4, 1964, pp. 629-690.

\bibitem{Anderson-Positive-Real-Lemma} B. D. O. Anderson, ``A System
Theory Criterion for Positive Real Matrices'', SIAM Journal of Control,
Vol. 5, No. 2, May 1967, pp. 171-182.

\bibitem{Devoret-Les-Houches}Michel H. Devoret, in \emph{Quantum
fluctuations,} Les Houches, Elsevier, Amsterdam, (1997).

\bibitem{Burkard-Brito}Guido Burkard and Frederico Brito, ``Non-additivity
of decoherence rates in superconducting qubits'', Phys. Rev. B 72,
054528 (2005).

\bibitem{Brito} D. P. DiVincenzo, Frederico Brito, and Roger H. Koch,
``Efficient evaluation of decoherence rates in complex Josephson
circuits,'' Phys. Rev. B 74, 014514 (2006).

\bibitem{moreHcalculations} Joshua Dempster, Bo Fu, David G. Ferguson,
D. I. Schuster, and Jens Koch, ``Understanding degenerate ground
states of a protected quantum circuit in the presence of disorder,''
Phys. Rev. B 90, 094518, (2014).

\bibitem{Newcomb-Resistance-Extraction} Newcomb, R. W., ``On the
n-port Brune resistance extraction'', Trans. IEEE Circuit Theory,
vol. CT-10, no. 1, p. 125, March, 1963.

\bibitem{Newcomb-simultaneous-diagonalization}Newcomb, R. W., ``On
the simultaneous diagonalization of two semi-definite matrices,''
Quart. Appl. Math., vol. 19, pp. 144-146; July, 1961.

\bibitem{Pozar} D. Pozar, {\em Microwave Engineering}, third edition
(Wiley VCH, 2005).

\bibitem{Reagor}M. Reagor, Hanhee Paik, G. Catelani, L. Sun, C. Axline,
E. Holland, I.M. Pop, N.A. Masluk, T. Brecht, L. Frunzio, M.H. Devoret,
L.I. Glazman, R.J. Schoelkopf, ``Reaching 10 ms single photon lifetimes
for superconducting aluminum cavities'', Appl. Phys. Lett. 102, 192604
(2013).

\bibitem{Cauer-1}Cauer, W., ``\emph{Theorie der linearen Wechselstromschaltungen}'',
2nd ed., Akademie-Verlag GmbH, Berlin, 1954.

\bibitem{Cauer-2}Cauer, W., ``Synthesis of Linear Communication
Networks'', vols. I and II, 2nd ed., McGraw-Hill Book Company, New
York, 1958.

\bibitem{Belevitch-Transformer}Belevitch, V., ``Theory of 2n-terminal
networks with applications to conference telephony'', Electrical
Communication, vol. 27, no. 3, p. 233, September, 1950.

\bibitem{Tellegen-gyrator}B. D. H. Tellegen, ``The gyrator, a new
electric network element'', Philips Res. Rep. 3, pp. 81-101, April,
1948.

\bibitem{Anderson-Gyrator}B. D. O. Anderson, ``Minimal Gyrator Lossless
Synthesis'', IEEE Transactions on Circuit Theory, vol. CT-20, no.
1, January 1973.

\bibitem{Bayard} Bayard, M., ``Résolution du problème de la synthèse
des réseaux de Kirchhoff par la détermination de réseaux purement
réactifs'', Câbles et Transmission, vol. 4, no. 4, pp. 281-296, October,
1950.

\bibitem{Leroy} Leroy, R., ``Synthèse de réseaux passifs à n-paires
de bornes'', Câbles et Transmission, vol. 4, pp. 234-247, July, 1950.

\bibitem{Bayard-summary-1952}Bayard, M., ``Synthesis of N-terminal
pair networks'', Proc. Brooklyn Polytech. Symp. Mod. Network Synth.,
vol. I, pp. 66-83, 1952.

\bibitem{Bayard-book} Bayard, M., \emph{Théorie des réseaux de Kirchhoff},
Éditions de la Revue d'Optique, Paris, 1954.

\bibitem{Newcomb-multiport-Brune-notes-1} Newcomb, R. W., ``A Nonreciprocal
n-Port Brune Synthesis'', Stanford Electronics Laboratories Tech.
Rept. 2254-5, November, 1962.

\bibitem{Newcomb-multiport-Brune-notes-2} Newcomb, R. W., ``The
n-Port Brune Section Detailed'', Stanford Electronics Laboratories
Tech. Rept. 6554-7, December, 1963.

\bibitem{Oono} Oono, Y., ``Synthesis of a finite 2n-terminal network
as the extension of Brune's two-terminal network theory'', The Journal
of the Institute of Electrical Communication Engineers of Japan, vol.
31, no. 9, pp. 163-181, August, 1948 (in Japanese).

\bibitem{McMillan} McMillan, B., ``Introduction to formal realizability
theory, II'', Bell System Tech. J., vol. 31, no. 3, pp. 541-600,
May, 1952.

\bibitem{Tellegen} Tellegen, B. D. H., ``Synthesis of 2n-poles by
networks containing the minimum number of elements'', J. Math. Phys.,
vol. 32, no. 1, pp. 1-18, April, 1953.

\bibitem{Duffin} Duffin, R. J., D. Hazony and N. Morrison, ``Network
Synthesis Through Hybrid Matrices'', Case Institute of Technology
Tech. Rept. AFCRL 63-568, August 18, 1964.

\bibitem{Youla} Youla, D. C. and G. I. Zysman, ``Synthesis of Passive
Reciprocal n-Ports'', Polytechnic Institute of Brooklyn Electrophysics
Memo PIBMRI-1297-65, p. 37, November 8, 1965.

\bibitem{Belevitch-Brune} Belevitch, V., ``On the Brune process
for n-ports'', Trans. IRE Circuit Theory, vol. CT-7, no. 3, pp. 280-296,
September, 1960.

\bibitem{Kitaev-toric-code} Alexei Y. Kitaev, ``Fault-tolerant quantum
computation by anyons'', Ann. Phys., 303, 2-30, (2003).

\bibitem{DiVincenzo-skew-lattice}DiVincenzo, D. P. Fault-tolerant
architectures for superconducting qubits. Phys. Scr. T137, 014020
(2009).

\bibitem{T1-Houck}A. A. Houck, J. A. Schreier, B. R. Johnson, J.
M. Chow, Jens Koch, J. M. Gambetta, D. I. Schuster, L. Frunzio, M.
H. Devoret, S. M. Girvin, R. J. Schoelkopf, ``Controlling the spontaneous
emission of a superconducting transmon qubit'', Phys. Rev. Lett.
101, 080502 (2008).

\bibitem{high-constrast}David C. McKay, Ravi Naik, Philip Reinhold,
Lev S. Bishop, David I. Schuster, ``High contrast qubit interactions
using multimode cavity QED'', Phys. Rev. Lett. 114, 080501 (2015).

\bibitem{Bronn-Purcell-Filter}Nicholas T. Bronn, Easwar Magesan,
Nicholas A. Masluk, Jerry M. Chow, Jay M. Gambetta, Matthias Steffen,
``Reducing Spontaneous Emission in Circuit Quantum Electrodynamics
by a Combined Readout/Filter Technique'', arXiv:1504.04353.

\bibitem{Eyob-Purcell-Filter}Eyob A. Sete, John M. Martinis, Alexander
N. Korotkov, ``Quantum theory of a bandpass Purcell filter for qubit
readout'', arXiv:1504.06030.\end{thebibliography}
\end{document}